\documentclass[%
 reprint,
 amsmath,amssymb,
 aps,
]{revtex4-2}

\usepackage{graphicx}
\usepackage{dcolumn}
\usepackage{bm}

\usepackage{makecell}
\usepackage{wasysym}
\usepackage{braket}
\usepackage{mathrsfs}
\usepackage{pifont}
\usepackage{float}
\usepackage{placeins}

\usepackage{hyperref}
\hypersetup{colorlinks=true, linkcolor=blue, citecolor=blue, filecolor=magenta, urlcolor=blue}

\newcommand{\angleaverage}[1]{\ensuremath{\left\langle #1 \right\rangle}}

\newcommand{\figRef}[2]{\ref{#1}\hyperref[#1]{#2}}

\ExplSyntaxOn
\NewDocumentCommand \colVec { m }
  {
    \begin{pmatrix}
      \clist_use:nn {#1} { \\ }
    \end{pmatrix}
  }
\ExplSyntaxOff

\allowdisplaybreaks

\begin{document}

\title{Twisted Kagome Bilayers: High-Order Van Hove Singularities, Sublattice Interference, Magic Angles, and Possible Topology}

\author{David T. S. Perkins}
\email{d.perkins@lboro.ac.uk}
\affiliation{Department of Physics, Loughborough University, Loughborough LE11 3TU, England, United Kingdom}

\author{Joseph J. Betouras}
\email{J.Betouras@lboro.ac.uk}
\affiliation{Department of Physics, Loughborough University, Loughborough LE11 3TU, England, United Kingdom}


\begin{abstract}

We develop a low-energy continuum model to describe the moir\'{e} physics of heterostructures, which is a generalization of the celebrated Bistritzer-MacDonald (BM) method \cite{Bistritzer2011}. We take as an example the moir\'{e} physics of electrons in twisted bilayer kagome metals near $1/3$ filling where monolayer Dirac cones lie. We demonstrate the emergence of magic angles where significant local band flattening occurs as a high-order Van Hove singularity appears and find a momentum-dependent anti-unitary particle-hole symmetry potentially enabling stable topology. We, furthermore, show that while sublattice interference effects are present, their role is not as prominent as in monolayer kagome.

\end{abstract}

\maketitle

\textit{Introduction.} Interaction effects get boosted in correlated electronic systems when the kinetic energy of electrons is considerably reduced in comparison to the characteristic interaction energy scales. When the Brillouin zone (BZ) hosts flat bands or regions with diverging density of states (DOS), in the form of regular or high-order Van Hove singularities (HOVHS), near the Fermi surface, then the corresponding system is a natural place for correlation effects to dominate; different susceptibilities compete, leading to certain electronic instabilities and, therefore, emergent electronic phases \cite{Classen2025}. Very recently, there has been a surge of interest in systems with vanishing Fermi velocities, HOVHS, and flat bands, which can drastically alter transport, thermodynamic properties, and the phase diagram of a quantum material \cite{Efremov2019, Classen2025}. A HOVHS occurs when both the divergence and Hessian determinant of the energy dispersion vanish at some point in the BZ \cite{Chandrasekaran_catastrophe_2020, Yuan2019}. Experiments have become increasingly complicated and extremely accurate providing access to measurements not accessible before \cite{Kang2022, Hu2022, Cho2021, Chandrasekaran_NatComms_2024}.

A widely used method to achieve this Fermi velocity reduction is moir\'{e} band engineering, wherein twisting between layers of two-dimensional (2D) van der Waals (vdW) materials is used to alter the band structure and yield emergent moir\'{e} patterns. Graphene-based vdW heterostructures have epitomised the power of moir\'{e} physics: twisted graphene multilayers, most notably twisted bilayer graphene (TBG), exhibit magic angles hosting flat bands, unconventional superconductivity, strange metallicity, and non-trivial topology \cite{Cao2018,Song2019,Cao2021,Park2022}; bilayer graphene quasicrystals host localized states exhibiting 12-fold rotational order \cite{Ahn2018,Moon2019,Yu2019,Pezzini2020}; twisting between graphene and transition metal dichalcogenides enables the tuning of spin-orbit coupling and spin-charge interconversion in spintronic devices \cite{Li2019,Sousa2022,Peterfalvi2022,Veneri2022,Perkins2024,Wojciechowska2025}; and twisting in graphene on hexagonal Boron Nitride (hBN) grants control over gapping graphene's Dirac cones \cite{Moon2014}.

An exemplary system for probing the physics of strong correlations is the kagome lattice, which possesses a flat band across the entire BZ, while hosting Dirac electrons around the BZ corners and VHSs at the BZ edge. The kagome lattice with nearest-neighbor (NN) hopping is the line graph of a honeycomb lattice; graph theory guarantees that such lattices possess at least one flat band \cite{Lieb1989,Kollar2020,Ma2020}. Both monolayer and bilayer variants of the kagome lattice appear in several materials \cite{Sante2026}, with intense numerical efforts having identified a vast list of relevant materials \cite{Regnault2022}. In previous work on commensurate twisted bilayer kagome (TBK) systems we demonstrated how the band structure can be tuned to host HOVHSs, band gaps, and giant Chern numbers of order 10 through twisting, dimerization, and time-reversal symmetry (TRS) breaking \cite{Perkins2025b}, complementary to a recent study on the topology and electronic structure of monolayer kagome under TRS breaking \cite{Wang2025}. The role of twisting in kagome multilayers has also become of central interest to study quantum geometries detached from their monolayer constituents \cite{Hung2026arxiv}. However, the lack of a general method to study small incommensurate twisted systems where only a subset of bands are located near the Fermi surface has prevented the study of small twist TBK.

In this Letter, we introduce a generalized formulation of the widely applied Bistritzer-MacDonald (BM) method \cite{Lopes_dos_Santos2007,Bistritzer2011,Lopes_dos_Santos2012,Koshino2018,Song2019,Catarina2019,Koshino2019,Wu2019,Bernevig2021,Scheer2022,Ma2024,Zhou2024,Calugaru2025} used to describe misaligned heterostructures for small incommensurate twists, and apply it to the moir\'{e} physics of electrons in TBK near $1/3$ filling where Dirac cones exist within the monolayers. Specifically, we reformulate the BM construction of the moir\'{e} continuum Hamiltonian by isolating the active bands within each monolayer, allowing for the consideration of arbitrary twisted heterostructures formed of monolayers possessing several sublattice sites but whose (non-extended) Fermi surface, and thus resultant behavior, is defined only by a small subset of bands. Our method reveals TBK to possess magic angles akin to TBG corresponding to the onset of HOVHSs potentially hosting stable topology. We prove that TBK hosts an approximate unitary particle-hole symmetry (PHS), resulting in a \textit{momentum-dependent} anti-unitary PHS, $\mathscr{P}_{\mathbf{k}}$, with the property $\mathscr{P}_{-\mathbf{k}}\mathscr{P}_{\mathbf{k}} = -1$. We also show that although sublattice interference effects are present, their role is not as prominent as in monolayer kagome.

\begin{figure}
    \centering
    \includegraphics[width=\linewidth]{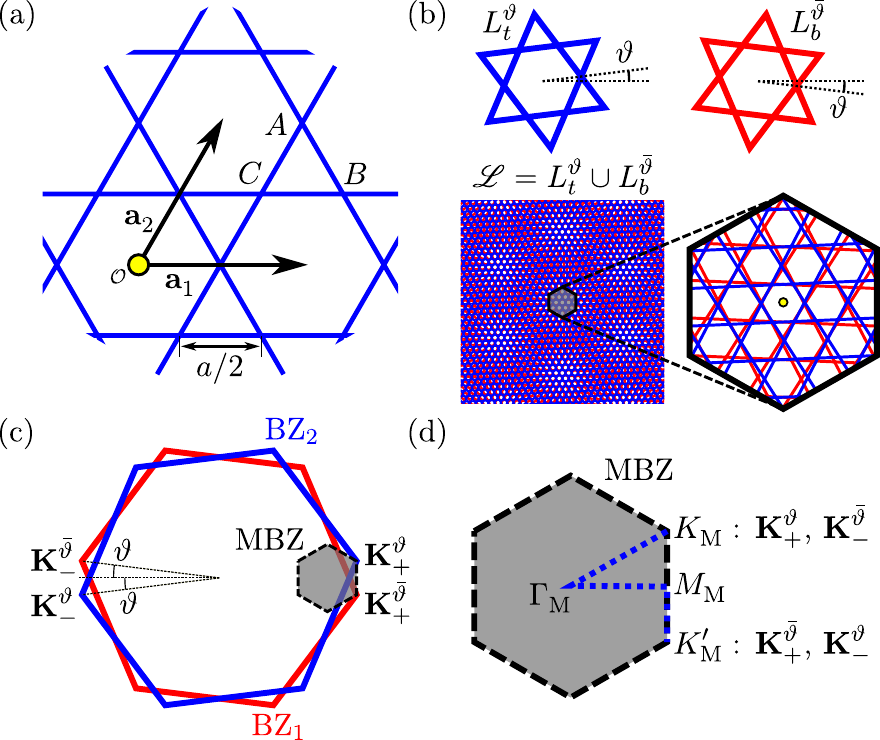}
    \caption{(a): Kagome lattice with sublattice, lattice vectors, and NN distance labelled. The yellow dot corresponds to the unit cell origin and the centre of rotation used to define the twisted bilayer. (b): Construction of the TBK system in the twist symmetric frame. (c): Definition of the MBZ (shaded region) by connection of two Dirac points belonging to the BZs of the neighboring layers. (d): Labelling of the MBZ high-symmetry points and the high-symmetry path.}
    \label{Setup_schematic}
\end{figure}

\textit{Model.} We define the TBK system starting from an $AA$-stacked bilayer with an interlayer separation $d_{\perp}$ and then rotating the top (bottom) layer clockwise by $\vartheta = \theta/2$ ($\bar{\vartheta} = -\vartheta$) about a hexagon centre, see Fig. \figRef{Setup_schematic}{a}. This is the twist-symmetric frame with a total twist of $\theta$. To construct a continuum Hamiltonian for TBK, we first define the Hamiltonian for a kagome monolayer with NN tunneling $t_{0} > 0$. The monolayer hosts a flat band at $2t_{0}$ and a Dirac cone at $-t_{0}$: $1/3$ filling places the Fermi surface at the Dirac point. Near $1/3$ filling, the emergent physics of the kagome lattice will be determined solely by the Dirac states hosting the Dirac cones, with the Fermi surface being confined around the BZ corners. We capture the low-energy physics by introducing a constant energy shift of $t_{0}$ and projecting out the flat band state to create an effective Hamiltonian, $\mathscr{H}_{\mathbf{k}}$. In the eigenbasis of the Dirac states, $\mathscr{H}_{\mathbf{k}} = -v_{F} k \sigma_{z}$ ($\hbar = 1$), where $v_{F} = \sqrt{3}at_{0}/2$ is the Fermi velocity, $\{\sigma_{i}\}$ are the Pauli matrices supplemented with identity, and $\mathbf{k}$ is the momentum local to a Dirac point. The low-energy continuum Hamiltonian for TBK near $1/3$ filling may then be written using a generalized formulation of the BM model \cite{Bistritzer2011,Koshino2018,Song2019,Catarina2019,Koshino2019,Wu2019,Zhou2024},
\begin{equation}
    \mathscr{H}_{\text{TBK},\mathbf{k}}^{\tau} = \begin{pmatrix}
        \mathscr{H}_{\mathbf{k}^{\vartheta}}^{(b)} & \mathscr{H}_{\text{T},\mathbf{k}\mathbf{p}}^{\tau} \\
        \mathscr{H}_{\text{T},\mathbf{k}\mathbf{p}}^{\tau \, \dagger} & \mathscr{H}_{\mathbf{p}^{\bar{\vartheta}}}^{(t)}
    \end{pmatrix},
    \label{TBK_Hamiltonian_main}
\end{equation}
where $\mathscr{H}_{\mathbf{q}^{\varphi}}^{(l)}$ is the monolayer Hamiltonian for layer $l$, $\tau = \pm1$ is the monolayer valley index corresponding to expansion around $\mathbf{K}_{\tau}^{\null} = (\tau \frac{4\pi}{3a},0)$ in the monolayer Hamiltonian, and $\mathscr{H}_{\text{T},\mathbf{k}\mathbf{p}}^{\tau}$ is the interlayer tunneling Hamiltonian. We truncate the tunneling Hamiltonian to the dominant set of terms yielding three possible tunneling processes, $\mathscr{H}_{\text{T},\mathbf{k}\mathbf{p}}^{\tau} = \frac{4\omega_{0}}{3} \sum_{j\in\{b,tr,tl\}} T_{j,\mathbf{k}\mathbf{p}}^{\tau} \delta_{\mathbf{p}-\mathbf{k},\tau \mathbf{q}_{j}}^{\null}$, where $\mathbf{q}_{\text{b}}^{\null} = (R_{\bar{\vartheta}}^{\null} - R_{\vartheta}^{\null}) \mathbf{K}_{+}^{\null}$, $\mathbf{q}_{\text{tr}}^{\null} = R_{2\pi/3}^{\null} \mathbf{q}_{\text{b}}^{\null}$, $\mathbf{q}_{\text{tl}}^{\null} = R_{2\pi/3}^{-1} \mathbf{q}_{\text{b}}^{\null}$,
\begin{subequations}
\begin{gather}
    T_{j,\mathbf{k}\mathbf{p}}^{\tau} = \tilde{c} \, \sigma_{0}^{\null} + \bar{s} \, \sigma_{x}^{\null} + i \tau \tilde{s} \, \sigma_{y}^{\null} - \tau \bar{c} \, \sigma_{z}^{\null}
    \\
    \tilde{c} = \cos \tilde{\phi}_{\mathbf{k}\mathbf{p}}^{\theta}, \, \bar{c} = \cos \bar{\phi}_{\mathbf{k}\mathbf{p}}^{j}, \, \tilde{s} = \sin \tilde{\phi}_{\mathbf{k}\mathbf{p}}^{\theta}, \, \bar{s} = \sin \bar{\phi}_{\mathbf{k}\mathbf{p}}^{j},
\end{gather}
\label{TBK_tunnelling_matrices_vectors}%
\end{subequations}
$\tilde{\phi}_{\mathbf{k}\mathbf{p}}^{\theta} = \frac{1}{2}(\phi_{\mathbf{k}} - \phi_{\mathbf{p}} + \theta)$, $\bar{\phi}_{\mathbf{k}\mathbf{p}}^{j} = \frac{1}{2}(\phi_{\mathbf{k}} + \phi_{\mathbf{p}}) + \frac{2\pi}{3} (\delta_{j,tl}^{\null} - \delta_{j,tr}^{\null})$, $\phi_{\mathbf{k}}$ is the azimuthal angle of momentum $\mathbf{k}$, $\delta_{i,j}$ is the Kronecker delta, and $\omega_{0}$ is the energy characterizing the tunneling between layers at the Dirac point. This truncation is suitable for systems whose Fourier transformed overlap integrals decay sufficiently quickly, as is the case in TBG. This condition depends upon the details of the tunneling model including the interlayer separation, electron orbital type, and decay rate. The dimension of $\mathscr{H}_{\text{TBK},\mathbf{k}}^{\tau}$ is formally infinite, requiring another level of truncation by limiting the number of shells included in the $q$-lattice. The Hamiltonian's considered here follow a ``tripod'' construction for calculating band structures and Fermi velocities and a $\Gamma_{\text{M}}$-centric construction for testing the quality of PHS \cite{Bernevig2021}. For a derivation of Eqs. \ref{TBK_Hamiltonian_main} and \ref{TBK_tunnelling_matrices_vectors}, the additional terms beyond truncation in tunneling processes, specific shell truncations for different twist angles, and details of the tunneling model, see Ref. \cite{SMref}.

\begin{figure*}
    \centering
    \includegraphics[width=\linewidth]{./Fig2.pdf}
    \caption{(a): Band structure of TBK near $1/3$ filling for various twist angles with the energy zeroed average energy of the first conduction band (solid blue line) and first valence band at $K_{M}'$. Solid (dashed) lines indicate the monolayer valley $\tau = +1$ ($\tau = -1$), while $n$ labels the band number. (b): Contour plots of the first conduction (left) and valence (right) bands for $\theta = 0.95^{\circ}$. Dashed lines mark the MBZ boundary. Shaded regions indicate the corner of interest in panel (e). (c): Band structure for $\theta = 0.95^{\circ}$ with PHS imposed. (d): Variation of the renormalized Dirac velocity at $K_{M}'$ as the twist angle is varied with (dashed blue) and without (solid red) PHS. The inset shows a close-up on the variation of $v_{F}^{*}$ around $\theta_{3}^{*}$. (e): Surface plots of the first conduction and valence bands around the $K_{\text{M}}'$ point (black dot) near $\theta_{3}^{*}$ and $\theta_{2}^{*}$ (top/bottom, respectively) showing the local flattening of the Dirac cone and its persistence. The MBZ boundary is marked by the dashed line.}
    \label{Electronic_structure}
\end{figure*}

\textit{Electronic Structure and Symmetry.} Given the recently reported rare-earth based materials, $X$Pb${}_{3}$ and LaTl${}_{3}$, hosting a kagome lattice on their surface \cite{Mihalyuk2022,Vekovshinin2024,Denisov2025,Vekovshinin2025,Denisov2026}, we note that the outermost electrons of the atoms forming the kagome lattice, Pb and Tl, occupy $p$-orbitals. With this in mind, to illustrate our ideas we consider a Slater-Koster-type model based on $p_{z}$-orbitals to acquire $\omega_{0} \simeq -0.0819 t_{0}$ \cite{SMref}. We note the features of our following results are robust over a range of $\omega_{0}/t_{0} \in [-0.1,-0.005]$ \cite{SMref}. We show the TBK band structure for various twist angles in Fig. \figRef{Electronic_structure}{a}, with the first conduction band (i.e. first above $1/3$ filling) highlighted in blue, and note that Dirac cones persist around the MBZ corners with some renormalized Fermi velocity, $v_{F}^{*}$. For $\theta = 0.95^{\circ}$, the Dirac cones and their local region flatten drastically, suggesting the existence of a magic angle where symmetry allowed HOVHSs emerge with the form $\varepsilon_{\mathbf{k}} = c_{1} k + c_{2} k^{2} + c_{3,0} k^{3} + c_{3,x} k^{3} \cos \phi + c_{3,y} k^{3} \sin \phi + \mathcal{O}(k^{4})$ about the MBZ corners. Additionally, we find significant flattening of other bands across large regions of the MBZ occurs for generically small twists, $\theta \lesssim 1^{\circ}$ (Fig. \figRef{Electronic_structure}{a}) \cite{SMref}.

The TBK systems possesses a $D_{6}$ point group symmetry \cite{Perkins2025a}, TRS ($\mathcal{T}$), and an approximate PHS, $\Sigma$. However, the valley-resolved Hamiltonian in Eq. \ref{TBK_Hamiltonian_main} instead possesses a $D_{3}$ point group, magnetic symmetries $C_{2z}\mathcal{T}$ and $C_{2y}\mathcal{T}$ ($k_{y} \rightarrow -k_{y}$), and $\Sigma$, but lacks TRS. The PHS is approximate since it is only broken by small contributions $\mathcal{O}(\theta)$ to the $\mathscr{H}_{\text{TBK},\mathbf{k}}^{\tau}$: by setting $\theta = 0$ in all but the $\mathbf{q}$ vectors, the PHS becomes exact with $E_{n,\mathbf{k}} = -E_{-n,-\mathbf{k}}$, where $n$ is the band number such that $n = \pm m$ for the $m^{\text{th}}$ conduction/valence band. We illustrate this approximate PHS in Fig. \figRef{Electronic_structure}{b} where the $|n| = 1$ bands appear to qualitatively adhere to $\Sigma$ but differ quantitatively. An example with $\Sigma$ enforced is given in Fig. \figRef{Electronic_structure}{c}. We provide a proof of this approximate symmetry in Ref. \cite{SMref}.

From symmetry alone, we may determine if the Dirac cones appearing around the MBZ corners can be flattened \cite{Sheffer2023}. Since TBK and TBG possess identical symmetries, then, without $\Sigma$, the exact Hamiltonian will yield velocity operators of the form $\hat{v}_{x,y} = f_{1}(\theta) \sigma_{x} \pm f_{2}(\theta) \sigma_{y}$ \cite{Sheffer2023}. The functions $f_{1,2}(\theta)$ are not necessarily zero simultaneously, and so a second parameter is required to find a critical point where both functions vanish. Imposing $\Sigma$ to be exact guarantees $f_{2}(\theta) = 0$, and the corresponding Hamiltonian permits $v_{F}^{*} \rightarrow 0$ using only a single parameter, as is the case for TBG \cite{Sheffer2023}. We plot the variation of $v_{F}^{*}$ in Fig. \figRef{Electronic_structure}{d}, comparing TBK with and without $\Sigma$, and observe several magic angles including $\theta_{3}^{*} \simeq 0.95^{\circ}$ and $\theta_{2}^{*} \simeq 1.35^{\circ}$ ($\theta_{i}^{*}$ is the $i^{\text{th}}$ magic angle), where the Dirac cones flatten drastically, see Fig. \figRef{Electronic_structure}{e}. For the case with $\theta \simeq 0.946^{\circ}$ and without $\Sigma$, we obtain $c_{1}/c_{2} \sim c_{2}/c_{3,0} \sim c_{2}/ c_{3,x} \sim 0.01$ and $c_{3,x}/c_{3,y} \sim 0.1$, emphasizing the emergence of HOVHS physics. These values are found from a least squares fitting of the minimal surface described by the aforementioned expansion to a local square region of side length $10^{-2} |\mathbf{K}_{\text{M}}|$ centred on $K_{M}'$. The discontinuities in $v_{F}^{*}$ without $\Sigma$ in Fig. \figRef{Electronic_structure}{d} results from the degeneracy between the $|n| = 1$ bands lifting at $K_{\text{M}}^{\null}$ and $K_{\text{M}}'$ and being replaced by a Dirac cone formed of the $n = 1$ and $n = 2$ bands. However, new degeneracies between the $|n|=1$ bands form elsewhere in the MBZ \cite{SMref}. Moreover, the $|n|=1$ eigenstates are equally distributed across both layers, indicating that the gap between the $|n|=1$ bands cannot be controlled by the application of an out-of-plane electric field. Lastly, we note that the band structures for $\theta \simeq \theta_{3}^{*}, \theta_{2}^{*}$, and $\theta = 2.85^{\circ} \simeq \theta_{1}^{*}$ appear to be the same as the third, second, and first magic angles in TBG, respectively \cite{Song2019}.

\textit{Topology.} Akin to TBG, TBK possesses an approximate anti-unitary PHS defined by the combination of $\Sigma$ and $C_{2z}\mathcal{T}$ symmetries but does not host non-trivial topology via the usual Chern number. However, what sets this symmetry apart from TBG is its explicit momentum-dependence. For TBG, stable topology results from a \textit{momentum-independent} anti-unitary PHS with the property $\mathscr{P}^{2} = -1$ \cite{Fu2006}, and the demonstration that $\mathscr{P}$ is a good approximate symmetry at the level of the eigenstates, introducing an error of order 1\% \cite{Song2021}. In TBK, due to the anti-unitary PHS gaining momentum-dependence, we must generalise $\mathscr{P}^{2} \rightarrow \mathscr{P}_{-\mathbf{k}} \mathscr{P}_{\mathbf{k}}$ \cite{SMref}. We calculate the quality of $\mathscr{P}_{\mathbf{k}}$ in TBK and find errors of $\sim 3$\% and $\sim 15.5$\% near the first and third magic angles, respectively \cite{Song2021,SMref}. We thus propose TBK to be a potential host of stable topology, appreciating that the quality of PHS is sensitive to both twist angle and material details. A definitive answer depends on each particular experimental realization.

\begin{figure}
    \centering
    \includegraphics[width=\linewidth]{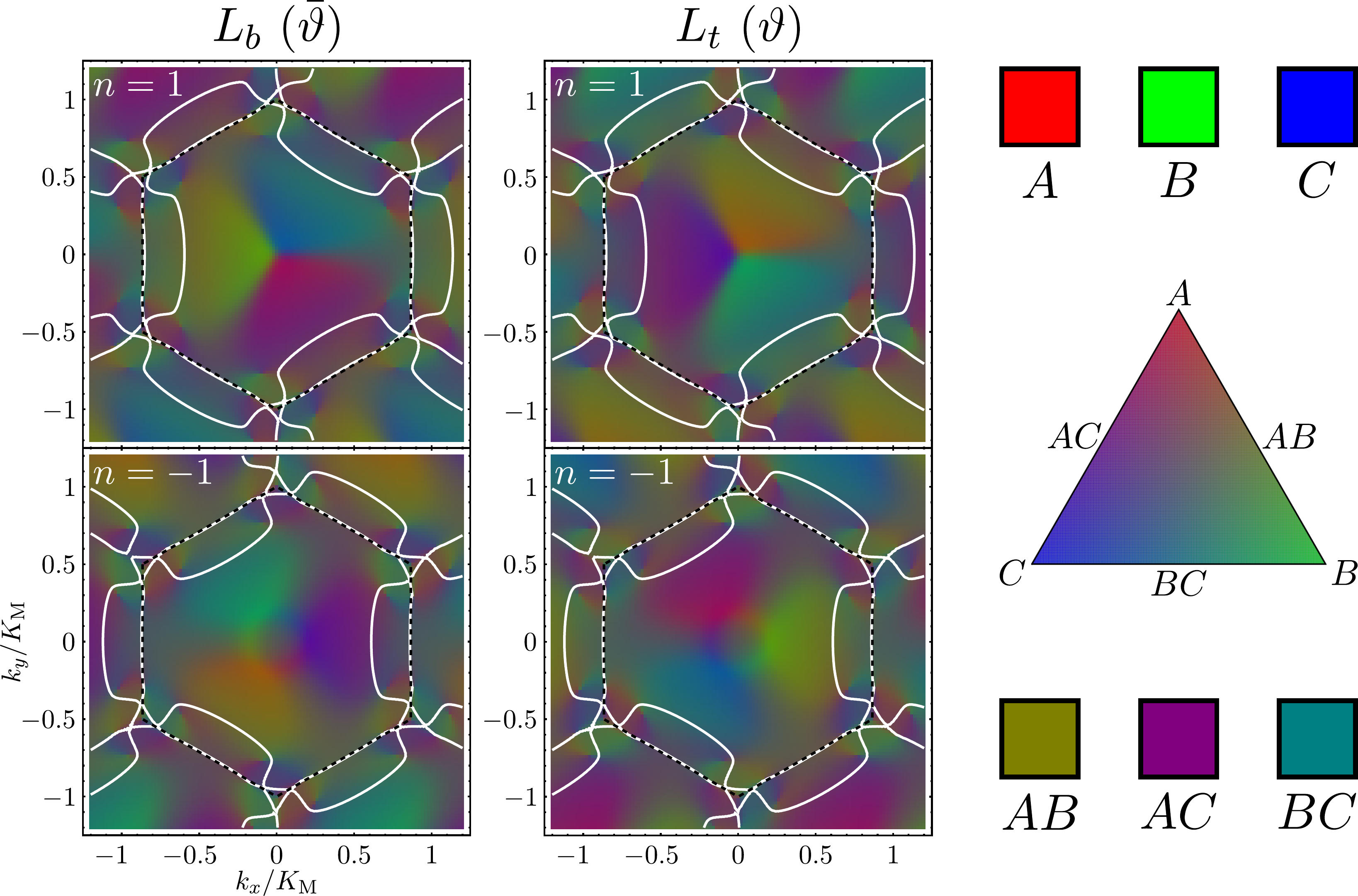}
    \caption{Layer-resolved sublattice projection of the first conduction (top) and valence (bottom) bands for the $\theta = 0.95^{\circ}$ system. Contours in the flat region near the moir\'{e} Dirac point energy, $(\varepsilon_{\text{D}} \times 10^{3} )t_{0}$, are highlighted: $E_{c} = (\varepsilon_{\text{D}} + 0.02) \times10^{3} \, t_{0}$ for $n = 1$ and $E_{c} = (\varepsilon_{\text{D}} - 0.03) \times10^{3} \, t_{0}$ for $n = -1$.}
    \label{Sublattice_proj}
\end{figure}

\textit{Sublattice Interference.} At the monolayer level, the kagome eigenstates play a central role in determining the ordered phases arising from strong correlations near $5/12$ filling (Van Hove) via sublattice interference, wherein the Fermi surface exhibits strong sublattice polarizations and require Fermi surface nesting to connect regions with the same sublattice projections \cite{Kiesel2012}, an effect not observed in graphene. Given the flattening of bands observed in TBK, electronic correlations will play a prominent role and potentially enable valley-polarized phases. The material-specific details of flattening are therefore of prime importance in determining the possible electronic phases. A detailed analysis of interactions in TBK is beyond the scope of this work. However,  sublattice interference should play a key role in phase formation for TBK: large, commensurate twist TBK has already been shown to display strong sublattice polarization \cite{Perkins2025b}. We plot the layer-resolved sublattice projections for $\theta \simeq \theta_{3}^{*}$ in Fig. \ref{Sublattice_proj}. We see that the transition between regions with different projections is significantly smoother than in large twist TBK \cite{Perkins2025b} and the sublattice projections are not as strongly polarized with no scattering inherently forbidden by sublattice projection.

The reduction of sublattice interference for TBK near $1/3$ filling can be understood by considering the underlying monolayer eigenstates. The sublattice interference commonly discussed in monolayer kagome happens at $5/12$ filling, corresponding to the VHS of the second band, with the Fermi surface being located far from the Dirac points. However, for the case of $1/3$ filling considered here, the sublattice interference of the TBK system will be related to the monolayer sublattice projections around the Dirac points. Both monolayer Dirac states possess projections that are never strongly localized to a single sublattice near the Dirac points. Hence, it is natural that the sublattice projection of the bands nearest $1/3$ filling in TBK will possess mixtures of the two Dirac states' sublattice projections, and thus not be heavily biased towards any one sublattice.

Regarding nesting, the Fermi surface for a doping close to the moir\'{e} Dirac cone (Fig. \ref{Sublattice_proj}), does not possess any natural zero-temperature nesting vectors due to the curving of the Fermi surface arising with the onset of a HOVHS \cite{Nag2024arxiv,Perkins2025b}. Nonetheless, HOVHSs typically enhance the transition temperature and so nesting occurs between the thermally broadened regions of the Fermi surface via \textit{diffusive nesting} vectors \cite{Beck2026}. Therefore, while sublattice interference will be present and act to reduce the influence of local interactions, an effect not reported in TBG, its overall role will not be as significant as in monolayer kagome since diffuse nesting is not forbidden by incompatible sublattice projections.

\textit{Discussion.} We have developed a method, generalizing the BM model typically used for honeycomb systems, to study generic twisted bilayer heterostructures at small incommensurate twist angles when the Fermi surface is not extended and not all bands play an active role in the Fermi surface physics. As an example, we then applied it to the kagome lattice near $1/3$ filling to study the band structure, possible topology, and prevalence of sublattice interference in TBK. Our analysis demonstrates the emergent magic angles akin to TBG where the first conduction and valence bands flatten drastically around the MBZ corners to allow for enhanced power-law divergences in the DOS due to the onset of a HOVHS; perfect HOVHSs prevented by a weak breaking of PHS and small, yet negligible, quadratic terms.  The anti-unitary PHS of TBK exhibits $\mathscr{P}_{-\mathbf{k}}\mathscr{P}_{\mathbf{k}} = -1$, consistent with the properties enabling Euler class topology in TBG \cite{SMref,Fu2006,Song2021}, with the momentum-dependence of this symmetry generalizing the momentum-independent property $\mathscr{P}^{2} = -1$ seen in TBG. Real material parameter will dictate the details of these topological properties. Moreover, without the need for additional tuning parameters, we found bands with large flat regions beyond the first conduction and valence band to emerge for small twists, $\theta \lesssim 1^{\circ}$. We emphasize that these results are robust to variations in the characteristic energy scale, $\omega_{0}$, of interlayer tunnelling in TBK, such the that magic angles and bandwidths are reduced for smaller $\omega_{0}$.

A comment is in order on the role of spin-orbit coupling (SOC), filling factor, and hopping range and what aspects of our results are expected in realistic kagome-based vdW heterostructures. In accounting for second- and third-NN intralayer tunnelling ($t_{2}$ and $t_{3}$, respectively), the flat band becomes dispersive and the energetic separation of the Dirac point from the VHSs in the first and second bands is reduced in an asymmetric fashion for $t_{2,3} \leq 0.5 t_{0}$. Hence, the range of Fermi energies that can be captured by the model we present here will be reduced, requiring the inclusion of higher-order momentum terms in the monolayer Hamiltonian and $T$-matrices to consider wider ranges of the Fermi energy. Likewise, for filling factors with considerable deviations from $1/3$, higher-order momentum terms will also need to be accounted for. Regarding SOC, we appreciate that this will result in some form of spin-splitting, yielding bands with non-trivial spin character. The origin (intrinsic or proximity-induced) and form of SOC will determine how the bands will split and their corresponding spin polarizations, as is known for proximitized graphene \cite{Peterfalvi2022,Veneri2022,Rao2023,Sun2023,Perkins2024}. Nonetheless, SOC can act to lift degeneracies and generate band inversions \cite{Offidani2018}, therefore enabling non-trivial abelian valley Chern numbers. A detailed analysis of SOC in TBK is left for future work.

Finally, we established that the sublattice interference of the kagome lattice will play a notably less impactful role for TBK near $1/3$ filling, arising from a lack of strong sublattice polarization around the monolayer Dirac cones at this filling. Nonetheless, sublattice polarization is still present and so we expect sublattice interference to still affect the emergent ordered phases due to electron-electron interactions in TBK. The strength of hybridization between the monolayer orbitals may also play a key role in TBK, with stronger interlayer tunneling amplitudes having been shown to yield unique quantum geometries differing from the underlying monolayer for commensurate TBK systems with loop-current order \cite{Hung2026arxiv}. Further work on TBK systems will benefit from the rapidly expanding field of kagome materials. The specifics of the overlap integrals can then be informed by experiment and lattice relaxation can also be accounted for. We expect that the qualitative features we describe in this work will remain intact. The main message is that using the generalized BM model developed here, we are able to identify a system hosting HOVHSs, magic angles akin to TBG with potentially the same topology, but with Fermi surfaces and nesting effects subject to a weakened sublattice interference.

\begin{acknowledgements}
    We would like to thank Bob Joynt for discussions and Tianhong Lu and Luiz H. Santos for providing the code necessary to calculate non-abelian Chern numbers and their helpful comments and discussions on this work. The work was supported by the UK Engineering and Physical Sciences Research Council (EPSRC) through grants EP/X012557/1 and EP/T034351/1.
\end{acknowledgements}

%


\clearpage


\begin{widetext}

\section*{Supplemental Material for ``Twisted Kagome Bilayers: High-Order Van Hove Singularities, Sublattice Interference, Magic Angles, and Possible Topology''}

\FloatBarrier

\setcounter{section}{0}
\setcounter{equation}{0}
\setcounter{figure}{0}
\setcounter{table}{0}
\setcounter{page}{1}

\renewcommand{\theequation}{S\arabic{equation}}
\renewcommand{\thesection}{S\arabic{section}}
\renewcommand{\thesubsection}{S\arabic{section}.\arabic{subsection}}
\renewcommand{\thesubsubsection}{S\arabic{section}.\arabic{subsection}.\arabic{subsubsection}}
\renewcommand{\thefigure}{S\arabic{figure}}
 \renewcommand{\thetable}{S\arabic{table}}

\section{Monolayer Kagome}

\subsection{Geometry, Hamiltonian, and Eigenstates}

The kagome lattice is the line graph of honeycomb lattice and may be viewed as the collection of points created by the set of corner-sharing triangles that differs from the triangular Bravais lattice, see Fig. 1 of the main text. Like the honeycomb lattice of graphene, the kagome lattice is a variant of the triangular lattice with three sublattice sites, $\{A,B,C\}$, per trinagular lattice site (i.e. it hosts three atomic sites per unit cell). We define a unit cell for this system whose origin (lattice site) is positioned at the centre of an upwards pointing triangle, and label the set of lattice site positions by $\{\mathbf{R}_{i}\}$. We then label the lattice vectors connecting unit cells as $\mathbf{a}_{1,2}$ and the sublattice positions within a unit cell unit cell as $\boldsymbol{\delta}_{\eta}$ ($\eta \in \{A,B,C\}$). The particular geometry we focus on is defined by,
\begin{equation}
    \mathbf{a}_{1} = a \begin{pmatrix}
        1 \\ 0
    \end{pmatrix},
    \qquad
    \mathbf{a}_{2} = \frac{a}{2} \begin{pmatrix}
        \sqrt{3} \\ 1
    \end{pmatrix},
    \qquad 
    \boldsymbol{\delta}_{A} = \frac{a}{2\sqrt{3}} \begin{pmatrix}
        0 \\ 1
    \end{pmatrix}, \qquad \boldsymbol{\delta}_{B} = \frac{a}{4\sqrt{3}} \begin{pmatrix}
        \sqrt{3} \\ -1
    \end{pmatrix}, \qquad \boldsymbol{\delta}_{C} = -\frac{a}{4\sqrt{3}} \begin{pmatrix}
        \sqrt{3} \\ 1
    \end{pmatrix},
\end{equation}
where $a$ is the lattice constant. The set of all atomic sites on the kagome lattice may thus be wrttien as $L = \{n_{1} \mathbf{a}_{1} + n_{2} \mathbf{a}_{2} + \boldsymbol{\delta}_{\eta}; n_{1}, n_{2} \in \mathbb{Z}, \eta \in \{A,B,C\} \}$.

The Hamiltonian describing a single-orbital kagome lattice with nearest-neighbour hopping is
\begin{equation}
    H_{0}^{\null} = - t_{0}^{\null} \sum_{\angleaverage{i,j}} (b_{j}^{\dagger} a_{i}^{\null} + c_{j}^{\dagger} a_{i}^{\null} + c_{j}^{\dagger} b_{i}^{\null}) + \text{h.c.},
    \label{Kagome_monolayer_TB}
\end{equation}
where $t$ is the nearest-neighbour tunnelling energy, $\chi_{i}^{\dagger}$ ($\chi_{i}^{\null}$) is the creation (annihilation) operator for an electron on the $i^{\text{th}}$ site of sublattice $\chi$, and $\angleaverage{i,j}$ denotes the sum over unit cells that yield nearest-neighbour atomic sites. We can extract the continuum Hamiltonian, $\mathcal{H}_{0,\mathbf{p}}^{\null}$, from $H_{0}^{\null}$ by rewriting the real-space creation/annihilation operators in terms of their momentum-space counterparts,
\begin{equation}
    a_{i} = \frac{1}{\sqrt{N}} \sum_{\mathbf{p}} e^{i\mathbf{p} \cdot (\mathbf{R}_{i} + \boldsymbol{\delta}_{A})} a_{\mathbf{p}},
    \qquad
    b_{i} = \frac{1}{\sqrt{N}} \sum_{\mathbf{p}} e^{i\mathbf{p} \cdot (\mathbf{R}_{i} + \boldsymbol{\delta}_{B})} b_{\mathbf{p}},
    \qquad
    c_{i} = \frac{1}{\sqrt{N}} \sum_{\mathbf{p}} e^{i\mathbf{p} \cdot (\mathbf{R}_{i} + \boldsymbol{\delta}_{C})} c_{\mathbf{p}}.
\end{equation}
By definition, $H_{0}^{\null} = \sum_{\mathbf{p}} (a_{\mathbf{p}}^{\dagger}, b_{\mathbf{p}}^{\dagger}, c_{\mathbf{p}}^{\dagger}) \mathcal{H}_{0,\mathbf{p}}^{\null} (a_{\mathbf{p}}^{\null}, b_{\mathbf{p}}^{\null}, c_{\mathbf{p}}^{\null})^{\text{T}}$, and we find
\begin{equation}
    \mathcal{H}_{0,\mathbf{p}}^{\null} = \begin{pmatrix}
        0 & -t_{0}^{\null} \gamma_{ab}^{\null} (\mathbf{p}) & -t_{0}^{\null} \gamma_{ac}^{\null} (\mathbf{p}) \\
        -t_{0}^{\null} \gamma_{ab}^{\null} (\mathbf{p}) & 0 & -t_{0}^{\null} \gamma_{bc}^{\null} (\mathbf{p}) \\
        -t_{0}^{\null} \gamma_{ac}^{\null} (\mathbf{p}) & -t_{0}^{\null} \gamma_{bc}^{\null} (\mathbf{p}) & 0
    \end{pmatrix}, \qquad \gamma_{\eta\chi}(\mathbf{p}) = 2 \cos[\mathbf{p} \cdot (\boldsymbol{\delta}_{\eta}^{\null} - \boldsymbol{\delta}_{\chi}^{\null})].
    \label{Kagome_monolayer_continuum_Hamiltonian}
\end{equation}
The eigenvalues of this Hamiltonian are then found to be
\begin{equation}
    E_{0} = 2t_{0}, \qquad E_{\pm,\mathbf{p}} = -t_{0} \pm \mathcal{E}_{\mathbf{p}}, \qquad \frac{\mathcal{E}_{\mathbf{p}}}{t_{0}} = \sqrt{3 + 2 \cos(p_{x}a) + 4 \cos\left(\frac{p_{x}a}{2}\right) \cos\left(\frac{p_{y}a\sqrt{3}}{2}\right)},
\end{equation}
whose corresponding eigenstates are $\ket{\Psi_{f,\mathbf{p}}}$, $\ket{\Psi_{+,\mathbf{p}}}$, and $\ket{\Psi_{-,\mathbf{p}}}$, respectively. The kagome lattice therefore hosts a completely flat band across its Brillouin zone (BZ) in the form of $E_{0}$, whilst its dispersive bands possess Van Hove singularities at the BZ edges ($M$-points) and Dirac cones at the BZ corners ($K$ and $K'$ valleys). We note that the Dirac points located at the BZ corners carry a valley index, $\tau = \pm 1$, and define the vectors of the Dirac points situated on the $p_{x}$-axis as $\mathbf{K}_{\tau} = (\tau\frac{4\pi}{3a},0)^{\text{T}}$. Additionally, we write the reciprocal lattice vectors as $\mathbf{b}_{1} = \frac{2\pi}{\sqrt{3}a}(\sqrt{3},-1)^{\text{T}}$ and $\mathbf{b}_{2} = \frac{4\pi}{\sqrt{3}a} (0,1)^{\text{T}}$.

We choose to write the diagonalising matrix for the monolayer kagome Hamiltonian as $\mathcal{A}_{\mathbf{p}}^{K} = (\ket{\Psi_{-,\mathbf{p}}}, \ket{\Psi_{+,\mathbf{p}}}, \ket{\Psi_{f,\mathbf{p}}})$, so that
\begin{equation}
    \widetilde{\mathcal{H}}_{0,\mathbf{p}}^{\null} = \mathcal{A}_{\mathbf{p}}^{K\dagger} \mathcal{H}_{0,\mathbf{p}}^{\null} \mathcal{A}_{\mathbf{p}}^{K} = \begin{pmatrix}
        E_{-,\mathbf{p}} & 0 & 0 \\
        0 & E_{+,\mathbf{p}} & 0 \\
        0 & 0 & E_{0,\mathbf{p}}
    \end{pmatrix}.
\end{equation}
Written explicitly, the diagonalisation matrix takes on a rather cumbersome form ($\pi_{\mathbf{p}}^{(\pm)} = p_{x}^{\null} \pm \sqrt{3} p_{y}^{\null}$)
\begin{equation}
    \mathcal{A}_{\mathbf{p}}^{K} = \frac{1}{\sqrt{3}} \begin{pmatrix}
         -\frac{[1 + \varepsilon_{\mathbf{p}}^{\null} + 2\cos(p_{x}/2) \cos(\sqrt{3}p_{y}/2)] \sec(\pi_{\mathbf{p}}^{(-)}/4)}{\mathcal{N}_{-}^{\null} [2 + \varepsilon_{\mathbf{p}}^{\null} + \cos(\sqrt{3}\pi_{\mathbf{p}}^{(-)}/4) \sec(\pi_{\mathbf{p}}^{(-)}/4)] } & -\frac{[\varepsilon_{\mathbf{p}}^{\null} - 1 - 2\cos(p_{x}/2) \cos(\sqrt{3}p_{y}/2)] \sec(\pi_{\mathbf{p}}^{(-)}/4)}{\mathcal{N}_{+}^{\null} [\varepsilon_{\mathbf{p}}^{\null} - 2 - \cos(\sqrt{3}\pi_{\mathbf{p}}^{(-)}/4) \sec(\pi_{\mathbf{p}}^{(-)}/4)]} & -\frac{\sin(p_{x}/2)}{\mathcal{N}_{f}^{\null} \sin(\pi_{\mathbf{p}}^{(-)}/4)}
        \\
        -\frac{(2+\varepsilon_{\mathbf{p}}^{\null}) \cos(\pi_{\mathbf{p}}^{(-)}/4) + \cos(\sqrt{3} \pi_{\mathbf{p}}^{(+)}/4)}{\mathcal{N}_{-}^{\null}[(2+\varepsilon_{\mathbf{p}}^{\null}) \cos(\pi_{\mathbf{p}}^{(+)}/4) + \cos(\sqrt{3} \pi_{\mathbf{p}}^{(-)}/4)]} & -\frac{(\varepsilon_{\mathbf{p}}^{\null} - 2) \cos(\pi_{\mathbf{p}}^{(-)}/4) + \cos(\sqrt{3} \pi_{\mathbf{p}}^{(+)}/4)}{\mathcal{N}_{+}^{\null}[(\varepsilon_{\mathbf{p}}^{\null} - 2) \cos(\pi_{\mathbf{p}}^{(+)}/4) + \cos(\sqrt{3} \pi_{\mathbf{p}}^{(-)}/4)]} & \frac{\sin(\pi_{\mathbf{p}}^{(+)}/4)}{\mathcal{N}_{f}^{\null} \sin(\pi_{\mathbf{p}}^{(-)}/4)}
        \\
        -\mathcal{N}_{-}^{-1} & -\mathcal{N}_{+}^{-1} & \mathcal{N}_{f}^{-1}
    \end{pmatrix},
    \label{Kagome_A_matrix}
\end{equation}
where $\mathcal{N}_{i}$ is the normalisation factors ensuring $\braket{\Psi_{i}|\Psi_{i}} = 1$. Since we will later consider the kagome lattice near $1/3$ filling, where the physics will be primarily governed by the Dirac cones of the dispersive bands, and thus collapse the Hamiltonian down to the Dirac subspace formed of $\ket{\Psi_{\pm,\mathbf{p}}}$,
\begin{equation}
    \widetilde{\mathcal{H}}_{0}^{\null} \rightarrow \widetilde{\mathcal{H}}_{0}^{\text{D}} = \begin{pmatrix}
        E_{-,\mathbf{p}} & 0 \\
        0 & E_{+,\mathbf{p}}
    \end{pmatrix},
\end{equation}
we will approximate the system using first-order perturbation theory to expand $\widetilde{\mathcal{H}}_{0}^{D}$ around the Dirac points. Specifically, in letting $\mathbf{p} = \mathbf{K}_{\tau} + \mathbf{k}$, we may expand the Hamiltonian to first-order in the local radial momentum, $k$, and the eigenstates to zeroth-order. Naturally, the diagonalised Hamiltonian takes on the trivial form, $\widetilde{\mathcal{H}}_{0,\mathbf{p}}^{\text{D}} \simeq \widetilde{\mathscr{H}}_{\mathbf{k}}^{\text{D}\tau} = -v_{\text{F}}^{\null} k \sigma_{z}^{\null}$, where $v_{\text{F}} = \sqrt{3}at/2$ is the Fermi velocity, $\{\sigma_{i}\}$ ($i \in \{0,x,y,z\}$) are the Pauli matrices supplemented with identity, and we appreciate that time-reversal symmetry imposes $\widetilde{\mathscr{H}}_{\mathbf{k}}^{\text{D}+} = \widetilde{\mathscr{H}}_{-\mathbf{k}}^{\text{D}-}$, which allows us to write $\widetilde{\mathscr{H}}_{\mathbf{k}}^{\text{D}\tau} = \widetilde{\mathscr{H}}_{\mathbf{k}}^{\text{D}}$. Moreover, $\mathcal{A}_{\text{K},\mathbf{p}}^{\null}$ expanded around a Dirac point takes on a fairly simple form,
\begin{equation}
\begin{gathered}
   \mathcal{A}_{\text{K},\mathbf{p}}^{\null}\simeq \mathscr{A}_{\text{K},\mathbf{k}}^{\tau} = \begin{pmatrix}
            \psi_{-,\mathbf{k}}^{\tau(A)} & \psi_{+,\mathbf{k}}^{\tau(A)} & -\frac{1}{\sqrt{3}}
            \\
            \psi_{-,\mathbf{k}}^{\tau(B)} & \psi_{+,\mathbf{k}}^{\tau(B)} & \frac{1}{\sqrt{3}}
            \\
            \psi_{-,\mathbf{k}}^{\tau(C)} & \psi_{+,\mathbf{k}}^{\tau(C)} & \frac{1}{\sqrt{3}}
        \end{pmatrix},
        \,\,
        \psi_{\mu,\mathbf{k}}^{\tau(\eta)} = (1 - 2 \delta_{\eta,A}^{\null}) \sqrt{\frac{2}{3}} \times
        \begin{cases}
            \sin \left( \frac{\phi_{\mathbf{k}}}{2} + S_{\eta} \frac{2\pi}{3} + \frac{(1 - \tau)\pi}{4} \right), \quad &\mu = -
            \\
            \cos \left( \frac{\phi_{\mathbf{k}}}{2} + S_{\eta} \frac{2\pi}{3} + \frac{(\tau - 1)\pi}{4} \right), \quad &\mu = +,
        \end{cases}
\end{gathered} \label{Kagome_A_matrix_expanded}
\end{equation}
where $S_{\eta}^{\null} = \delta_{\eta,C}^{\null} - \delta_{\eta,B}^{\null}$, $\delta_{i,j}$ is the Kronecker delta, $\phi_{\mathbf{k}} = \arctan(k_{y}/k_{x})$ is the local azimuthal angle, and $\psi_{\pm,\mathbf{k}}^{\tau(\eta)}$ are the sublattice projections of the Dirac states near $\mathbf{K}_{\tau}$. When the lattice is rotated anticlockwise by some angle $\vartheta$, the sublattice projections change according to $\phi_{\mathbf{k}} \rightarrow \phi_{\mathbf{k}} -\vartheta$.

\subsection{Gauge-Fixing and Berry Phase}

The expressions for the low-energy eigenstates of the dispersive bands in Eq. (\ref{Kagome_A_matrix_expanded}), whilst valid, are not single-valued upon a $2\pi$ rotation about the Dirac point. Rather, they are anti-periodic of this range. To ensure the eigenstates are single-valued and well-defined, as necessary for determining the Berry phase and hence winding number, we introduce $e^{i\phi_{\mathbf{k}}/2}$ as a gauge-fixing phase factor in the Dirac states. Specifically, the gauge-fixed eigenstates are,
\begin{equation}
    \ket{\psi_{\pm,\mathbf{k}}^{\tau}} = e^{i \tau \phi_{\mathbf{k}}/2} (\psi_{\pm,\mathbf{k}}^{\tau(A)}, \, \psi_{\pm,\mathbf{k}}^{\tau(B)}, \, \psi_{\pm,\mathbf{k}}^{\tau(C)})^{\text{T}}.
\end{equation}
The Berry connection for the Dirac states is then readily obtained as,
\begin{equation}
    \boldsymbol{\mathcal{A}} = i \braket{\psi_{\pm,\mathbf{k}}^{\tau}|\nabla_{\mathbf{k}}\psi_{\pm,\mathbf{k}}^{\tau}} = \frac{\tau}{2} \mathbf{e}_{\phi},
\end{equation}
which yields a Berry phase of $\gamma_{\pm}^{(\tau)} = \tau\pi$, or, equivalently, a winding number of $\mathcal{W}_{\tau}^{\null} = \tau = \pm 1$. This result can also be obtained by choosing instead to work in the basis of irreducible representations \cite{Zhou2025}.

\subsection{Sublattice Projection Operators}

 We start by defining the sublattice projection operators for the kagome monolayer in its natural sublattice basis, $\{A,B,C\}$,
\begin{equation}
    \mathcal{P}_{A,\mathbf{p}}^{K} = \ket{A_{\mathbf{p}}}\bra{A_{\mathbf{p}}} = \begin{pmatrix}
        1 & 0 & 0 \\
        0 & 0 & 0 \\
        0 & 0 & 0
    \end{pmatrix},
    \quad
    \mathcal{P}_{B,\mathbf{p}}^{K} = \ket{B_{\mathbf{p}}}\bra{B_{\mathbf{p}}} = \begin{pmatrix}
        0 & 0 & 0 \\
        0 & 1 & 0 \\
        0 & 0 & 0
    \end{pmatrix},
    \quad
    \mathcal{P}_{C,\mathbf{p}}^{K} = \ket{C_{\mathbf{p}}}\bra{C_{\mathbf{p}}} = \begin{pmatrix}
        0 & 0 & 0 \\
        0 & 0 & 0 \\
        0 & 0 & 1
    \end{pmatrix}.
\end{equation}
The eigenstates of $\mathcal{H}_{0,\mathbf{p}}$ can be decomposed into linear combinations of the kagome sublattice (KSL) states,
\begin{equation}
    \ket{\Psi_{\mu,\mathbf{p}}^{\null}} = \sum_{\eta} \Psi_{\mu,\mathbf{p}}^{(\eta)} \ket{\eta_{\mathbf{p}}^{\null}},
\end{equation}
with their sublattice projections $\Psi_{\mu,\mathbf{p}}^{(\eta)}$ given in Eq. (\ref{Kagome_A_matrix}). We may then rewrite the kagome sublattice projection operators in the eignbasis and collapse them onto the Dirac subspace,
\begin{equation}
    \widetilde{\mathcal{P}}_{\eta,\mathbf{p}}^{K} = \begin{pmatrix}
        |\Psi_{-,\mathbf{p}}^{(\eta)}|^{2} & \Psi_{-,\mathbf{p}}^{(\eta)*} \Psi_{+,\mathbf{p}}^{(\eta)} & \Psi_{-,\mathbf{p}}^{(\eta)*} \Psi_{f,\mathbf{p}}^{(\eta)} \\
        \Psi_{+,\mathbf{p}}^{(\eta)*} \Psi_{-,\mathbf{p}}^{(\eta)} & |\Psi_{+,\mathbf{p}}^{(\eta)}|^{2} & \Psi_{+,\mathbf{p}}^{(\eta)*} \Psi_{f,\mathbf{p}}^{(\eta)} \\
        \Psi_{f,\mathbf{p}}^{(\eta)*} \Psi_{-,\mathbf{p}}^{(\eta)} & \Psi_{f,\mathbf{p}}^{(\eta)*} \Psi_{+,\mathbf{p}}^{(\eta)} & |\Psi_{f,\mathbf{p}}^{(\eta)}|^{2}
    \end{pmatrix}
    \longrightarrow
    \widetilde{\mathcal{P}}_{\eta,\mathbf{p}}^{D} = \begin{pmatrix}
        |\Psi_{-,\mathbf{p}}^{(\eta)}|^{2} & \Psi_{-,\mathbf{p}}^{(\eta)*} \Psi_{+,\mathbf{p}}^{(\eta)} \\
        \Psi_{+,\mathbf{p}}^{(\eta)*} \Psi_{-,\mathbf{p}}^{(\eta)} & |\Psi_{+,\mathbf{p}}^{(\eta)}|^{2}
    \end{pmatrix}.
\end{equation}
Given our eventual focus on $1/3$ filling, we expand these operators around the Dirac points as
\begin{equation}
    \widetilde{\mathcal{P}}_{\eta,\mathbf{p}}^{D} \simeq \widetilde{\mathscr{P}}_{\eta,\mathbf{p}}^{D} = \begin{pmatrix}
        |\psi_{-,\mathbf{k}}^{\tau(\eta)}|^{2} & \psi_{-,\mathbf{k}}^{\tau(\eta)*} \psi_{+,\mathbf{k}}^{\tau(\eta)} \\
        \psi_{+,\mathbf{k}}^{\tau(\eta)*} \psi_{-,\mathbf{k}}^{\tau(\eta)} & |\psi_{+,\mathbf{k}}^{(\eta)}|^{2}
    \end{pmatrix}.
    \label{KSLproj_downfold_expansion_general}
\end{equation}
which, upon recalling Eq. (\ref{Kagome_A_matrix_expanded}), simplifies to
\begin{equation}
    \widetilde{\mathscr{P}}_{\eta,\mathbf{p}}^{D} = \frac{1}{3} \begin{pmatrix}
        1 - \tau \cos \left( \phi_{\mathbf{k}}  - S_{\eta} \frac{2\pi}{3} \right) & \sin \left( \phi_{\mathbf{k}} - S_{\eta} \frac{2\pi}{3} \right) \\
        \sin \left( \phi_{\mathbf{k}} - S_{\eta} \frac{2\pi}{3} \right) & 1 + \tau \cos \left( \phi_{\mathbf{k}} - S_{\eta} \frac{2\pi}{3} \right)
    \end{pmatrix}.
    \label{KSL_downfold_expansion}
\end{equation}

\section{Derivation of the Tunnelling Matrices for Twisted Bilayer Kagome Near 1/3 Filling}

A twisted kagome bilayer is created when one layer of a bilayer kagome system is rotationally misaligned with the other. Our focus will be on creating such a misalignment starting from any one of the high-symmetry stackings. In doing so, we shall choose to work in the twist-symmetric frame, wherein the top layer is rotated by $\vartheta = \theta/2$ and the bottom layer by $\bar{\vartheta} = -\theta/2$ to give a total twist angle of $\theta$. In other words, by defining $\mathbf{v}^{\vartheta} = R_{\vartheta} \mathbf{v}$ with $R_{\vartheta}$ as the 2D rotation matrix, the individual layer lattices become $L_{b}^{\bar{\vartheta}} = \{n_{1} \mathbf{a}_{1}^{\bar{\vartheta}} + n_{2} \mathbf{a}_{2}^{\bar{\vartheta}} + \boldsymbol{\delta}_{\chi}^{\bar{\vartheta}}; n_{1}, n_{2} \in \mathbb{Z}, \chi \in \{A,B,C\} \}$ and $L_{t}^{\vartheta} = \{n_{1} \mathbf{a}_{1}^{\vartheta} + n_{2} \mathbf{a}_{2}^{\vartheta} + \boldsymbol{\delta}_{\chi}^{\vartheta} + \mathbf{d}_{s}^{\vartheta} + d_{\perp} \mathbf{e}_{z}; n_{1}, n_{2} \in \mathbb{Z}, \chi \in \{A,B,C\} \}$, which combine to give $\mathscr{L} = L_{1}^{\bar{\vartheta}} \cup L_{2}^{\vartheta}$. In this section, we shall focus on deriving the matrices describing tunnelling between the two layers of this twisted heterostrcture due to its differences compared with TBG.

\subsection{General Tunnelling Between Twisted Kagome Layers}

To construct the Hamiltonian for TBK near $1/3$ filling, we will construct a generalised Bistritzer-MacDonald (BM) model in terms of the Dirac states expanded around the monolayer BZ corners which capture the low-energy physics of the kagome system in this scenario, noting that the interlayer tunnelling matrices will differ greatly compared to the usual TBG system. Our below derivation closely follows the approach of Ref. \cite{Catarina2019} for determining the interlayer tunnelling matrices.

The general Hamiltonian for any twisted bilayer system is
\begin{equation}
    H = H_{1}^{\null} + H_{2}^{\null} + H_{\text{T}}^{\null} + H_{\text{T}}^{\dagger} = \begin{pmatrix}
        H_{1}^{\null} & H_{\text{T}}^{\null} \\
        H_{\text{T}}^{\dagger} & H_{2}^{\null}
    \end{pmatrix},
\end{equation}
with $H_{l}$ being the monolayer Hamiltonian of layer $l$ and $H_{T}$ being the interlayer tunnelling Hamiltonian describing tunnelling from layer 2 to layer 1. As detailed in the main text, near $1/3$ filling, we may project the kagome monolayer Hamiltonian onto the Dirac subspace of its eigenbasis and approximate it close to the a Dirac point. In otherwords, we may view the relevant monolayer Hamiltonian as $\widetilde{\mathscr{H}}_{\mathbf{k}}^{\text{D},\tau}$ within a given monolayer valley, $\tau$, with $\mathbf{k}$ being the local momentum relative to the Dirac point. Let us take one step back from this and consider the Dirac projected kagome Hamiltonian before expanding about a Dirac point, $\widetilde{\mathcal{H}}_{\mathbf{k}_{1}}^{\text{D}}$. We may then write the Dirac projected Hamiltonian for TBK near $1/3$ filling in momentum-space as
\begin{equation}
    \mathcal{H}_{\text{TBK}} = \begin{pmatrix}
        \widetilde{\mathcal{H}}_{\mathbf{k}_{1}}^{\text{D}(1)} & \mathcal{H}_{\text{T},\mathbf{k}_{1}\mathbf{k}_{2}}^{\null} \\
        \mathcal{H}_{\text{T},\mathbf{k}_{1}\mathbf{k}_{2}}^{\dagger} & \widetilde{\mathcal{H}}_{\mathbf{k}_{2}}^{\text{D}(2)}
    \end{pmatrix},
    \qquad
    \mathcal{H}_{\text{T},\mathbf{k}_{1}\mathbf{k}_{2}}^{\null} = \begin{pmatrix}
        \mathcal{H}_{\text{T},\mathbf{k}_{1}\mathbf{k}_{2}}^{--} & \mathcal{H}_{\text{T},\mathbf{k}_{1}\mathbf{k}_{2}}^{-+} \\
        \mathcal{H}_{\text{T},\mathbf{k}_{1}\mathbf{k}_{2}}^{+-} & \mathcal{H}_{\text{T},\mathbf{k}_{1}\mathbf{k}_{2}}^{++}
    \end{pmatrix},
    \qquad
    \mathcal{H}_{\text{T},\mathbf{k}_{1}\mathbf{k}_{2}}^{\mu\nu} = \bra{\Psi_{\mu,\mathbf{k}_{1}}^{(1)}} H_{T} \ket{\Psi_{\nu,\mathbf{k}_{2}}^{(2)}}
    \label{TBK_Hamiltonian},
\end{equation}
where we use the superscript $l = 1,2$ as a layer index. To acquire the individual tunnelling matrices comprising the full interlayer tunnelling matrix, let us consider a generic Dirac subspace matrix element and aim to relate it back to a real-space description. However, unlike the usual BM model for TBG which works in the sublattice space of graphene, to expand the Dirac states in terms of Wannier states, we must first decompose the Dirac states into the kagome sublattice states,
\begin{equation}
    \ket{\Psi_{\mu,\mathbf{p}}^{(l)}} = \sum_{\eta} \Psi_{\mu,\mathbf{p}}^{(\eta_{l}^{\null})} \ket{l,\eta_{\mathbf{p}}^{\null}} = \frac{1}{\sqrt{N}} \sum_{\eta} \sum_{i} \Psi_{\mu,\mathbf{p}}^{(\eta_{l}^{\null})} e^{i \mathbf{p} \cdot (\mathbf{R}_{i}^{(l)} + \boldsymbol{\delta}_{\eta})} \ket{\mathbf{R}_{i}^{(l)},\eta}.
\end{equation}
The twisted tunnelling matrix elements in the Dirac eigenbasis may now be related back to the overlap integrals of the real-space atomic orbitals on the KSL, assuming the two-centre approximation, $t_{\perp}(\mathbf{R}_{i}^{(1)} + \boldsymbol{\delta}_{\chi}^{(1)} - \mathbf{R}_{j}^{(2)} - \boldsymbol{\delta}_{\eta}^{(2)}) = \bra{\mathbf{R}_{i}^{(1)},\chi} H_{\text{T}} \ket{\mathbf{R}_{j}^{(2)},\eta}$. Specifically, letting $\mathbf{v}_{\mu\nu}^{(mn)} = \mathbf{v}_{\mu}^{(m)} - \mathbf{v}_{\nu}^{(n)}$ for ease of notation, we obtain
\begin{equation}
    \mathcal{H}_{\text{T},\mathbf{k}_{1}\mathbf{k}_{2}}^{\mu\nu} = \frac{1}{N} \sum_{i,j} \sum_{\eta,\chi} \Psi_{\mu,\mathbf{k}_{1}}^{(\chi_{1}^{\null})*} \Psi_{\nu,\mathbf{k}_{2}}^{(\eta_{2}^{\null})} t_{\perp}(\mathbf{R}_{ij}^{(12)} + \boldsymbol{\delta}_{\chi\eta}^{(12)}) e^{-i \mathbf{k}_{1}^{\null} \cdot (\mathbf{R}_{i}^{(1)} + \boldsymbol{\delta}_{\chi}^{(1)})} e^{i \mathbf{k}_{2}^{\null} \cdot (\mathbf{R}_{j}^{(2)} + \boldsymbol{\delta}_{\eta}^{(2)})}.
    \label{HT_elements_general}
\end{equation}

To move to a purely momentum space description, we introduce the Fourier transform of the overlap integral ($A_{\text{s}}$ is the system area),
\begin{equation}
    t_{\perp}(\mathbf{r}) = \int \frac{d^{2}q}{(2\pi)^{2}} t_{\perp}(\mathbf{q}) e^{i\mathbf{q} \cdot \mathbf{r}} = \frac{1}{A_{\text{s}}} \sum_{\mathbf{q}} t_{\perp}(\mathbf{q}) e^{i\mathbf{q} \cdot \mathbf{r}}.
    \label{OI_FT}
\end{equation}
Substituting Eq. (\ref{OI_FT}) into Eq. (\ref{HT_elements_general}), we acquire ($A_{\text{uc}}$ is the area of a unit cell)
\begin{equation}
\begin{split}
    \mathcal{H}_{\text{T},\mathbf{k}_{1}\mathbf{k}_{2}}^{\mu\nu} &= \frac{1}{N^{2} A_{\text{uc}}}
    \sum_{\mathbf{q}} \sum_{\eta,\chi} \Psi_{\mu,\mathbf{k}_{1}}^{(\chi_{1}^{\null})*} \Psi_{\nu,\mathbf{k}_{2}}^{(\eta_{2}^{\null})} t_{\perp}(\mathbf{q}) e^{i (\mathbf{q} - \mathbf{k}_{1}) \cdot \boldsymbol{\delta}_{\chi}^{(1)}} e^{i (\mathbf{k}_{2} - \mathbf{q}) \cdot \boldsymbol{\delta}_{\eta}^{(2)}}
    \sum_{i,j} \, e^{i (\mathbf{q} - \mathbf{k}_{1}^{\null}) \cdot \mathbf{R}_{i}^{(1)}} e^{i (\mathbf{k}_{2}^{\null} - \mathbf{q}) \cdot \mathbf{R}_{j}^{(2)}}
    \\
    &= \frac{1}{A_{\text{uc}}} \sum_{i,j} \sum_{\mathbf{q}} \sum_{\eta,\chi} \Psi_{\mu,\mathbf{k}_{1}}^{(\chi_{1}^{\null})*} \Psi_{\nu,\mathbf{k}_{2}}^{(\eta_{2}^{\null})} t_{\perp}(\mathbf{k}_{1} + \mathbf{G}_{i}^{(1)}) e^{i \mathbf{G}_{i}^{(1)} \cdot \boldsymbol{\delta}_{\chi}^{(1)}} e^{-i \mathbf{G}_{j}^{(2)} \cdot \boldsymbol{\delta}_{\eta}^{(2)}} \delta_{\mathbf{k}_{2} - \mathbf{k}_{1}, \mathbf{G}_{i}^{(1)} - \mathbf{G}_{j}^{(2)}},
    \label{TBK_HT_general}
\end{split}
\end{equation}
where $\{\mathbf{G}_{i}\}$ is the set of reciprocal lattice vectors and we made use of $\sum_{i} e^{i \mathbf{k} \cdot \mathbf{R}_{i}} = N \sum_{j} \delta_{\mathbf{k},\mathbf{G}_{j}}$ to obtain the second line. In writing Eq. (\ref{TBK_HT_general}), we have not made any specific assumptions about the form of $t(\mathbf{q})$ nor the details of the individual layers. Let us now consider precisely the TBK system defined by $\mathscr{L}$ near $1/3$ filling and expand the monolayer Hamiltonians around their respective Dirac points. In this case, we define the local momenta according to $\mathbf{k}_{1} = \mathbf{K}_{\tau}^{\bar{\vartheta}} + \mathbf{k}$ and $\mathbf{k}_{2} = \mathbf{K}_{\tau}^{\vartheta} + \mathbf{p}$, such that $|\mathbf{k}|,|\mathbf{p}| \ll |\mathbf{K}_{\tau}|$,
\begin{subequations}
\begin{gather}
    \widetilde{\mathcal{H}}_{\mathbf{k}_{1}}^{\text{D}(1)} \simeq \widetilde{\mathscr{H}}_{\mathbf{k}^{\vartheta}}^{\text{D}},
    \qquad
    \widetilde{\mathcal{H}}_{\mathbf{k}_{2}}^{\text{D}(2)} \simeq \widetilde{\mathscr{H}}_{\mathbf{p}^{\bar{\vartheta}}}^{\text{D}},
    \qquad
    \mathcal{H}_{\text{T},\mathbf{k}_{1}\mathbf{k}_{2}}^{\mu\nu} \simeq \mathscr{H}_{\text{T},\mathbf{k}\mathbf{p}}^{\tau,\mu\nu}
    \\
    \mathscr{H}_{\text{T},\mathbf{k}\mathbf{p}}^{\tau,\mu\nu} = \frac{1}{A_{\text{uc}}} \sum_{i} \sum_{\eta,\chi} \psi_{\mu,\mathbf{k}}^{\tau(\chi_{1}^{\null})*} \psi_{\nu,\mathbf{p}}^{\tau(\eta_{2}^{\null})} t_{\perp}(\mathbf{K}_{\tau}^{\null} + \mathbf{G}_{i}^{\null}) e^{i \mathbf{G}_{i} \cdot (\boldsymbol{\delta}_{\chi} - \boldsymbol{\delta}_{\eta})}\delta_{\mathbf{p} - \mathbf{k}, (\mathbf{K}_{\tau}^{\bar{\vartheta}} + \mathbf{G}_{i}^{\bar{\vartheta}}) - (\mathbf{G}_{i}^{\vartheta}+\mathbf{K}_{\tau}^{\vartheta})},
\end{gather}
\label{TBK_onethird_LE_Hamiltonian}%
\end{subequations}
where we have noted that the Kronecker delta can only be satisfied when $\mathbf{G}_{i} = \mathbf{G}_{j}$ due to $\mathbf{k}$ and $\mathbf{p}$ being restricted to small momenta and not allowing for large momentum scattering. For our particular case of TBK, see Eq. (\ref{Kagome_A_matrix_expanded}), the Dirac state overlaps may be written as (suppressing gauge-fixing factors here for a moment)
\begin{equation}
\begin{split}
    \psi_{+,\mathbf{k}}^{\tau(\chi_{1}^{\null})*} \psi_{+,\mathbf{p}}^{\tau(\eta_{2}^{\null})} &= \frac{(1-2\delta_{\chi,A}) (1-2\delta_{\eta,A})}{3} \left[ \cos \left( \frac{\phi_{\mathbf{k}} - \phi_{\mathbf{p}} + \theta}{2} + (S_{\chi} - S_{\eta}) \frac{2\pi}{3} \right) + \tau \cos \left( \frac{\phi_{\mathbf{k}} + \phi_{\mathbf{p}}}{2} + (S_{\chi} + S_{\eta}) \frac{2\pi}{3} \right) \right],
    \\
    \psi_{+,\mathbf{k}}^{\tau(\chi_{1}^{\null})*} \psi_{-,\mathbf{p}}^{\tau(\eta_{2}^{\null})} &= \frac{(1-2\delta_{\chi,A}) (1-2\delta_{\eta,A})}{3} \left[ \sin \left( \frac{\phi_{\mathbf{k}} + \phi_{\mathbf{p}}}{2} + (S_{\chi} + S_{\eta}) \frac{2\pi}{3} \right) - \tau \sin \left( \frac{\phi_{\mathbf{k}} - \phi_{\mathbf{p}} + \theta}{2} + (S_{\chi} - S_{\eta}) \frac{2\pi}{3} \right) \right],
    \\
    \psi_{-,\mathbf{k}}^{\tau(\chi_{1}^{\null})*} \psi_{-,\mathbf{p}}^{\tau(\eta_{2}^{\null})} &= \frac{(1-2\delta_{\chi,A}) (1-2\delta_{\eta,A})}{3} \left[ \cos \left( \frac{\phi_{\mathbf{k}} - \phi_{\mathbf{p}} + \theta}{2} + (S_{\chi} - S_{\eta}) \frac{2\pi}{3} \right) - \tau \cos \left( \frac{\phi_{\mathbf{k}} + \phi_{\mathbf{p}}}{2} + (S_{\chi} + S_{\eta}) \frac{2\pi}{3} \right) \right].
\end{split}
\end{equation}
The only remaining task we have in constructing the continuum Hamiltonian for TBK is to truncate the sum over reciprocal lattice vectors. The number of terms retained is determined by how quickly $t_{\perp}(\mathbf{q})$ decays with $q$.

\begin{figure}
    \centering
    \includegraphics[width=0.8\linewidth]{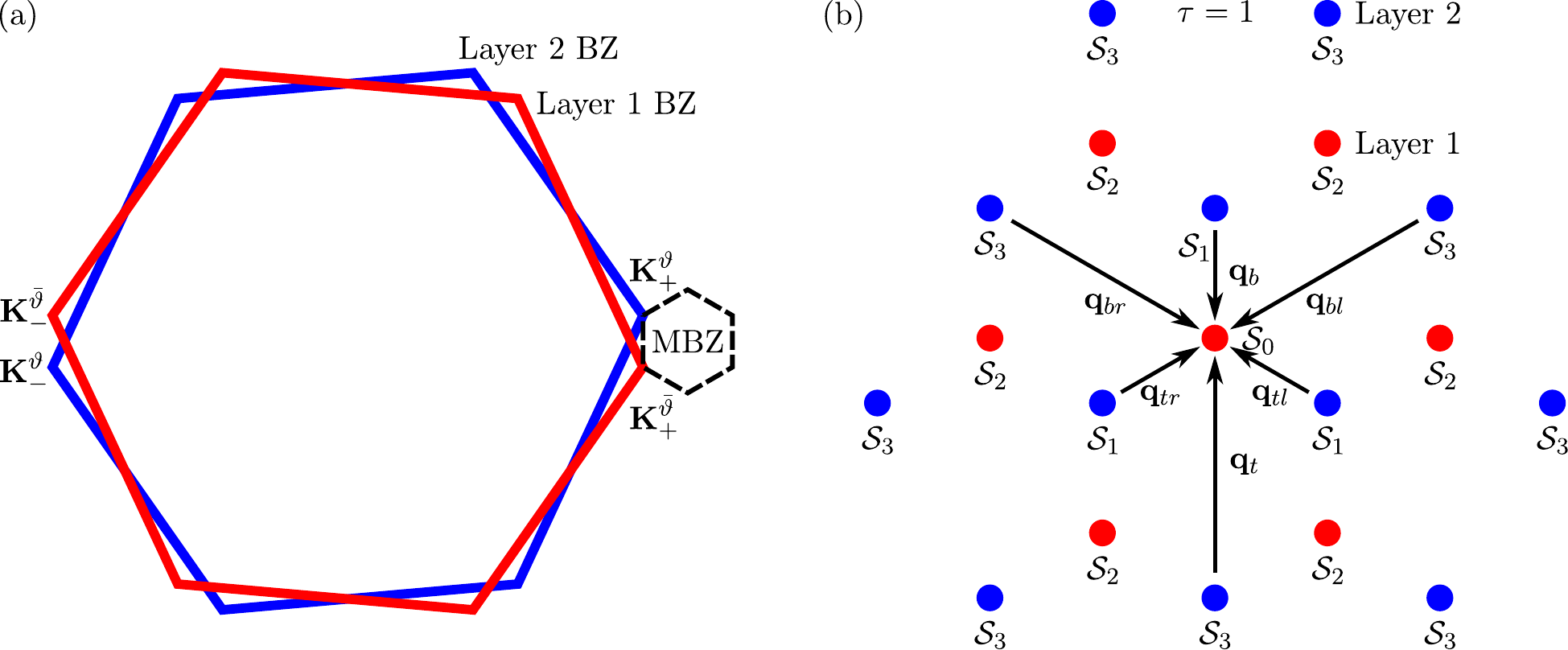}
    \caption{(a): Construction of the MBZ from the constituent monolayer BZs with layer 1 rotated by $\bar{\vartheta} = -\vartheta$ and layer 2 rotated by $\vartheta$. (b): Illustration of the $q$-lattice generated by the momentum conserving processes connecting the two layers for $\tau = 1$. We denote the hopping shell of a point by $\mathcal{S}_{n}$ with $n$ being the number of steps required to connect that point to the point with a Hamiltonian possessing no momentum shift, $\mathcal{S}_{0}$. The minimal model described by Eq. (\ref{BM_tunnelling_minimum}) which retains only terms situated at the set of Dirac points located closest to the origin, $\mathcal{D}_{1}$, connects nearest-neighbours of the $q$-lattice via $\mathbf{q}_{b}$, $\mathbf{q}_{tr}$, and $\mathbf{q}_{tl}$. Including the second-closest set of Dirac points, $\mathcal{D}_{2}$, described by Eq. (\ref{BM_tunnelling_2ndSet}) introduces three new vectors that third-nearest-neighbour sites of the $q$-lattice.}
    \label{q_lattice_illustration}
\end{figure}

Let us assume that $t_{\perp}(\mathbf{q})$ decays sufficiently rapidly that we may retain the minimum number of terms generated by the sum over reciprocal lattice vectors in Eq. (\ref{TBK_onethird_LE_Hamiltonian}), $\mathbf{G}_{0} = \mathbf{0}$, $\tau \mathbf{G}_{1} = -\tau \mathbf{b}_{1}$, and $\tau \mathbf{G}_{2} = -\tau(\mathbf{b}_{1} + \mathbf{b}_{2})$. This truncation corresponds to picking the three equivalent Dirac points within the monolayer's first BZ and so are equidistant from the origin. Other choices of $\mathbf{G}_{i}$ will place the argument of $t_{\perp}(\mathbf{K}_{\tau} + \mathbf{G}_{i})$ further from the origin and thus correspond to small corrections. The TBK interlayer tunnelling Hamiltonian may therefore be written as (including gauge-fixing factors)
\begin{subequations}
\begin{gather}
    \mathscr{H}_{\text{T},\mathbf{k}\mathbf{p}}^{\tau} = e^{-i\tau(\phi_{\mathbf{k}}-\phi_{\mathbf{p}} + \theta)/2} \left[ T_{b,\mathbf{k}\mathbf{p}}^{\tau} \delta_{\mathbf{p} - \mathbf{k}, \tau \mathbf{q}_{b}}^{\null} + T_{tr,\mathbf{k}\mathbf{p}}^{\tau} \delta_{\mathbf{p} - \mathbf{k}, \tau \mathbf{q}_{tr}}^{\null} + T_{tl,\mathbf{k}\mathbf{p}}^{\tau} \delta_{\mathbf{p} - \mathbf{k}, \tau \mathbf{q}_{tl}}^{\null} \right],
    \\
    \begin{split}
        T_{b,\mathbf{k}\mathbf{p}}^{\tau} &= \omega_{0} \sum_{\chi,\eta} \begin{pmatrix}
            \psi_{-,\mathbf{k}}^{\tau(\chi)*} \psi_{-,\mathbf{p}}^{\tau(\eta)} & \psi_{-,\mathbf{k}}^{\tau(\chi)*} \psi_{+,\mathbf{p}}^{\tau(\eta)} \\
            \psi_{+,\mathbf{k}}^{\tau(\chi)*} \psi_{-,\mathbf{p}}^{\tau(\eta)} & \psi_{+,\mathbf{k}}^{\tau(\chi)*} \psi_{+,\mathbf{p}}^{\tau(\eta)}
        \end{pmatrix}
        \\
        &=
        \frac{4 \omega_{0}}{3}
        \begin{pmatrix}
            \cos \left( \frac{\phi_{\mathbf{k}} - \phi_{\mathbf{p}} + \theta}{2} \right) - \tau \cos \left( \frac{\phi_{\mathbf{k}} + \phi_{\mathbf{p}}}{2} \right) & \sin \left( \frac{\phi_{\mathbf{k}} + \phi_{\mathbf{p}}}{2} \right) + \tau \sin \left( \frac{\phi_{\mathbf{k}} - \phi_{\mathbf{p}} + \theta}{2} \right) \\
            \sin \left( \frac{\phi_{\mathbf{k}} + \phi_{\mathbf{p}}}{2} \right) - \tau \sin \left( \frac{\phi_{\mathbf{k}} - \phi_{\mathbf{p}} + \theta}{2} \right) & \cos \left( \frac{\phi_{\mathbf{k}} - \phi_{\mathbf{p}} + \theta}{2} \right) + \tau \cos \left( \frac{\phi_{\mathbf{k}} + \phi_{\mathbf{p}}}{2} \right)
        \end{pmatrix},
        \end{split}
    \\
    \begin{split}
        T_{tr,\mathbf{k}\mathbf{p}}^{\tau} &= \omega_{0} \sum_{\chi,\eta} \begin{pmatrix}
            \psi_{-,\mathbf{k}}^{\tau(\chi)*} \psi_{-,\mathbf{p}}^{\tau(\eta)} & \psi_{-,\mathbf{k}}^{\tau(\chi)*} \psi_{+,\mathbf{p}}^{\tau(\eta)} \\
            \psi_{+,\mathbf{k}}^{\tau(\chi)*} \psi_{-,\mathbf{p}}^{\tau(\eta)} & \psi_{+,\mathbf{k}}^{\tau(\chi)*} \psi_{+,\mathbf{p}}^{\tau(\eta)}
        \end{pmatrix}
        e^{-i \tau \mathbf{b}_{1} \cdot (\boldsymbol{\delta}_{\chi}-\boldsymbol{\delta}_{\eta})}
        \\
        &=
        \frac{4 \omega_{0}}{3}
        \begin{pmatrix}
            \cos \left( \frac{\phi_{\mathbf{k}} - \phi_{\mathbf{p}} + \theta}{2} \right) - \tau \cos \left( \frac{\phi_{\mathbf{k}} + \phi_{\mathbf{p}}}{2} - \frac{2\pi}{3} \right) & \sin \left( \frac{\phi_{\mathbf{k}} + \phi_{\mathbf{p}}}{2} - \frac{2\pi}{3} \right) + \tau \sin \left( \frac{\phi_{\mathbf{k}} - \phi_{\mathbf{p}} + \theta}{2} \right) \\
            \sin \left( \frac{\phi_{\mathbf{k}} + \phi_{\mathbf{p}}}{2} - \frac{2\pi}{3} \right) - \tau \sin \left( \frac{\phi_{\mathbf{k}} - \phi_{\mathbf{p}} + \theta}{2} \right) & \cos \left( \frac{\phi_{\mathbf{k}} - \phi_{\mathbf{p}} + \theta}{2} \right) + \tau \cos \left( \frac{\phi_{\mathbf{k}} + \phi_{\mathbf{p}}}{2} - \frac{2\pi}{3} \right)
        \end{pmatrix},
    \end{split}
    \\
    \begin{split}
        T_{tl,\mathbf{k}\mathbf{p}}^{\tau} &= \omega_{0} \sum_{\chi,\eta} \begin{pmatrix}
            \psi_{-,\mathbf{k}}^{\tau(\chi)*} \psi_{-,\mathbf{p}}^{\tau(\eta)} & \psi_{-,\mathbf{k}}^{\tau(\chi)*} \psi_{+,\mathbf{p}}^{\tau(\eta)} \\
            \psi_{+,\mathbf{k}}^{\tau(\chi)*} \psi_{-,\mathbf{p}}^{\tau(\eta)} & \psi_{+,\mathbf{k}}^{\tau(\chi)*} \psi_{+,\mathbf{p}}^{\tau(\eta)}
        \end{pmatrix}
    e^{-i \tau (\mathbf{b}_{1} + \mathbf{b}_{2}) \cdot (\boldsymbol{\delta}_{\chi}-\boldsymbol{\delta}_{\eta})}
    \\
    &=
    \frac{4 \omega_{0}}{3}
        \begin{pmatrix}
            \cos \left( \frac{\phi_{\mathbf{k}} - \phi_{\mathbf{p}} + \theta}{2} \right) - \tau \cos \left( \frac{\phi_{\mathbf{k}} + \phi_{\mathbf{p}}}{2} + \frac{2\pi}{3} \right) & \sin \left( \frac{\phi_{\mathbf{k}} + \phi_{\mathbf{p}}}{2} + \frac{2\pi}{3} \right) + \tau \sin \left( \frac{\phi_{\mathbf{k}} - \phi_{\mathbf{p}} + \theta}{2} \right) \\
            \sin \left( \frac{\phi_{\mathbf{k}} + \phi_{\mathbf{p}}}{2} + \frac{2\pi}{3} \right) - \tau \sin \left( \frac{\phi_{\mathbf{k}} - \phi_{\mathbf{p}} + \theta}{2} \right) & \cos \left( \frac{\phi_{\mathbf{k}} - \phi_{\mathbf{p}} + \theta}{2} \right) + \tau \cos \left( \frac{\phi_{\mathbf{k}} + \phi_{\mathbf{p}}}{2} + \frac{2\pi}{3} \right)
        \end{pmatrix},
    \end{split}
\end{gather}
\label{BM_tunnelling_minimum}
\end{subequations}
where $\omega_{0} = t_{\perp}(\mathbf{K}_{\tau})/A_{\text{uc}}$ and
\begin{subequations}
\begin{gather}
    \mathbf{q}_{b} = (R_{\bar{\vartheta}} - R_{\vartheta}) \mathbf{K}_{+} = \frac{8\pi}{3a} \sin\left(\frac{\theta}{2}\right)
    \begin{pmatrix}
        0 \\ -1
    \end{pmatrix},
    \\
    \mathbf{q}_{tr}^{\null} = (R_{\bar{\vartheta}} - R_{\vartheta})(\mathbf{K}_{+} - \mathbf{b}_{1}) = \frac{4\pi}{3a} \sin\left(\frac{\theta}{2}\right)
    \begin{pmatrix}
        \sqrt{3} \\ 1
    \end{pmatrix},
    \\
    \mathbf{q}_{tl}^{\null} = (R_{\bar{\vartheta}} - R_{\vartheta})(\mathbf{K}_{+} - \mathbf{b}_{1} - \mathbf{b}_{2}) = \frac{4\pi}{3a} \sin\left(\frac{\theta}{2}\right)
    \begin{pmatrix}
        -\sqrt{3} \\ 1
    \end{pmatrix},
\end{gather}
\end{subequations}
generate a \textit{$q$-lattice} describing the allowed tunnelling processes between the layers subject to momentum conservation. We illustrate the emergent moir\'{e} BZ (MBZ) and resulting $q$-lattice in Fig. \ref{q_lattice_illustration}. By picturing each layer's Hamiltonian as a point on the $q$-lattice and assigning a momentum of $\mathbf{k}$ to the starting Hamiltonian of layer 1 located at $\mathcal{S}_{0}$, we determine the momentum appearing in a given Hamiltonian by adding the appropriate combination of $q$-vectors to connect it back to $\mathcal{S}_{0}$. For the tunnelling Hamiltonian described in Eq. (\ref{BM_tunnelling_minimum}), we can only achieve this through nearest-neighbour hopping around the $q$-lattice.

To show how to construct the effective continuum Hamiltonian for TBK from Eq. (\ref{BM_tunnelling_minimum}), let us write down the Hamiltonians as we increase the number of shells included in the $q$-lattice. The simplest Hamiltonian includes both $\mathcal{S}_{1}$ and $\mathcal{S}_{0}$,
\begin{equation}
    \mathscr{H}_{\text{TBK}} = \begin{pmatrix}
        \widetilde{\mathscr{H}}_{\mathbf{k}^{\vartheta}}^{\text{D}} & T_{b,\mathbf{k}}^{\tau} & T_{tr,\mathbf{k}}^{\tau} & T_{tl,\mathbf{k}}^{\tau} \\
        T_{b,\mathbf{k}}^{\tau\dagger} & \widetilde{\mathscr{H}}_{\mathbf{k}^{\bar{\vartheta}}+\tau\mathbf{q}_{b}^{\bar{\vartheta}}}^{\text{D}} & 0 & 0 \\
        T_{tr,\mathbf{k}}^{\tau\dagger} & 0 & \widetilde{\mathscr{H}}_{\mathbf{k}^{\bar{\vartheta}}+\tau\mathbf{q}_{tr}^{\bar{\vartheta}}}^{\text{D}} & 0 \\
        T_{tl,\mathbf{k}}^{\tau\dagger} & 0 & 0 & \widetilde{\mathscr{H}}_{\mathbf{k}^{\bar{\vartheta}}+\tau\mathbf{q}_{tl}^{\bar{\vartheta}}}^{\text{D}}
    \end{pmatrix}.
    \label{TBK_Hamiltonian_1shell}
\end{equation}
Adding in $\mathcal{S}_{2}$, replacing $\widetilde{\mathscr{H}}_{\mathbf{k}^{\vartheta}+n_{b}\tau\mathbf{q}_{b}^{\vartheta}+n_{tr}\tau\mathbf{q}_{tr}^{\vartheta}+n_{tl}\tau\mathbf{q}_{tl}^{\vartheta}}^{\text{D}}$ by $\mathscr{H}_{lmn}^{\vartheta,\tau}$, $T_{i,\mathbf{k}+n_{b}\tau\mathbf{q}_{b}^{\vartheta}+n_{tr}\tau\mathbf{q}_{tr}^{\vartheta}+n_{tl}\tau\mathbf{q}_{tl}^{\vartheta}}^{\tau}$ by $T_{i,lmn}^{\tau}$, and using $\bar{m} = -m$ for sake of notation,
\begin{equation}
    \mathscr{H}_{\text{TBK}} = \begin{pmatrix}
        \mathscr{H}_{000}^{\vartheta,\tau} & T_{b,000}^{\tau} & T_{tr,000}^{\tau} & T_{tl,000}^{\tau} & 0 & 0 & 0 & 0 & 0 & 0 \\
        T_{b,000}^{\tau\dagger} & \mathscr{H}_{100}^{\bar{\vartheta},\tau} & 0 & 0 & T_{tr,1\bar{1}0}^{\tau\dagger} & T_{tl,10\bar{1}}^{\tau\dagger} & 0 & 0 & 0 & 0 \\
        T_{tr,000}^{\tau\dagger} & 0 & \mathscr{H}_{010}^{\bar{\vartheta},\tau} & 0 & 0 & 0 & T_{b,\bar{1}10}^{\tau\dagger} & T_{tl,01\bar{1}}^{\tau\dagger} & 0 & 0 \\
        T_{tl,000}^{\tau\dagger} & 0 & 0 & \mathscr{H}_{001}^{\bar{\vartheta},\tau} & 0 & 0 & 0 & 0 & T_{b,\bar{1}01}^{\tau\dagger} & T_{tr,0\bar{1}1}^{\tau\dagger} \\
        0 & T_{tr,1\bar{1}0}^{\tau} & 0 & 0 & \mathscr{H}_{1\bar{1}0}^{\bar{\vartheta},\tau} & 0 & 0 & 0 & 0 & 0 \\
        0 & T_{tl,10\bar{1}}^{\tau} & 0 & 0 & 0 & \mathscr{H}_{10\bar{1}}^{\bar{\vartheta},\tau}  & 0 & 0 & 0 & 0 \\
        0 & 0 & T_{b,\bar{1}10}^{\tau} & 0 & 0 & 0 & \mathscr{H}_{\bar{1}10}^{\bar{\vartheta},\tau} & 0 & 0 & 0 \\
        0 & 0 & T_{tl,01\bar{1}}^{\tau} & 0 & 0 & 0 & 0 & \mathscr{H}_{01\bar{1}}^{\bar{\vartheta},\tau} & 0 & 0 \\
        0 & 0 & 0 & T_{b,\bar{1}01}^{\tau} & 0 & 0 & 0 & 0 & \mathscr{H}_{\bar{1}01}^{\bar{\vartheta},\tau} & 0 \\
        0 & 0 & 0 & T_{tr,0\bar{1}1}^{\tau} & 0 & 0 & 0 & 0 & 0 & \mathscr{H}_{0\bar{1}1}^{\bar{\vartheta},\tau}
    \end{pmatrix}.
\end{equation}
Lastly, we note that the integers labeling the sites of the $q$-lattice must always sum to 0 or 1. Specifically, when the block diagonal component of the Hamiltonian corresponds to $L_{b}^{\bar{\vartheta}}$, $n_{b} + n_{tr} + n_{tl} = 0$, whereas the case in which the component corresponds to $L_{t}^{\vartheta}$, $n_{b} + n_{tr} + n_{tl} = 1$. We note this due to its usefulness in constructing these Hamiltonians numerically and helps differentiate the two classes of $q$-lattice sites quickly.

\subsubsection{Compact Notation}

Before we consider the symmetries of TBK, let us note that the construction method of the moir\'{e} Hamiltonian outlined above is, using the nomenclature of Ref. \cite{Bernevig2021}, a ``triangular model'' centred on a corner of the MBZ. We may write the full Hamiltonian in a fairly compact form, analagous to Ref. \cite{Song2019}, as ($\Phi_{\mathbf{k},\mathbf{p}}^{\tau} = e^{-\tau i\tilde{\phi}_{\mathbf{k}\mathbf{p}}}$)
\begin{subequations}
    \begin{gather}
        \left[ \mathscr{H}_{\text{TBK},\mathbf{k}}^{\tau} \right]_{\mathbf{Q},\mathbf{Q}'} = \widetilde{\mathscr{H}}_{\mathbf{k} + \mathbf{Q}}^{\text{D},\tau} \delta_{\mathbf{Q},\mathbf{Q}'}^{\null}
        + \sum_{j \in \{b,tr,tl\}} \left( \mathscr{H}_{j,\mathbf{k},\mathbf{Q},\mathbf{Q}'}^{\tau} + \mathscr{H}_{j,\mathbf{k},\mathbf{Q},\mathbf{Q}'}^{\tau\dagger} \right),
        \\
        \mathscr{H}_{j,\mathbf{k},\mathbf{Q},\mathbf{Q}'}^{\tau} = \delta_{\mathbf{Q}'-\mathbf{Q},\tau \mathbf{q}_{j}}^{\null} \Phi_{\mathbf{k}+\mathbf{Q},\mathbf{k}+\mathbf{Q}'}^{\tau} T_{j,\mathbf{k}+\mathbf{Q},\mathbf{k}+\mathbf{Q}'}^{\tau},
        \qquad
        \mathscr{H}_{j,\mathbf{k},\mathbf{Q},\mathbf{Q}'}^{\tau\dagger} = \delta_{\mathbf{Q}-\mathbf{Q}', \tau \mathbf{q}_{j}}^{\null} \Phi_{\mathbf{k}+\mathbf{Q}',\mathbf{k}+\mathbf{Q}}^{\tau*} T_{j,\mathbf{k}+\mathbf{Q}',\mathbf{k}+\mathbf{Q}}^{\tau \, \dagger},
        \\
        \begin{gathered}
            \mathbf{Q} = n_{b} \mathbf{q}_{b} + n_{tr} \mathbf{q}_{tr} + n_{tl} \mathbf{q}_{tl},
            \\
            \Rightarrow \quad \begin{matrix}
                L_{t}^{\vartheta}: Q_{+}^{\tau} = \{ \tau\mathbf{Q}|n_{b},n_{tr},n_{tl} \in \mathbb{Z} \text{ s.t. } n_{b}+n_{tr}+n_{tl} = 1 \} = \{ \mathcal{S}_{2n+1}| n \in \mathbb{N}_{0} \},
                \\
                L_{b}^{\bar{\vartheta}}: Q_{-}^{\tau} = \{ \tau\mathbf{Q}|n_{b},n_{tr},n_{tl} \in \mathbb{Z} \text{ s.t. } n_{b}+n_{tr}+n_{tl} = 0 \} = \{ \mathcal{S}_{2n}| n \in \mathbb{N}_{0} \}.
            \end{matrix}
        \end{gathered}
    \end{gather}
\end{subequations}
The $\Gamma_{\text{M}}$-centric approach takes on a similar form to the above corner centric construction and can be quickly deduced by letting $\mathbf{k} = \boldsymbol{k} + \tau \mathbf{q}_{tl}$ and $\mathbf{Q} = \boldsymbol{\mathcal{Q}} - \tau \mathbf{q}_{tl}$, which ultimately leaves the Hamiltonian largely unchanged, but yields different sets for the $q$-lattice sites belonging to each layer. Specifically, analogous to Ref. \cite{Song2019},
\begin{subequations}
    \begin{gather}
        \left[ \mathscr{H}_{\text{TBK},\boldsymbol{k}}^{\tau} \right]_{\boldsymbol{\mathcal{Q}},\boldsymbol{\mathcal{Q}}'} = \widetilde{\mathscr{H}}_{\boldsymbol{k} + \boldsymbol{\mathcal{Q}}}^{\text{D},\tau} \delta_{\boldsymbol{\mathcal{Q}},\boldsymbol{\mathcal{Q}}'}^{\null}
        + \sum_{j \in \{b,tr,tl\}} \left( \mathscr{H}_{j,\boldsymbol{k},\boldsymbol{\mathcal{Q}},\boldsymbol{\mathcal{Q}}'}^{\tau} + \mathscr{H}_{j,\boldsymbol{k},\boldsymbol{\mathcal{Q}},\boldsymbol{\mathcal{Q}}'}^{\tau\dagger} \right),
        \\
        \mathscr{H}_{j,\boldsymbol{k},\boldsymbol{\mathcal{Q}},\boldsymbol{\mathcal{Q}}'}^{\tau} = \delta_{\boldsymbol{\mathcal{Q}}'-\boldsymbol{\mathcal{Q}},\tau \mathbf{q}_{j}}^{\null} \Phi_{\boldsymbol{k}+\boldsymbol{\mathcal{Q}},\boldsymbol{k}+\boldsymbol{\mathcal{Q}}'}^{\tau} T_{j,\boldsymbol{k}+\boldsymbol{\mathcal{Q}},\boldsymbol{k}+\boldsymbol{\mathcal{Q}}'}^{\tau},
        \qquad
        \mathscr{H}_{j,\boldsymbol{k},\boldsymbol{\mathcal{Q}},\boldsymbol{\mathcal{Q}}'}^{\tau\dagger} = \delta_{\boldsymbol{\mathcal{Q}}-\boldsymbol{\mathcal{Q}}', \tau \mathbf{q}_{j}}^{\null} \Phi_{\boldsymbol{k}+\boldsymbol{\mathcal{Q}}',\boldsymbol{k}+\boldsymbol{\mathcal{Q}}}^{\tau*} T_{j,\boldsymbol{k}+\boldsymbol{\mathcal{Q}}',\boldsymbol{k}+\boldsymbol{\mathcal{Q}}}^{\tau \, \dagger},
        \\
        \begin{gathered}
            \boldsymbol{\mathcal{Q}} = n_{b} \mathbf{q}_{b} + n_{tr} \mathbf{q}_{tr} + (n_{tl}+1) \mathbf{q}_{tl},
            \\
            \Rightarrow \quad \begin{matrix}
                L_{t}^{\vartheta}: \mathcal{Q}_{+}^{\tau} = \{ \tau \boldsymbol{\mathcal{Q}}|n_{b},n_{tr},n_{tl} \in \mathbb{Z} \text{ s.t. } n_{b}+n_{tr}+n_{tl} = 1 \} = \{ \mathcal{S}_{2n+1}| n \in \mathbb{N}_{0} \},
                \\
                L_{b}^{\bar{\vartheta}}: \mathcal{Q}_{-}^{\tau} = \{ \tau \boldsymbol{\mathcal{Q}}|n_{b},n_{tr},n_{tl} \in \mathbb{Z} \text{ s.t. } n_{b}+n_{tr}+n_{tl} = 0 \} = \{ \mathcal{S}_{2n}| n \in \mathbb{N}_{0} \}.
            \end{matrix}
        \end{gathered}
    \end{gather}
\end{subequations}
This $\Gamma_{\text{M}}$-centric phrasing of the system will be of particular importance for writing the symmetries of TBK in a compact manner and in constructing their proofs due to the relationship between the two sets of $q$-lattice sites, $\forall \boldsymbol{\mathcal{Q}}_{i} \in \mathcal{Q}_{\mp}^{\tau}$, $-\boldsymbol{\mathcal{Q}}_{i} \in \mathcal{Q}_{\pm}^{\tau}$.

The tunnelling matrices can be written in a compact form upon substituting the sublattice projections of the Dirac listed in Eq. (\ref{Kagome_A_matrix_expanded}) into their general expressions in Eq. (\ref{BM_tunnelling_minimum}) and performing the sublattice sums. Specifically, we find that the $T$-matrices can be written compactly in terms of the Pauli matrices (supplemented with identity) acting in the Dirac subspace
\begin{subequations}
\begin{gather}
    T_{j,\mathbf{k}\mathbf{p}}^{\tau} = \frac{4\omega_{0}}{3} \left[ \cos \tilde{\phi}_{\mathbf{k}\mathbf{p}}^{\null} \sigma_{0}^{\null} + \sin \bar{\phi}_{\mathbf{k}\mathbf{p}}^{j} \sigma_{x}^{\null} + i \tau \sin \tilde{\phi}_{\mathbf{k}\mathbf{p}}^{\null} \sigma_{y}^{\null} - \tau \cos \bar{\phi}_{\mathbf{k}\mathbf{p}}^{j} \sigma_{z}^{\null} \right],
    \\
    \tilde{\phi}_{\mathbf{k}\mathbf{p}}^{\null} = \frac{1}{2} (\phi_{\mathbf{k}}^{\null} - \phi_{\mathbf{p}}^{\null} + \theta), \qquad \bar{\phi}_{\mathbf{k}\mathbf{p}}^{j} = \frac{1}{2} \left(\phi_{\mathbf{k}}^{\null} + \phi_{\mathbf{p}}^{\null} + (\delta_{j,tl}^{\null} -\delta_{j,tr}) \frac{2\pi}{3} \right).
\end{gather}
\end{subequations}
We note that the $T$-matrices are real.

\subsection{Exact Symmetries}

Through our construction by twisting around the $D_{3}$ symmetry site (i.e. where two equivalent triangles overlap in the $AA$ stacked system), the point group symmetry of the TBK system will also be $D_{3}$. However, some symmetries will only be present in the full two-valley description when writing the full Hamiltonian as a pair of disconnected single-valley Hamiltonians, $\mathscr{H}_{\text{TBK},\boldsymbol{k}} = \text{diag}(\mathscr{H}_{\text{TBK},\boldsymbol{k}}^{+},\mathscr{H}_{\text{TBK},\boldsymbol{k}}^{-})$, whilst new symmetries emerge within the single-valley description. For the full two-valley picture, $\mathscr{H}_{\text{TBK},\boldsymbol{k}}$ possesses $C_{3z}$, $C_{2y}$, and time-reversal symmetry, $\mathcal{T}$. In contrast, we note that the single-valley model, $\mathscr{H}_{\text{TBK},\boldsymbol{k}}^{\tau}$, only retains the $C_{3z}$ symmetry of the full system since this transformation preserves the valley quantum number, whereas $C_{2y}$ $(p_{x} \rightarrow p_{x})$ and $\mathcal{T}$ ($\mathbf{p} \rightarrow - \mathbf{p}$) exchange the monolayer valleys ($K \leftrightarrow K'$). The symmetry can be further reduced by distinguishing the layers, e.g. by application of an out-of-plane electric field, to break the dihedral $C_{2y}$ symmetry. However, two new symmetries emerge within the single-valley picture: $C_{2z}\mathcal{T}$ and $C_{2y} \mathcal{T}$ ($k_{y} \rightarrow -k_{y}$, $L_{t}^{\vartheta} \leftrightarrow L_{b}^{\bar{\vartheta}} \equiv \theta \rightarrow -\theta$), assuming the layers are identical for the latter. Given we do not include any terms that break $\mathcal{T}$, we shall focus on the symmetries of the single-valley model, which may be written as
\begin{subequations}
\begin{alignat}{2}
    C_{3z}^{\null} \, &: \quad \left[ \mathscr{H}_{\text{TBK},\boldsymbol{k}}^{\tau} \right]_{\boldsymbol{\mathcal{Q}},\boldsymbol{\mathcal{Q}}'} = R_{3z,\boldsymbol{\mathcal{Q}},\boldsymbol{\mathcal{P}}}^{\dagger} \left[ \mathscr{H}_{\text{TBK},R_{\frac{2\pi}{3}}\boldsymbol{k}}^{\tau} \right]_{\boldsymbol{\mathcal{P}},\boldsymbol{\mathcal{P}}'} R_{3z,\boldsymbol{\mathcal{P}}',\boldsymbol{\mathcal{Q}}'}^{\null},
    \quad
    &&R_{3z,\boldsymbol{\mathcal{Q}},\boldsymbol{\mathcal{Q}}'}^{\null} = \delta_{\boldsymbol{\mathcal{Q}},R_{\frac{2\pi}{3}}\boldsymbol{\mathcal{Q}}'},
    \label{C3z_operator_relation}
    \\
    C_{2x}^{\null} \, &: \quad \left[ \mathscr{H}_{\text{TBK},\boldsymbol{k}}^{\tau} \right]_{\boldsymbol{\mathcal{Q}},\boldsymbol{\mathcal{Q}}'} = R_{2x,\boldsymbol{\mathcal{Q}},\boldsymbol{\mathcal{P}}}^{\dagger} \left[ \mathscr{H}_{\text{TBK},\tilde{\boldsymbol{k}}}^{\tau (\theta \rightarrow -\theta)} \right]_{\boldsymbol{\mathcal{P}},\boldsymbol{\mathcal{P}}'} R_{2x,\boldsymbol{\mathcal{P}}',\boldsymbol{\mathcal{Q}}'}^{\null},
    && R_{2x,\boldsymbol{\mathcal{Q}},\boldsymbol{\mathcal{Q}}'}^{\null} = \delta_{\boldsymbol{\mathcal{Q}},\widetilde{\boldsymbol{\mathcal{Q}}}'}^{\null} \sigma_{z}^{\null} e^{i \tau (\phi_{\mathbf{k}+\boldsymbol{\mathcal{Q}}'}  + \xi_{\boldsymbol{\mathcal{Q}}'} \frac{\theta}{2})},
    \label{C2x_operator_relation}
    \\
    C_{2z}^{\null} \mathcal{T} \, &: \quad \left[ \mathscr{H}_{\text{TBK},\boldsymbol{k}}^{\tau} \right]_{\boldsymbol{\mathcal{Q}},\boldsymbol{\mathcal{Q}}'} = \mathscr{T}_{2z,\boldsymbol{\mathcal{Q}},\boldsymbol{\mathcal{P}}}^{\tau} \left[ \mathscr{H}_{\text{TBK},\boldsymbol{k}}^{\tau} \right]_{\boldsymbol{\mathcal{P}},\boldsymbol{\mathcal{P}}'} \mathscr{T}_{2z,\boldsymbol{\mathcal{P}},\boldsymbol{\mathcal{Q}}'}^{\tau},
    \quad
    &&\mathscr{T}_{2z,\boldsymbol{\mathcal{Q}},\boldsymbol{\mathcal{Q}}'}^{\tau} = \delta_{\boldsymbol{\mathcal{Q}},\boldsymbol{\mathcal{Q}}'}^{\null} e^{-i \tau (\phi_{\boldsymbol{k}+\boldsymbol{\mathcal{Q}}} + \xi_{\boldsymbol{\mathcal{Q}}} \frac{\theta}{2})} \mathcal{K},
    \label{C2zT_operator_relation}
    \\
    C_{2y}^{\null} \mathcal{T} \, &: \quad \left[ \mathscr{H}_{\text{TBK},\boldsymbol{k}}^{\tau} \right]_{\boldsymbol{\mathcal{Q}},\boldsymbol{\mathcal{Q}}'} = \mathscr{T}_{2y,\boldsymbol{\mathcal{Q}},\boldsymbol{\mathcal{P}}}^{\null} \left[ \mathscr{H}_{\text{TBK},\tilde{\boldsymbol{k}}}^{\tau (\theta \rightarrow -\theta)} \right]_{\boldsymbol{\mathcal{P}},\boldsymbol{\mathcal{P}}'} \mathscr{T}_{2y,\boldsymbol{\mathcal{P}}',\boldsymbol{\mathcal{Q}}'}^{\null},
    \quad
    &&\mathscr{T}_{2y,\boldsymbol{\mathcal{Q}},\boldsymbol{\mathcal{Q}}'}^{\null} = \delta_{\boldsymbol{\mathcal{Q}},\widetilde{\boldsymbol{\mathcal{Q}}}'}^{\null} \sigma_{z}^{\null} \mathcal{K},
    \label{C2yT_operator_relation}
\end{alignat}
\end{subequations}
where we introduce a tilde on vectors ($\tilde{\boldsymbol{k}} = (k_{x},-k_{y})$) to indicate reversal of their $y$-components, $\xi_{\boldsymbol{\mathcal{Q}} \in \mathcal{Q}_{\mp}^{\tau}} = \pm 1 = -\text{sgn}\left(n_{b} + n_{tr} + n_{tl} - \frac{1}{2} \right)$, $\mathcal{K}$ denotes complex conjugation, $\delta_{(a,b),(c,d)} = \delta_{a,c} \delta_{b,d}$, and we noted that $C_{2z}^{\null} \mathcal{T}$ and $C_{2y}^{\null} \mathcal{T}$ are their own respective inverses in the gauge-fixed monolayer eigenbasis. Moreover, we emphasise that the operators involving time-reversal must be anti-unitary. Below we provide proofs for each of these symmetry operations.

\subsubsection{Proof of $C_{3z}$ Operator}

We start by writing the operators out explicitly,
\begin{equation}
\begin{split}
    R_{3,\boldsymbol{\mathcal{Q}},\boldsymbol{\mathcal{P}}}^{\dagger} \bigg[ \mathscr{H}_{\text{TBK},R_{\frac{2\pi}{3}}\boldsymbol{k}}^{\tau} &\bigg]_{\boldsymbol{\mathcal{P}},\boldsymbol{\mathcal{P}}'} R_{3,\boldsymbol{\mathcal{P}}',\boldsymbol{\mathcal{Q}}'}^{\null}
    =
    \delta_{R_{\frac{2\pi}{3}} \boldsymbol{\mathcal{Q}},\boldsymbol{\mathcal{P}}}^{\null}
    \widetilde{\mathscr{H}}_{R_{\frac{2\pi}{3}}\boldsymbol{k} + \boldsymbol{\mathcal{P}}}^{\text{D},\tau} \delta_{\boldsymbol{\mathcal{P}},\boldsymbol{\mathcal{P}}'}^{\null}
    \delta_{\boldsymbol{\mathcal{P}}',R_{\frac{2\pi}{3}}\boldsymbol{\mathcal{Q}}'}^{\null}
    \\
    &+ \sum_{j \in \{b,tr,tl\}} \left( \delta_{R_{\frac{2\pi}{3}} \boldsymbol{\mathcal{Q}},\boldsymbol{\mathcal{P}}}^{\null} \mathscr{H}_{j,R_{\frac{2\pi}{3}}\boldsymbol{k},\boldsymbol{\mathcal{P}},\boldsymbol{\mathcal{P}}'}^{\tau}
    \delta_{\boldsymbol{\mathcal{P}}',R_{\frac{2\pi}{3}}\boldsymbol{\mathcal{Q}}'}^{\null}
    +
    \delta_{R_{\frac{2\pi}{3}} \boldsymbol{\mathcal{Q}},\boldsymbol{\mathcal{P}}}^{\null} \mathscr{H}_{j,R_{\frac{2\pi}{3}}\boldsymbol{k},\boldsymbol{\mathcal{P}},\boldsymbol{\mathcal{P}}'}^{\tau\dagger}
    \delta_{\boldsymbol{\mathcal{P}}',R_{\frac{2\pi}{3}}\boldsymbol{\mathcal{Q}}'}^{\null} \right),
\end{split}
\end{equation}
and consider the three terms separately. The first term is handled straightforwarly by noting that the block diagonal contributions exhibit continuous rotational symmetry ($C_{\infty}$),
\begin{equation}
\begin{split}
    \delta_{R_{\frac{2\pi}{3}} \boldsymbol{\mathcal{Q}},\boldsymbol{\mathcal{P}}}^{\null}
    \widetilde{\mathscr{H}}_{R_{\frac{2\pi}{3}}\boldsymbol{k} + \boldsymbol{\mathcal{P}}}^{\text{D},\tau} \delta_{\boldsymbol{\mathcal{P}},\boldsymbol{\mathcal{P}}'}^{\null}
    \delta_{\boldsymbol{\mathcal{P}}',R_{\frac{2\pi}{3}}\boldsymbol{\mathcal{Q}}'}^{\null}
    =
    \delta_{\boldsymbol{\mathcal{Q}},\boldsymbol{\mathcal{Q}}'}^{\null} \widetilde{\mathscr{H}}_{R_{\frac{2\pi}{3}} (\boldsymbol{k} + \boldsymbol{\mathcal{Q}})}^{\text{D},\tau}
    =
    \delta_{\boldsymbol{\mathcal{Q}},\boldsymbol{\mathcal{Q}}'}^{\null} \widetilde{\mathscr{H}}_{\boldsymbol{k} + \boldsymbol{\mathcal{Q}}}^{\text{D},\tau},
\end{split}
\end{equation}
thus matching the relation of Eq. (\ref{C3z_operator_relation}). The interlayer tunnelling terms require a little more care in their treatment. Specifically, by defining all angles on the range $\phi_{\boldsymbol{k}} \in [-\pi,\pi)$, we will have to make use of
\begin{equation}
    \phi_{R_{\frac{2\pi}{3}}\boldsymbol{k}} = \phi_{\boldsymbol{k}} + \frac{2\pi}{3} - 2\pi s_{\boldsymbol{k}}^{\Theta},
    \qquad
    s_{\boldsymbol{k}}^{\Theta} = \begin{cases}
        1, \quad & \phi_{\boldsymbol{k}} \geq \pi/3 \\
        0, \quad & \phi_{\boldsymbol{k}} < \pi/3
    \end{cases},
\end{equation}
which leads us to write
\begin{equation}
    \tilde{\phi}_{R_{\frac{2\pi}{3}}\boldsymbol{k},R_{\frac{2\pi}{3}}\boldsymbol{p}}^{\null} = \tilde{\phi}_{\boldsymbol{k},\boldsymbol{p}}^{\null} - \pi (s_{\boldsymbol{k}}^{\Theta} - s_{\boldsymbol{p}}^{\Theta}),
    \qquad
    \bar{\phi}_{R_{\frac{2\pi}{3}}\boldsymbol{k},R_{\frac{2\pi}{3}}\boldsymbol{p}}^{j} = \bar{\phi}_{\boldsymbol{k},\boldsymbol{p}}^{j} + \frac{2\pi}{3} - \pi (s_{\boldsymbol{k}}^{\Theta} + s_{\boldsymbol{p}}^{\Theta}).
\end{equation}
From these relations, we see that
\begin{equation}
    \colVec{\exp \left[ -i\tilde{\phi}_{R_{\frac{2\pi}{3}}\boldsymbol{k},R_{\frac{2\pi}{3}}\boldsymbol{p}}^{\null} \right] \\ \cos \tilde{\phi}_{R_{\frac{2\pi}{3}}\boldsymbol{k},R_{\frac{2\pi}{3}}\boldsymbol{p}}^{\null} \\ \sin \tilde{\phi}_{R_{\frac{2\pi}{3}}\boldsymbol{k},R_{\frac{2\pi}{3}}\boldsymbol{p}}^{\null}}
    =
    (1 - 2\delta_{s_{\boldsymbol{k}}^{\Theta}+s_{\boldsymbol{p}}^{\Theta},1})
    \colVec{ \exp[-i\tilde{\phi}_{\boldsymbol{k},\boldsymbol{p}}^{\null}] \\ \cos \tilde{\phi}_{\boldsymbol{k},\boldsymbol{p}}^{\null} \\ \sin \tilde{\phi}_{\boldsymbol{k},\boldsymbol{p}}^{\null}},
\end{equation}
whilst
\begin{equation}
    \colVec{\cos \bar{\phi}_{R_{\frac{2\pi}{3}}\boldsymbol{k},R_{\frac{2\pi}{3}}\boldsymbol{p}}^{j} \\ \sin \bar{\phi}_{R_{\frac{2\pi}{3}}\boldsymbol{k},R_{\frac{2\pi}{3}}\boldsymbol{p}}^{j}}
    =
    (1 - 2\delta_{s_{\boldsymbol{k}}^{\Theta}+s_{\boldsymbol{p}}^{\Theta},1})
    \colVec{\cos \bar{\phi}_{\boldsymbol{k},\boldsymbol{p}}^{j-1} \\ \sin \bar{\phi}_{\boldsymbol{k},\boldsymbol{p}}^{j-1}},
\end{equation}
where $j-1$ indicates cyclic permutation of the $q$-vectors according to the ordering $\{\mathbf{q}_{tr},\mathbf{q}_{b},\mathbf{q}_{tl}\}$. In otherwords, the $C_{3z}$ operator transforms $\{\bar{\phi}_{\boldsymbol{k},\boldsymbol{p}}^{tr},\bar{\phi}_{\boldsymbol{k},\boldsymbol{p}}^{b},\bar{\phi}_{\boldsymbol{k},\boldsymbol{p}}^{tl}\} \rightarrow \{\bar{\phi}_{\boldsymbol{k},\boldsymbol{p}}^{b},\bar{\phi}_{\boldsymbol{k},\boldsymbol{p}}^{tl},\bar{\phi}_{\boldsymbol{k},\boldsymbol{p}}^{tr}\}$, which we can interpret as a rotation given by $R_{\frac{2\pi}{3}}^{-1}$ of the labels.

Now let us consider the first of the interlayer tunnelling terms,
\begin{equation}
\begin{split}
    \sum_{j} \delta_{R_{\frac{2\pi}{3}} \boldsymbol{\mathcal{Q}},\boldsymbol{\mathcal{P}}}^{\null} &\mathscr{H}_{j,R_{\frac{2\pi}{3}}\boldsymbol{k},\boldsymbol{\mathcal{P}},\boldsymbol{\mathcal{P}}'}^{\tau}
    \delta_{\boldsymbol{\mathcal{P}}',R_{\frac{2\pi}{3}}\boldsymbol{\mathcal{Q}}'}^{\null}
    \\
    &= \sum_{j} \delta_{R_{\frac{2\pi}{3}} \boldsymbol{\mathcal{Q}},\boldsymbol{\mathcal{P}}}^{\null}
    \delta_{\boldsymbol{\mathcal{P}}'-\boldsymbol{\mathcal{P}},\tau \mathbf{q}_{j}} \exp \left[ -i\tau \tilde{\phi}_{R_{\frac{2\pi}{3}}\boldsymbol{k} + \boldsymbol{\mathcal{P}},R_{\frac{2\pi}{3}}\boldsymbol{k} + \boldsymbol{\mathcal{P}}'}^{\null} \right] T_{j,R_{\frac{2\pi}{3}}\boldsymbol{k} + \boldsymbol{\mathcal{P}},R_{\frac{2\pi}{3}}\boldsymbol{k} + \boldsymbol{\mathcal{P}}'}^{\tau}
    \delta_{\boldsymbol{\mathcal{P}}',R_{\frac{2\pi}{3}}\boldsymbol{\mathcal{Q}}'}^{\null}
    \\
    &=
    \sum_{j} \delta_{R_{\frac{2\pi}{3}}(\boldsymbol{\mathcal{Q}}'-\boldsymbol{\mathcal{Q}}),\tau \mathbf{q}_{j}}
    \exp \left[ -i\tau \tilde{\phi}_{R_{\frac{2\pi}{3}}(\boldsymbol{k} + \boldsymbol{\mathcal{Q}}),R_{\frac{2\pi}{3}} (\boldsymbol{k} + \boldsymbol{\mathcal{Q}}')}^{\null} \right]
    T_{j,R_{\frac{2\pi}{3}}(\boldsymbol{k} + \boldsymbol{\mathcal{Q}}),R_{\frac{2\pi}{3}}(\boldsymbol{k} + \boldsymbol{\mathcal{Q}}')}^{\tau}
    \\
    &= \sum_{j} \delta_{\boldsymbol{\mathcal{Q}}'-\boldsymbol{\mathcal{Q}},\tau R_{\frac{2\pi}{3}}^{-1} \mathbf{q}_{j}}
    (1-2\delta_{s_{\boldsymbol{k}+\boldsymbol{\mathcal{Q}}}^{\Theta}+s_{\boldsymbol{k}+\boldsymbol{\mathcal{Q}'}}^{\Theta},1})^{2} \exp \left[ -i\tau \tilde{\phi}_{\boldsymbol{k} + \boldsymbol{\mathcal{Q}},\boldsymbol{k} + \boldsymbol{\mathcal{Q}}'}^{\null} \right]
    T_{j-1,\boldsymbol{k} + \boldsymbol{\mathcal{Q}},\boldsymbol{k} + \boldsymbol{\mathcal{Q}}'}^{\tau}
    \\
    &= \sum_{j} \delta_{\boldsymbol{\mathcal{Q}}'-\boldsymbol{\mathcal{Q}},\tau \mathbf{q}_{j-1}} e^{-i\tau \tilde{\phi}_{\boldsymbol{k} + \boldsymbol{\mathcal{Q}},\boldsymbol{k} + \boldsymbol{\mathcal{Q}}'}^{\null}}
    T_{j-1,\boldsymbol{k} + \boldsymbol{\mathcal{Q}},\boldsymbol{k} + \boldsymbol{\mathcal{Q}}'}^{\tau},
\end{split}
\end{equation}
which, upon a relabeling of $j \rightarrow j+1$, clearly obeys Eq. (\ref{C3z_operator_relation}). The same can be seen for the second interlayer tunnelling term.

\subsubsection{Proof of $C_{2x}$ Operator}

In this subsection, for the sake of brevity, we introduce $\widetilde{\boldsymbol{\mathcal{Q}}} = (\mathcal{Q}_{x},-\mathcal{Q}_{y})$ as a shorthand notation for vectors reflected in the $x$-axis. Writing out the $C_{2x}$ symmetry transformation explicitly yields
\begin{equation}
\begin{split}
    &R_{2x,\boldsymbol{\mathcal{Q}},\boldsymbol{\mathcal{P}}}^{\dagger} \bigg[ \mathscr{H}_{\text{TBK},\tilde{\boldsymbol{k}}}^{\tau (\theta \rightarrow -\theta)} \bigg]_{\boldsymbol{\mathcal{P}},\boldsymbol{\mathcal{P}}'} R_{2x,\boldsymbol{\mathcal{P}}',\boldsymbol{\mathcal{Q}}'}^{\null}
    \\
    &\quad=
    R_{2x,\boldsymbol{\mathcal{Q}},\boldsymbol{\mathcal{P}}}^{\dagger} \widetilde{\mathscr{H}}_{\tilde{\boldsymbol{k}} + \boldsymbol{\mathcal{P}}}^{\text{D},\tau} \delta_{\boldsymbol{\mathcal{P}},\boldsymbol{\mathcal{P}}'}^{\null} R_{2x,\boldsymbol{\mathcal{Q}},\boldsymbol{\mathcal{P}}}^{\null}
    +
    \sum_{j \in \{b,tr,tl\}} \left( R_{2x,\boldsymbol{\mathcal{Q}},\boldsymbol{\mathcal{P}}}^{\dagger} \mathscr{H}_{j,\tilde{\boldsymbol{k}},\boldsymbol{\mathcal{P}},\boldsymbol{\mathcal{P}}'}^{\tau (\theta \rightarrow -\theta)}
    R_{2x,\boldsymbol{\mathcal{P}}'\boldsymbol{\mathcal{Q}}'}^{\null} + R_{2x,\boldsymbol{\mathcal{Q}},\boldsymbol{\mathcal{P}}}^{\dagger} \mathscr{H}_{j,\tilde{\boldsymbol{k}},\boldsymbol{\mathcal{P}},\boldsymbol{\mathcal{P}}'}^{\tau\dagger (\theta \rightarrow -\theta)} R_{2x,\boldsymbol{\mathcal{P}}'\boldsymbol{\mathcal{Q}}'}^{\null} \right).
\end{split}
\end{equation}
The first term is readily seen to be
\begin{equation}
    R_{2x,\boldsymbol{\mathcal{Q}},\boldsymbol{\mathcal{P}}}^{\dagger} \widetilde{\mathscr{H}}_{\tilde{\boldsymbol{k}} + \boldsymbol{\mathcal{P}}}^{\text{D},\tau} \delta_{\boldsymbol{\mathcal{P}},\boldsymbol{\mathcal{P}}'}^{\null} R_{2x,\boldsymbol{\mathcal{P}}',\boldsymbol{\mathcal{Q}}'}^{\null}
    =
    \widetilde{\mathscr{H}}_{\tilde{\boldsymbol{k}} + \widetilde{\boldsymbol{\mathcal{Q}}}}^{\text{D},\tau} \delta_{\boldsymbol{\mathcal{Q}},\boldsymbol{\mathcal{Q}}'}^{\null}
    =
    \widetilde{\mathscr{H}}_{\boldsymbol{k} + \boldsymbol{\mathcal{Q}}}^{\text{D},\tau} \delta_{\boldsymbol{\mathcal{Q}},\boldsymbol{\mathcal{Q}}'}^{\null}
\end{equation}
where we noted that $\widetilde{\mathscr{H}}_{\tilde{\boldsymbol{k}}}^{\text{D},\tau} = \widetilde{\mathscr{H}}_{\boldsymbol{k}}^{\text{D},\tau}$.

The challenge begins when we consider the interlayer tunnelling terms. Let us first consider the first tunnelling contribution (suppressing the $4\omega_{0}/3$ factor), where we appreciate that the $C_{2x}$ operation exchanges the layers which may may be reinterpreted as letting $\theta \rightarrow -\theta$ and hence $\mathbf{q}_{j} \rightarrow - \mathbf{q}_{j}$,
\begin{equation}
\begin{alignedat}{2}
    R_{2x,\boldsymbol{\mathcal{Q}},\boldsymbol{\mathcal{P}}}^{\dagger} \mathscr{H}_{j,\tilde{\boldsymbol{k}},\boldsymbol{\mathcal{P}},\boldsymbol{\mathcal{P}}'}^{\tau (\theta \rightarrow -\theta)} R_{2x,\boldsymbol{\mathcal{P}}',\boldsymbol{\mathcal{Q}}'}^{\null}
    &= \,
    &&\delta_{\boldsymbol{\mathcal{Q}},\widetilde{\boldsymbol{\mathcal{P}}}}^{\null}
    \delta_{\boldsymbol{\mathcal{P}}'-\boldsymbol{\mathcal{P}},-\tau\mathbf{q}_{j}^{\null}}
    \delta_{\boldsymbol{\mathcal{P}}',\widetilde{\boldsymbol{\mathcal{Q}}}'}^{\null}
    e^{-i \tau (\phi_{\mathbf{k}+\boldsymbol{\mathcal{Q}}} + \xi_{\boldsymbol{\mathcal{Q}}} \frac{\theta}{2})}
    e^{-i \tau \tilde{\phi}_{\tilde{\boldsymbol{k}}+\boldsymbol{\mathcal{P}},\tilde{\boldsymbol{k}}+\boldsymbol{\mathcal{P}}'}^{(\theta \rightarrow -\theta)}}
    e^{i \tau (\phi_{\mathbf{k}+\boldsymbol{\mathcal{Q}}'} + \xi_{\boldsymbol{\mathcal{Q}}'} \frac{\theta}{2})}
    \\& &&\times \sigma_{z}^{\null }T_{j,\tilde{\boldsymbol{k}}+\boldsymbol{\mathcal{P}},\tilde{\boldsymbol{k}}+\boldsymbol{\mathcal{P}}'}^{\tau (\theta \rightarrow -\theta)} \sigma_{z}^{\null}
    \\
    &= \,
    &&\delta_{\widetilde{\boldsymbol{\mathcal{Q}}}'-\widetilde{\boldsymbol{\mathcal{Q}}},-\tau\mathbf{q}_{j}^{\null}}
    e^{-i \tau (\phi_{\mathbf{k}+\boldsymbol{\mathcal{Q}}} + \xi_{\boldsymbol{\mathcal{Q}}} \frac{\theta}{2})}
    e^{-i \tau \tilde{\phi}_{\tilde{\boldsymbol{k}}+\widetilde{\boldsymbol{\mathcal{Q}}},\tilde{\boldsymbol{k}}+\widetilde{\boldsymbol{\mathcal{Q}}}'}^{(\theta \rightarrow -\theta)}}
    e^{i \tau (\phi_{\mathbf{k}+\boldsymbol{\mathcal{Q}}'} + \xi_{\boldsymbol{\mathcal{Q}}'} \frac{\theta}{2})}
    \sigma_{z}^{\null} T_{j,\tilde{\boldsymbol{k}}+\widetilde{\boldsymbol{\mathcal{Q}}},\tilde{\boldsymbol{k}}+\widetilde{\boldsymbol{\mathcal{Q}}}'}^{\tau (\theta \rightarrow -\theta)} \sigma_{z}^{\null}.
\end{alignedat}
\end{equation}
Next, we recall that $\phi_{\boldsymbol{k}} = \arctan(k_{y}/k_{x})$ and $\phi_{\boldsymbol{k}} \in [-\pi,\pi)$, leading us to appreciate $\phi_{\tilde{\boldsymbol{k}}} = -\phi_{\boldsymbol{k}}$. Moveover, we note that $\tilde{\mathbf{q}}_{b} = -\mathbf{q}_{b}$, $\tilde{\mathbf{q}}_{tr} = -\mathbf{q}_{tl}$, $\tilde{\mathbf{q}}_{tl} = -\mathbf{q}_{tr}$. Therefore, by writing $\tilde{\phi}_{\tilde{\boldsymbol{k}}+\widetilde{\boldsymbol{\mathcal{Q}}},\tilde{\boldsymbol{k}}+\widetilde{\boldsymbol{\mathcal{Q}}}'}^{(\theta \rightarrow -\theta)} = (\phi_{\boldsymbol{k}+\boldsymbol{\mathcal{Q}}'}^{\null} - \phi_{\boldsymbol{k}+\boldsymbol{\mathcal{Q}}}^{\null} - \theta)/2$, the above expression may therefore be simplified to
\begin{equation}
    R_{2x,\boldsymbol{\mathcal{Q}},\boldsymbol{\mathcal{P}}}^{\dagger} \mathscr{H}_{j,\tilde{\boldsymbol{k}},\boldsymbol{\mathcal{P}},\boldsymbol{\mathcal{P}}'}^{\tau (\theta \rightarrow -\theta)} R_{2x,\boldsymbol{\mathcal{P}}',\boldsymbol{\mathcal{Q}}'}^{\null}
    =
    \delta_{\boldsymbol{\mathcal{Q}}'-\boldsymbol{\mathcal{Q}},\tau\mathbf{q}_{\bar{j}}^{\null}}
    e^{-\frac{i}{2} \tau (\phi_{\boldsymbol{k}+\boldsymbol{\mathcal{Q}}} - \phi_{\boldsymbol{k}+\boldsymbol{\mathcal{Q}}'})}
    e^{-i \tau (\xi_{\boldsymbol{\mathcal{Q}}} - \xi_{\boldsymbol{\mathcal{Q}}'} - 1) \frac{\theta}{2}}
    \sigma_{z}^{\null} T_{j,\tilde{\boldsymbol{k}}+\widetilde{\boldsymbol{\mathcal{Q}}},\tilde{\boldsymbol{k}}+\widetilde{\boldsymbol{\mathcal{Q}}}'}^{\tau (\theta \rightarrow -\theta)} \sigma_{z}^{\null},
\end{equation}
where $\{\bar{b},\bar{tr},\bar{tl}\} = \{b,tl,tr\}$. The Kronecker delta ensures that $\boldsymbol{\mathcal{Q}} \in \mathcal{Q}_{-}^{\tau}$ and $\boldsymbol{\mathcal{Q}}' \in \mathcal{Q}_{+}^{\tau}$, and hence $\xi_{\boldsymbol{\mathcal{Q}}} = - \xi_{\boldsymbol{\mathcal{Q}}'} = 1$, thus yielding
\begin{equation}
    R_{2x,\boldsymbol{\mathcal{Q}},\boldsymbol{\mathcal{P}}}^{\dagger} \mathscr{H}_{j,\tilde{\boldsymbol{k}},\boldsymbol{\mathcal{P}},\boldsymbol{\mathcal{P}}'}^{\tau (\theta \rightarrow -\theta)} R_{2x,\boldsymbol{\mathcal{P}}',\boldsymbol{\mathcal{Q}}'}^{\null}
    =
    \delta_{\boldsymbol{\mathcal{Q}}'-\boldsymbol{\mathcal{Q}},\tau\mathbf{q}_{\bar{j}}^{\null}}
    e^{-i \tau \tilde{\phi}_{\boldsymbol{k}+\boldsymbol{\mathcal{Q}},\boldsymbol{k}+\boldsymbol{\mathcal{Q}}'}}
    \sigma_{z}^{\null} T_{j,\tilde{\boldsymbol{k}}+\widetilde{\boldsymbol{\mathcal{Q}}},\tilde{\boldsymbol{k}}+\widetilde{\boldsymbol{\mathcal{Q}}}'}^{\tau (\theta \rightarrow -\theta)} \sigma_{z}^{\null},
\end{equation}

Following this, the tunnelling matrix will need to be written explicitly in terms of its Pauli matrix decomposition,
\begin{equation}
\begin{split}
    \sigma_{z}^{\null} T_{j,\tilde{\boldsymbol{k}}+\widetilde{\boldsymbol{\mathcal{Q}}},\tilde{\boldsymbol{k}}+\widetilde{\boldsymbol{\mathcal{Q}}}'}^{\tau (\theta \rightarrow -\theta)} \sigma_{z}^{\null}
    &=
    \cos \tilde{\phi}_{\tilde{\boldsymbol{k}}+\widetilde{\boldsymbol{\mathcal{Q}}},\tilde{\boldsymbol{k}}+\widetilde{\boldsymbol{\mathcal{Q}}}'}^{(\theta \rightarrow -\theta)} \sigma_{0}^{\null}
    - \sin \bar{\phi}_{\tilde{\boldsymbol{k}}+\widetilde{\boldsymbol{\mathcal{Q}}},\tilde{\boldsymbol{k}}+\widetilde{\boldsymbol{\mathcal{Q}}}'}^{j} \sigma_{x}^{\null}
    - i \tau \sin \tilde{\phi}_{\tilde{\boldsymbol{k}}+\widetilde{\boldsymbol{\mathcal{Q}}},\tilde{\boldsymbol{k}}+\widetilde{\boldsymbol{\mathcal{Q}}}'}^{(\theta \rightarrow -\theta)} \sigma_{y}^{\null}
    - \tau \cos \bar{\phi}_{\tilde{\boldsymbol{k}}+\widetilde{\boldsymbol{\mathcal{Q}}},\tilde{\boldsymbol{k}}+\widetilde{\boldsymbol{\mathcal{Q}}}'}^{j} \sigma_{z}^{\null}
    \\
    &=
    \cos \tilde{\phi}_{\boldsymbol{k}+\boldsymbol{\mathcal{Q}},\boldsymbol{k}+\boldsymbol{\mathcal{Q}}'}^{\null} \sigma_{0}^{\null}
    + \sin \bar{\phi}_{\boldsymbol{k}+\boldsymbol{\mathcal{Q}},\boldsymbol{k}+\boldsymbol{\mathcal{Q}}'}^{\bar{j}} \sigma_{x}^{\null}
    + i \tau \sin \tilde{\phi}_{\boldsymbol{k}+\boldsymbol{\mathcal{Q}},\boldsymbol{k}+\boldsymbol{\mathcal{Q}}'}^{\null} \sigma_{y}^{\null}
    - \tau \cos \bar{\phi}_{\boldsymbol{k}+\boldsymbol{\mathcal{Q}},\boldsymbol{k}+\boldsymbol{\mathcal{Q}}'}^{\bar{j}} \sigma_{z}^{\null}
    \\
    &=
    T_{\bar{j},\boldsymbol{k}+\boldsymbol{\mathcal{Q}},\boldsymbol{k}+\boldsymbol{\mathcal{Q}}'}^{\tau},
    \label{T-matrix_transformation_C2x}
\end{split}
\end{equation}
where in the third line we made use of $\phi_{\tilde{\boldsymbol{k}}} = -\phi_{\boldsymbol{k}}$. Therefore we have shown that
\begin{equation}
    \sum_{j} R_{2x,\boldsymbol{\mathcal{Q}},\boldsymbol{\mathcal{P}}}^{\dagger} \mathscr{H}_{j,\tilde{\boldsymbol{k}},\boldsymbol{\mathcal{P}},\boldsymbol{\mathcal{P}}'}^{\tau (\theta \rightarrow -\theta)} R_{2x,\boldsymbol{\mathcal{P}}',\boldsymbol{\mathcal{Q}}'}^{\null}
    =
    \sum_{j} \mathscr{H}_{\bar{j},\boldsymbol{k},\boldsymbol{\mathcal{Q}},\boldsymbol{\mathcal{Q}}'}^{\tau}
    =
    \sum_{j} \mathscr{H}_{j,\boldsymbol{k},\boldsymbol{\mathcal{Q}},\boldsymbol{\mathcal{Q}}'}^{\tau}.
\end{equation}
The same will naturally be true for the $\mathscr{H}_{j,\tilde{\boldsymbol{k}},\boldsymbol{\mathcal{P}},\boldsymbol{\mathcal{Q}}'}^{\tau \dagger (\theta \rightarrow -\theta)}$ piece.

\subsubsection{Proof of $C_{2z}\mathcal{T}$ Operator}

We start by writing the operators out explicitly,
\begin{equation}
\begin{alignedat}{2}
    \mathscr{T}_{2z,\boldsymbol{\mathcal{Q}},\boldsymbol{\mathcal{P}}}^{\tau} \left[ \mathscr{H}_{\text{TBK},\boldsymbol{k}}^{\tau} \right]_{\boldsymbol{\mathcal{P}},\boldsymbol{\mathcal{P}}'} \mathscr{T}_{2z,\boldsymbol{\mathcal{P}},\boldsymbol{\mathcal{Q}}'}^{\tau}
    &= \mathscr{T}_{2z,\boldsymbol{\mathcal{Q}},\boldsymbol{\mathcal{P}}}^{\tau} \widetilde{\mathscr{H}}_{\boldsymbol{k} + \boldsymbol{\mathcal{P}}}^{\text{D},\tau} \delta_{\boldsymbol{\mathcal{P}},\boldsymbol{\mathcal{P}}'}^{\null} \mathscr{T}_{2z,\boldsymbol{\mathcal{P}},\boldsymbol{\mathcal{Q}}'}^{\tau}
    \\
    &+ \sum_{j \in \{b,tr,tl\}} \left( \mathscr{T}_{2z,\boldsymbol{\mathcal{Q}},\boldsymbol{\mathcal{P}}}^{\tau} \mathscr{H}_{j,\boldsymbol{k},\boldsymbol{\mathcal{P}},\boldsymbol{\mathcal{P}}'}^{\tau} \mathscr{T}_{2z,\boldsymbol{\mathcal{P}}',\boldsymbol{\mathcal{Q}}'}^{\tau}
    + \mathscr{T}_{2z,\boldsymbol{\mathcal{Q}},\boldsymbol{\mathcal{P}}}^{\tau} \mathscr{H}_{j,\boldsymbol{k},\boldsymbol{\mathcal{P}},\boldsymbol{\mathcal{P}}'}^{\tau\dagger} \mathscr{T}_{2z,\boldsymbol{\mathcal{P}}',\boldsymbol{\mathcal{Q}}'}^{\tau} \right),
\end{alignedat}
\end{equation}
and consider the three terms separately. The easiest part to handle is the block diagonal piece in the first term,
\begin{equation}
\begin{split}
    \mathscr{T}_{2z,\boldsymbol{\mathcal{Q}},\boldsymbol{\mathcal{P}}}^{\tau} \widetilde{\mathscr{H}}_{\boldsymbol{k} + \boldsymbol{\mathcal{P}}}^{\text{D},\tau} \delta_{\boldsymbol{\mathcal{P}},\boldsymbol{\mathcal{P}}'}^{\null} \mathscr{T}_{2z,\boldsymbol{\mathcal{P}},\boldsymbol{\mathcal{Q}}'}^{\tau}
    &=
    \delta_{\boldsymbol{\mathcal{Q}},\boldsymbol{\mathcal{P}}}^{\null} e^{-i \tau (\phi_{\boldsymbol{k}+\boldsymbol{\mathcal{Q}}} + \xi_{\boldsymbol{\mathcal{Q}}} \frac{\theta}{2})} \mathcal{K}
    \widetilde{\mathscr{H}}_{\boldsymbol{k} + \boldsymbol{\mathcal{P}}}^{\text{D},\tau} \delta_{\boldsymbol{\mathcal{P}},\boldsymbol{\mathcal{P}}'}^{\null} 
    \delta_{\boldsymbol{\mathcal{P}}',\boldsymbol{\mathcal{Q}}'}^{\null} e^{-i \tau (\phi_{\boldsymbol{k}+\boldsymbol{\mathcal{P}}'} + \xi_{\boldsymbol{\mathcal{P}}'} \frac{\theta}{2})} \mathcal{K}
    \\
    &=
    \delta_{\boldsymbol{\mathcal{Q}},\boldsymbol{\mathcal{Q}}'}^{\null} e^{-i \tau (\phi_{\boldsymbol{k}+\boldsymbol{\mathcal{Q}}} + \xi_{\boldsymbol{\mathcal{Q}}} \frac{\theta}{2})}
    \widetilde{\mathscr{H}}_{\boldsymbol{k} + \boldsymbol{\mathcal{Q}}}^{\text{D},\tau} e^{i \tau (\phi_{\boldsymbol{k}+\boldsymbol{\mathcal{Q}}'} + \xi_{\boldsymbol{\mathcal{Q}}'} \frac{\theta}{2})}
    \\
    &=
    \delta_{\boldsymbol{\mathcal{Q}},\boldsymbol{\mathcal{Q}}'}^{\null} \widetilde{\mathscr{H}}_{\boldsymbol{k} + \boldsymbol{\mathcal{Q}}}^{\text{D},\tau},
\end{split}
\end{equation}
which clearly matches the relation in Eq. (\ref{C2zT_operator_relation}). The interlayer tunnelling terms require slightly more care in their handling, but are quickly established to also obey Eq. (\ref{C2zT_operator_relation}), as we illustrate below (suppressing the factor of $4\omega_{0}/3$),
\begin{equation}
\begin{split}
    \mathscr{T}_{2z,\boldsymbol{\mathcal{Q}},\boldsymbol{\mathcal{P}}}^{\tau} \mathscr{H}_{j,\boldsymbol{k},\boldsymbol{\mathcal{P}},\boldsymbol{\mathcal{P}}'}^{\tau} \mathscr{T}_{2z,\boldsymbol{\mathcal{P}},\boldsymbol{\mathcal{Q}}'}^{\tau}
    &=
    \delta_{\boldsymbol{\mathcal{Q}},\boldsymbol{\mathcal{P}}}^{\null} e^{-i \tau (\phi_{\boldsymbol{k}+\boldsymbol{\mathcal{Q}}} + \xi_{\boldsymbol{\mathcal{Q}}} \frac{\theta}{2})} \mathcal{K}
    \\
    &\qquad\times
    \delta_{\boldsymbol{\mathcal{P}}'-\boldsymbol{\mathcal{P}},\tau\mathbf{q}_{j}} e^{-\frac{i}{2} \tau (\phi_{\boldsymbol{k}+\boldsymbol{\mathcal{P}}} - \phi_{\boldsymbol{k}+\boldsymbol{\mathcal{P}}'} + \theta)} T_{j,\boldsymbol{k}+\boldsymbol{\mathcal{P}},\boldsymbol{k}+\boldsymbol{\mathcal{P}}'}^{\tau}
    \delta_{\boldsymbol{\mathcal{P}}',\boldsymbol{\mathcal{Q}}'}^{\null} e^{-i \tau (\phi_{\boldsymbol{k}+\boldsymbol{\mathcal{P}}'} + \xi_{\boldsymbol{\mathcal{P}}'} \frac{\theta}{2})} \mathcal{K}
    \\
    &= e^{-i \tau (\phi_{\boldsymbol{k}+\boldsymbol{\mathcal{Q}}} + \xi_{\boldsymbol{\mathcal{Q}}} \frac{\theta}{2})}
    \delta_{\boldsymbol{\mathcal{Q}}'-\boldsymbol{\mathcal{Q}},\tau\mathbf{q}_{j}} e^{\frac{i}{2} \tau (\phi_{\boldsymbol{k}+\boldsymbol{\mathcal{Q}}} - \phi_{\boldsymbol{k}+\boldsymbol{\mathcal{Q}}'} + \theta)} T_{j,\boldsymbol{k}+\boldsymbol{\mathcal{Q}},\boldsymbol{k}+\boldsymbol{\mathcal{Q}}'}^{\tau}
    e^{i \tau (\phi_{\boldsymbol{k}+\boldsymbol{\mathcal{Q}}'} + \xi_{\boldsymbol{\mathcal{Q}}'} \frac{\theta}{2})}
    \\
    &= e^{-\frac{i}{2} \tau (\phi_{\boldsymbol{k}+\boldsymbol{\mathcal{Q}}} - \phi_{\boldsymbol{k}+\boldsymbol{\mathcal{Q}}'})} e^{-\frac{i}{2} \tau (\xi_{\boldsymbol{\mathcal{Q}}} - \xi_{\boldsymbol{\mathcal{Q}}'} - 1)\theta}
    \delta_{\boldsymbol{\mathcal{Q}}'-\boldsymbol{\mathcal{Q}},\tau\mathbf{q}_{j}} T_{j,\boldsymbol{k}+\boldsymbol{\mathcal{Q}},\boldsymbol{k}+\boldsymbol{\mathcal{Q}}'}^{\tau}.
\end{split}
\end{equation}
The Kronecker delta here guarantees that $\boldsymbol{\mathcal{Q}}$ and $\boldsymbol{\mathcal{Q}}'$ belong to opposite $q$-lattice sublattices. Moreover, it specifically enforces $\boldsymbol{\mathcal{Q}} \in \mathcal{Q}_{-}$ and $\boldsymbol{\mathcal{Q}}' \in \mathcal{Q}_{+}$, and hence $\xi_{\boldsymbol{\mathcal{Q}}} = -\xi_{\boldsymbol{\mathcal{Q}}'} = 1$, therefore yielding
\begin{equation}
\begin{split}
    \mathscr{T}_{2z,\boldsymbol{\mathcal{Q}},\boldsymbol{\mathcal{P}}}^{\tau} \mathscr{H}_{j,\boldsymbol{k},\boldsymbol{\mathcal{P}},\boldsymbol{\mathcal{P}}'}^{\tau} \mathscr{T}_{2z,\boldsymbol{\mathcal{P}},\boldsymbol{\mathcal{Q}}'}^{\tau}
    = \delta_{\boldsymbol{\mathcal{Q}}'-\boldsymbol{\mathcal{Q}},\tau\mathbf{q}_{j}} e^{-\frac{i}{2} \tau (\phi_{\boldsymbol{k}+\boldsymbol{\mathcal{Q}}} - \phi_{\boldsymbol{k}+\boldsymbol{\mathcal{Q}}'} + \theta)} T_{j,\boldsymbol{k}+\boldsymbol{\mathcal{Q}},\boldsymbol{k}+\boldsymbol{\mathcal{Q}}'}^{\tau}
    =
    \mathscr{H}_{j,\boldsymbol{k},\boldsymbol{\mathcal{Q}},\boldsymbol{\mathcal{Q}}'}^{\tau},
\end{split}
\end{equation}
thus illustrating that the interlayer tunnelling terms describing electrons moving from $L_{t}^{\vartheta}$ to $L_{b}^{\bar{\vartheta}}$ satisfy Eq. (\ref{C2zT_operator_relation}). Likewise, we see that
\begin{equation}
\begin{split}
    \mathscr{T}_{2z,\boldsymbol{\mathcal{Q}},\boldsymbol{\mathcal{P}}}^{\tau} \mathscr{H}_{j,\boldsymbol{k},\boldsymbol{\mathcal{P}},\boldsymbol{\mathcal{P}}'}^{\tau \dagger} \mathscr{T}_{2z,\boldsymbol{\mathcal{P}},\boldsymbol{\mathcal{Q}}'}^\tau
    &=
    \delta_{\boldsymbol{\mathcal{Q}},\boldsymbol{\mathcal{P}}}^{\null} e^{-i \tau (\phi_{\boldsymbol{k}+\boldsymbol{\mathcal{Q}}} + \xi_{\boldsymbol{\mathcal{Q}}} \frac{\theta}{2})} \mathcal{K}
    \\
    &\qquad\times
    \delta_{\boldsymbol{\mathcal{P}}-\boldsymbol{\mathcal{P}}',\tau\mathbf{q}_{j}} e^{\frac{i}{2} \tau (\phi_{\boldsymbol{k}+\boldsymbol{\mathcal{P}}'} - \phi_{\boldsymbol{k}+\boldsymbol{\mathcal{P}}} + \theta)} T_{j,\boldsymbol{k}+\boldsymbol{\mathcal{P}}',\boldsymbol{k}+\boldsymbol{\mathcal{P}}}^{\tau \dagger}
    \delta_{\boldsymbol{\mathcal{P}}',\boldsymbol{\mathcal{Q}}'}^{\null} e^{-i\tau (\phi_{\boldsymbol{k}+\boldsymbol{\mathcal{P}}'} + \xi_{\boldsymbol{\mathcal{P}}'} \frac{\theta}{2})} \mathcal{K}
    \\
    &= e^{-i\tau (\phi_{\boldsymbol{k}+\boldsymbol{\mathcal{Q}}} + \xi_{\boldsymbol{\mathcal{Q}}} \frac{\theta}{2})}
    \delta_{\boldsymbol{\mathcal{Q}}-\boldsymbol{\mathcal{Q}}',\tau\mathbf{q}_{j}} e^{-\frac{i}{2} \tau (\phi_{\boldsymbol{k}+\boldsymbol{\mathcal{Q}}'} - \phi_{\boldsymbol{k}+\boldsymbol{\mathcal{Q}}} + \theta)} T_{j,\boldsymbol{k}+\boldsymbol{\mathcal{Q}}',\boldsymbol{k}+\boldsymbol{\mathcal{Q}}}^{\tau \dagger}
    e^{i\tau (\phi_{\boldsymbol{k}+\boldsymbol{\mathcal{Q}}'} + \xi_{\boldsymbol{\mathcal{Q}}'} \frac{\theta}{2})}
    \\
    &= e^{\frac{i}{2} \tau (\phi_{\boldsymbol{k}+\boldsymbol{\mathcal{Q}}'} - \phi_{\boldsymbol{k}+\boldsymbol{\mathcal{Q}}})} e^{\frac{i}{2} \tau (\xi_{\boldsymbol{\mathcal{Q}}'} - \xi_{\boldsymbol{\mathcal{Q}}} - 1)\theta}
    \delta_{\boldsymbol{\mathcal{Q}}'-\boldsymbol{\mathcal{Q}},\tau\mathbf{q}_{j}} T_{j,\boldsymbol{k}+\boldsymbol{\mathcal{Q}},\boldsymbol{k}+\boldsymbol{\mathcal{Q}}'}^{\tau \dagger},
\end{split}
\end{equation}
where the Kronecker delta guarantees that $\xi_{\boldsymbol{\mathcal{Q}}'} = -\xi_{\boldsymbol{\mathcal{Q}}} = 1$, and hence
\begin{equation}
    \mathscr{T}_{2z,\boldsymbol{\mathcal{Q}},\boldsymbol{\mathcal{P}}}^{\tau} \mathscr{H}_{j,\boldsymbol{k},\boldsymbol{\mathcal{P}},\boldsymbol{\mathcal{P}}'}^{\tau \dagger} \mathscr{T}_{2z,\boldsymbol{\mathcal{P}},\boldsymbol{\mathcal{Q}}'}^{\tau}
    =
    \delta_{\boldsymbol{\mathcal{Q}}'-\boldsymbol{\mathcal{Q}},\tau\mathbf{q}_{j}} e^{\frac{i}{2} \tau (\phi_{\boldsymbol{k}+\boldsymbol{\mathcal{Q}}'} - \phi_{\boldsymbol{k}+\boldsymbol{\mathcal{Q}}} + \theta)} T_{j,\boldsymbol{k}+\boldsymbol{\mathcal{Q}},\boldsymbol{k}+\boldsymbol{\mathcal{Q}}'}^{\tau \dagger}
    =
    \mathscr{H}_{j,\boldsymbol{k},\boldsymbol{\mathcal{Q}},\boldsymbol{\mathcal{Q}}'}^{\tau \dagger}.
\end{equation}
The above therefore verifies our representation of the $C_{2z}\mathcal{T}$ operator.

\subsubsection{Proof of $C_{2y}\mathcal{T}$ Operator}

The demonstration that the representation of the $C_{2y}\mathcal{T}$ operator in Eq. (\ref{C2yT_operator_relation}) is more involved than the $C_{2z}\mathcal{T}$ due to the presence of a Pauli matrix acting on the Dirac state subspace and the Kronecker deltas imposing different relations between the $\boldsymbol{\mathcal{Q}}$ vectors. However, unsurprisingly, the $C_{2y}\mathcal{T}$ symmetry acts on the Hamiltonian in a similar manner to $C_{2x}$, albeit as an anti-unitary operator as opposed to a unitary one. We quickly see that $\mathscr{T}_{2y,\boldsymbol{\mathcal{Q}},\boldsymbol{\mathcal{P}}^{\null}}^{\null} \mathscr{H}_{\tilde{\boldsymbol{k}}+\boldsymbol{\mathcal{P}}} \delta_{\boldsymbol{\mathcal{P}},\boldsymbol{\mathcal{P}}'} \mathscr{T}_{2y,\boldsymbol{\mathcal{P}}',\boldsymbol{\mathcal{Q}}'}^{\null} = \mathscr{H}_{\boldsymbol{k} + \boldsymbol{\mathcal{Q}}} \delta_{\boldsymbol{\mathcal{Q}},\boldsymbol{\mathcal{Q}}'}$, whilst the $T$-matrices transform identically to Eq. (\ref{T-matrix_transformation_C2x}). The only difference regarding the $C_{2y}\mathcal{T}$ symmetry is its complex conjugation affecting the gauge-fixing phase factors. To see this, let us consider the interlayer tunnelling term
\begin{equation}
\begin{split}
    \mathscr{T}_{2y,\boldsymbol{\mathcal{Q}},\boldsymbol{\mathcal{P}}}^{\null} \mathscr{H}_{j,\tilde{\boldsymbol{k}},\boldsymbol{\mathcal{P}},\boldsymbol{\mathcal{P}}'}^{\tau (\theta \rightarrow -\theta)} \mathscr{T}_{2y,\boldsymbol{\mathcal{P}},\boldsymbol{\mathcal{Q}}'}^{\null}
    &= \delta_{\boldsymbol{\mathcal{Q}},\widetilde{\boldsymbol{\mathcal{P}}}}^{\null} \sigma_{z}^{\null} \mathcal{K} \delta_{\boldsymbol{\mathcal{P}}'-\boldsymbol{\mathcal{P}},\tau\mathbf{q}_{j}^{\null}}
    e^{-i \tau \tilde{\phi}_{\tilde{\boldsymbol{k}}+\boldsymbol{\mathcal{P}},\tilde{\boldsymbol{k}}+\boldsymbol{\mathcal{P}}'}^{(\theta \rightarrow -\theta)}}
    T_{j,\tilde{\boldsymbol{k}}+\boldsymbol{\mathcal{P}},\tilde{\boldsymbol{k}}+\boldsymbol{\mathcal{P}}'}^{\tau (\theta \rightarrow -\theta)}
    \delta_{\boldsymbol{\mathcal{P}}',\widetilde{\boldsymbol{\mathcal{Q}}}'}^{\null} \sigma_{z}^{\null} \mathcal{K}
    \\
    &= \delta_{\widetilde{\boldsymbol{\mathcal{Q}}}'-\widetilde{\boldsymbol{\mathcal{Q}}},-\tau\mathbf{q}_{j}^{\null}}
    e^{i \tau \tilde{\phi}_{\tilde{\boldsymbol{k}}+\widetilde{\boldsymbol{\mathcal{Q}}},\tilde{\boldsymbol{k}}+\widetilde{\boldsymbol{\mathcal{Q}}}'}^{(\theta \rightarrow -\theta)}}
    \sigma_{z}^{\null} T_{j,\tilde{\boldsymbol{k}}+\widetilde{\boldsymbol{\mathcal{Q}}},\tilde{\boldsymbol{k}}+\widetilde{\boldsymbol{\mathcal{Q}}}'}^{\tau (\theta \rightarrow -\theta)} \sigma_{z}^{\null}
    \\
    &= \delta_{\boldsymbol{\mathcal{Q}}'-\boldsymbol{\mathcal{Q}},\tau\mathbf{q}_{\bar{j}}^{\null}}
    e^{i \tau \tilde{\phi}_{\tilde{\boldsymbol{k}}+\widetilde{\boldsymbol{\mathcal{Q}}},\tilde{\boldsymbol{k}}+\widetilde{\boldsymbol{\mathcal{Q}}}'}^{(\theta \rightarrow -\theta)}}
    T_{\bar{j},\boldsymbol{k}+\boldsymbol{\mathcal{Q}},\boldsymbol{k}+\boldsymbol{\mathcal{Q}}'}^{\tau}.
\end{split}
\end{equation}
By noting that $\tilde{\phi}_{\tilde{\boldsymbol{k}}+\widetilde{\boldsymbol{\mathcal{Q}}},\tilde{\boldsymbol{k}}+\widetilde{\boldsymbol{\mathcal{Q}}}'}^{(\theta \rightarrow -\theta)} = -\tilde{\phi}_{\boldsymbol{k}+\boldsymbol{\mathcal{Q}},\boldsymbol{k}+\boldsymbol{\mathcal{Q}}'}^{\null}$, we arrive at
\begin{equation}
    \mathscr{T}_{2y,\boldsymbol{\mathcal{Q}},\boldsymbol{\mathcal{P}}}^{\null} \mathscr{H}_{j,\tilde{\boldsymbol{k}},\boldsymbol{\mathcal{P}},\boldsymbol{\mathcal{P}}'}^{\tau (\theta \rightarrow -\theta)} \mathscr{T}_{2y,\boldsymbol{\mathcal{P}},\boldsymbol{\mathcal{Q}}'}^{\null}
    =
    \delta_{\boldsymbol{\mathcal{Q}}'-\boldsymbol{\mathcal{Q}},\tau\mathbf{q}_{\bar{j}}^{\null}}
    e^{-i \tau \tilde{\phi}_{\boldsymbol{k}+\boldsymbol{\mathcal{Q}},\boldsymbol{k}+\boldsymbol{\mathcal{Q}}'}}
    T_{\bar{j},\boldsymbol{k}+\boldsymbol{\mathcal{Q}},\boldsymbol{k}+\boldsymbol{\mathcal{Q}}'}^{\tau}
    =
    \mathscr{H}_{\bar{j},\boldsymbol{k},\boldsymbol{\mathcal{Q}},\boldsymbol{\mathcal{Q}}'}^{\tau},
\end{equation}
which upon summation over $j$ recovers the full interlayer tunnelling contribution to $\mathscr{H}_{\text{TBK},\boldsymbol{k}}^{\tau}$.

\subsection{Approximate Particle-Hole Symmetry}

Our proof of approximate particle-hole symmetry closely follows that of Ref. \cite{Song2019}. Let us start by shifting to the ``hexagon model'' picture of the TBK system, which corresponds to centring the momentum of the $\Gamma_{\text{M}}$ point. This can be achieved by letting $\mathbf{k} = \boldsymbol{k} + \mathbf{q}_{tl}$ and defining $\boldsymbol{\mathcal{Q}} = \mathbf{Q} + \mathbf{q}_{tl}$. The sublattices of the $q$-lattice now yield the two sets
\begin{equation}
    \boldsymbol{\mathcal{Q}} = n_{b} \mathbf{q}_{b} + n_{tr} \mathbf{q}_{tr} + (n_{tl}+1) \mathbf{q}_{tl} \quad \Rightarrow \quad \begin{matrix}
            L_{t}^{\vartheta}: \mathcal{Q}_{+}^{\tau} = \{\tau\boldsymbol{\mathcal{Q}}|n_{b},n_{tr},n_{tl} \in \mathbb{Z} \text{ s.t. } n_{b}+n_{tr}+n_{tl} = 1 \},
            \\
            L_{b}^{\bar{\vartheta}}: \mathcal{Q}_{-}^{\tau} = \{\tau\boldsymbol{\mathcal{Q}}|n_{b},n_{tr},n_{tl} \in \mathbb{Z} \text{ s.t. } n_{b}+n_{tr}+n_{tl} = 0 \},
        \end{matrix}
\end{equation}
which bare the nice relationship that $\forall \boldsymbol{\mathcal{Q}}_{i} \in \mathcal{Q}_{-}^{\tau}$, $-\boldsymbol{\mathcal{Q}}_{i} \in \mathcal{Q}_{+}^{\tau}$. We may then write the continuum Hamiltonian for the approximate TBK system in the $\Gamma_{\text{M}}$-centric frame as
\begin{equation}
    \left[ \mathscr{H}_{\text{aTBK},\boldsymbol{k}}^{\tau} \right]_{\boldsymbol{\mathcal{Q}},\boldsymbol{\mathcal{Q}}'}
    =
    \widetilde{\mathscr{H}}_{\boldsymbol{k} + \boldsymbol{\mathcal{Q}}}^{\text{D},\tau} \delta_{\boldsymbol{\mathcal{Q}},\boldsymbol{\mathcal{Q}}'}^{\null}
    +
    \sum_{j} \left( \delta_{\boldsymbol{\mathcal{Q}}'-\boldsymbol{\mathcal{Q}},\mathbf{q}_{j}}^{\null} \bar{\Phi}_{\boldsymbol{k}+\boldsymbol{\mathcal{Q}},\mathbf{k}+\boldsymbol{\mathcal{Q}}'}^{\tau} \bar{T}_{j,\boldsymbol{k}+\boldsymbol{\mathcal{Q}},\boldsymbol{k}+\boldsymbol{\mathcal{Q}}'}^{\tau}
    +
    \delta_{\boldsymbol{\mathcal{Q}}-\boldsymbol{\mathcal{Q}}',\mathbf{q}_{j}}^{\null} \bar{\Phi}_{\boldsymbol{k}+\boldsymbol{\mathcal{Q}}',\boldsymbol{k}+\boldsymbol{\mathcal{Q}}}^{\tau*} \bar{T}_{j,\boldsymbol{k}+\boldsymbol{\mathcal{Q}}',\boldsymbol{k}+\boldsymbol{\mathcal{Q}}}^{\tau \, \dagger} \right),
    \label{PHsymm_Hamiltonian}
\end{equation}
where the bar on the $T$-matrices and $\bar{\Phi}_{\mathbf{k},\mathbf{p}}^{\tau}$ indicates that $\theta$ has been set to zero where it appears explicitly in their forms, $\tilde{\phi}_{\mathbf{k}\mathbf{p}}^{\theta=0} = \frac{1}{2}(\phi_{\mathbf{k}}^{\null} - \phi_{\mathbf{p}}^{\null})$,
\begin{equation}
    \bar{\Phi}_{\mathbf{k},\mathbf{p}}^{\tau} = e^{-\tau i\tilde{\phi}_{\mathbf{k}\mathbf{p}}^{\theta = 0}},
    \qquad
    \bar{T}_{j,\mathbf{k}\mathbf{p}}^{\tau} = \cos \tilde{\phi}_{\mathbf{k}\mathbf{p}}^{\theta=0} \, \sigma_{0}^{\null} + \sin \bar{\phi}_{\mathbf{k}\mathbf{p}}^{j} \, \sigma_{x}^{\null} + i \tau \sin \tilde{\phi}_{\mathbf{k}\mathbf{p}}^{\theta=0} \, \sigma_{y}^{\null} - \tau \cos \bar{\phi}_{\mathbf{k}\mathbf{p}}^{j} \, \sigma_{z}^{\null}.
\end{equation}
It then becomes a matter of algebraic manipulation and summation to show that the unitary transformation, $\Sigma_{\boldsymbol{\mathcal{Q}},\boldsymbol{\mathcal{Q}}'} = \sigma_{y} \delta_{\boldsymbol{\mathcal{Q}},-\boldsymbol{\mathcal{Q}}'} \xi_{\boldsymbol{\mathcal{Q}}'}$ with $\xi_{\boldsymbol{\mathcal{Q}} \in \mathcal{Q}_{\pm}^{\tau}} = \mp 1$, yields
\begin{equation}
    \left[ \Sigma^{-1} \mathscr{H}_{\text{aTBK},\boldsymbol{k}}^{\tau} \Sigma \right]_{\boldsymbol{\mathcal{Q}},\boldsymbol{\mathcal{Q}}'} = -\left[ \mathscr{H}_{\text{aTBK},-\boldsymbol{k}}^{\tau} \right]_{\boldsymbol{\mathcal{Q}},\boldsymbol{\mathcal{Q}}'}.
\end{equation}
We now prove this relation.

\subsubsection{Properties of the Particle-Hole Symmetry Operator}

The particle-hole operator can be shown to be unitary by considering $\Sigma_{\boldsymbol{\mathcal{Q}},\boldsymbol{\mathcal{Q}}'}^{\dagger} = \sigma_{y} \delta_{\boldsymbol{\mathcal{Q}}',-\boldsymbol{\mathcal{Q}}} \xi_{\boldsymbol{\mathcal{Q}}}$. The product $[\Sigma^{\dagger} \Sigma]_{\boldsymbol{\mathcal{Q}},\boldsymbol{\mathcal{Q}}'}$ can be simplified in a couple of steps,
\begin{equation}
    [\Sigma^{\dagger} \Sigma]_{\boldsymbol{\mathcal{Q}},\boldsymbol{\mathcal{Q}}'} = \Sigma_{\boldsymbol{\mathcal{Q}},\boldsymbol{\mathcal{P}}}^{\dagger} \Sigma_{\boldsymbol{\mathcal{P}},\boldsymbol{\mathcal{Q}}'}^{\null}
    = \sigma_{y}^{2} \delta_{\boldsymbol{\mathcal{P}},-\boldsymbol{\mathcal{Q}}}^{\null} \delta_{\boldsymbol{\mathcal{P}},-\boldsymbol{\mathcal{Q}}'}^{\null} \xi_{\boldsymbol{\mathcal{Q}}}^{\null} \xi_{\boldsymbol{\mathcal{Q}}'}^{\null}
    = \sigma_{0}^{\null} \delta_{\boldsymbol{\mathcal{Q}},\boldsymbol{\mathcal{Q}}'}^{\null} \xi_{\boldsymbol{\mathcal{Q}}}^{\null} \xi_{\boldsymbol{\mathcal{Q}}'}^{\null}
    = \sigma_{0}^{\null} \delta_{\boldsymbol{\mathcal{Q}},\boldsymbol{\mathcal{Q}}'}^{\null},
\end{equation}
thus proving $\Sigma^{\dagger} = \Sigma^{-1}$. Moreover, it is quickly shown that $\Sigma^{2} = -1$ when working with the $\Gamma_{\text{M}}$-centric model. However, if we wish to instead use the corner-centric model, we must be careful in how we write the particle-hole operator. To be clear, for the $\Gamma_{\text{M}}$-centric model, we had previously written
\begin{equation}
    \Sigma_{\boldsymbol{\mathcal{Q}},\boldsymbol{\mathcal{Q}}'} = \sigma_{y} \delta_{\boldsymbol{\mathcal{Q}},-\boldsymbol{\mathcal{Q}}'} \xi_{\boldsymbol{\mathcal{Q}}'}, \qquad \xi_{\boldsymbol{\mathcal{Q}} \in \mathcal{Q}_{\pm}^{\tau}} = \mp 1,
\end{equation}
in the block basis of the $\Gamma_{\text{M}}$-centric $q$-lattice. We move back to the corner-centric model by recalling $\boldsymbol{\mathcal{Q}} = \mathbf{Q} + \mathbf{q}_{tl}$, allowing us to write, in the block basis corner-centric $q$-lattice,
\begin{equation}
    \Sigma_{\mathbf{Q},\mathbf{Q}'} = \sigma_{y} \delta_{\mathbf{Q}+\mathbf{Q}',-2\mathbf{q}_{tl}} \xi_{\mathbf{Q}'}, \qquad \xi_{\mathbf{Q} \in Q_{\pm}^{\tau}} = \mp 1 = -\text{sgn}\left(n_{b} + n_{tr} + n_{tl} - \frac{1}{2} \right).
\end{equation}

\subsubsection{Proof of Particle-Hole Symmetry}

The first term of Eq. \ref{PHsymm_Hamiltonian} under the particle-hole transformation changes according to
\begin{equation}
    \Sigma_{\boldsymbol{\mathcal{P}}\boldsymbol{\mathcal{Q}}}^{-1} \widetilde{\mathscr{H}}_{\boldsymbol{k} + \boldsymbol{\mathcal{Q}}}^{\text{D},\tau} \delta_{\boldsymbol{\mathcal{Q}},\boldsymbol{\mathcal{Q}}'}^{\null} \Sigma_{\boldsymbol{\mathcal{Q}}'\boldsymbol{\mathcal{P}}'}^{\null} = \sigma_{y}^{\null} \widetilde{\mathscr{H}}_{\boldsymbol{k} - \boldsymbol{\mathcal{P}}}^{\text{D},\tau} \sigma_{y}^{\null} \delta_{\boldsymbol{\mathcal{P}}\boldsymbol{\mathcal{P}}'}^{\null} \xi_{\boldsymbol{\mathcal{P}}}^{2} = - \widetilde{\mathscr{H}}_{-\boldsymbol{k} + \boldsymbol{\mathcal{P}}}^{\text{D},\tau} \delta_{\boldsymbol{\mathcal{P}}\boldsymbol{\mathcal{P}}'}^{\null},
\end{equation}
where we noted in the final equality that $\sigma_{i}^{\null} \sigma_{j}^{\null} = \delta_{ij}^{\null} - \sigma_{j}^{\null} \sigma_{i}^{\null}$ and that $\xi_{\boldsymbol{\mathcal{P}}}^{2} = 1$. Next, we consider the $\bar{T}_{j}^{\tau}$ piece of the interlayer tunnelling term. Under the particle-hole transformation,
\begin{equation}
    \Sigma_{\boldsymbol{\mathcal{P}}\boldsymbol{\mathcal{Q}}}^{-1} \delta_{\boldsymbol{\mathcal{Q}}-\boldsymbol{\mathcal{Q}}',\mathbf{q}_{j}}^{\null} \bar{T}_{j,\boldsymbol{k}+\boldsymbol{\mathcal{Q}},\boldsymbol{k}+\boldsymbol{\mathcal{Q}}'}^{\tau} \Sigma_{\boldsymbol{\mathcal{Q}}'\boldsymbol{\mathcal{P}}'}^{\null}
    =
    \delta_{\boldsymbol{\mathcal{P}}'-\boldsymbol{\mathcal{P}},\mathbf{q}_{j}}^{\null} \xi_{\boldsymbol{\mathcal{P}}}^{\null} \xi_{\boldsymbol{\mathcal{P}}'}^{\null} \bar{\Phi}_{\boldsymbol{k}-\boldsymbol{\mathcal{P}},\boldsymbol{k}-\boldsymbol{\mathcal{P}}'}^{\tau} \sigma_{y}^{\null} \bar{T}_{j,\boldsymbol{k}-\boldsymbol{\mathcal{P}},\boldsymbol{k}-\boldsymbol{\mathcal{P}}'}^{\tau} \sigma_{y}^{\null},
\end{equation}
so we focus on the transformation of the $T$-matrices and phase factors by making use of
\begin{equation}
    \begin{gathered}
        \tilde{\phi}_{\mathbf{k}\mathbf{p}}^{\theta=0} = -\tilde{\phi}_{\mathbf{p}\mathbf{k}}^{\theta=0}, \qquad \tilde{\phi}_{\mathbf{k}\mathbf{p}}^{\theta=0} = \tilde{\phi}_{(-\mathbf{k})(-\mathbf{p})}^{\theta=0} + \begin{cases}
            0, \quad &\text{sgn} \, \phi_{\mathbf{k}}^{\null} = \text{sgn} \, \phi_{\mathbf{p}}^{\null} \\
            \pi \, \text{sgn} \phi_{\mathbf{k}}^{\null}, \quad &\text{sgn} \, \phi_{\mathbf{k}}^{\null} = -\text{sgn} \, \phi_{\mathbf{p}}^{\null},
        \end{cases}
        \\
        \bar{\phi}_{\mathbf{k}\mathbf{p}}^{j} = \bar{\phi}_{\mathbf{p}\mathbf{k}}^{j}, \qquad \bar{\phi}_{\mathbf{k}\mathbf{p}}^{j} = \tilde{\phi}_{(-\mathbf{k})(-\mathbf{p})}^{j} + \begin{cases}
        0, \quad &\text{sgn} \, \phi_{\mathbf{k}}^{\null} = -\text{sgn} \, \phi_{\mathbf{p}}^{\null} \\
        \pi \, \text{sgn} \phi_{\mathbf{k}}^{\null}, \quad &\text{sgn} \, \phi_{\mathbf{k}}^{\null} = \text{sgn} \, \phi_{\mathbf{p}}^{\null},
    \end{cases}
    \end{gathered}
\end{equation}
where we take the angle of the momentum in the range $\phi_{\mathbf{k}} \in [-\pi,\pi)$. Denoting $s_{\mathbf{k}}^{\phi} = \text{sgn} \, \phi_{\mathbf{k}}^{\null}$, we write
\begin{equation}
\begin{split}
    \sigma_{y}^{\null} \bar{T}_{j,\boldsymbol{k}-\boldsymbol{\mathcal{P}},\boldsymbol{k}-\boldsymbol{\mathcal{P}}'}^{\tau} \sigma_{y}^{\null}
    &= \cos \tilde{\phi}_{(\boldsymbol{k}-\boldsymbol{\mathcal{P}})(\boldsymbol{k}-\boldsymbol{\mathcal{P}}')}^{\theta=0} \sigma_{0}^{\null}
    - \sin \bar{\phi}_{(\boldsymbol{k}-\boldsymbol{\mathcal{P}})(\boldsymbol{k}-\boldsymbol{\mathcal{P}}')}^{j} \sigma_{x}^{\null}
    + i \tau \sin \tilde{\phi}_{(\boldsymbol{k}-\boldsymbol{\mathcal{P}})(\boldsymbol{k}-\boldsymbol{\mathcal{P}}')}^{\theta=0} \sigma_{y}^{\null}
    + \tau \cos \bar{\phi}_{(\boldsymbol{k}-\boldsymbol{\mathcal{P}})(\boldsymbol{k}-\boldsymbol{\mathcal{P}}')}^{j} \sigma_{z}^{\null}
    \\
    &= \cos \tilde{\phi}_{(\boldsymbol{k}-\boldsymbol{\mathcal{P}}')(\boldsymbol{k}-\boldsymbol{\mathcal{P}})}^{\theta=0} \sigma_{0}^{\null}
    - \sin \bar{\phi}_{(\boldsymbol{k}-\boldsymbol{\mathcal{P}}')(\boldsymbol{k}-\boldsymbol{\mathcal{P}})}^{j} \sigma_{x}^{\null}
    - i \tau \sin \tilde{\phi}_{(\boldsymbol{k}-\boldsymbol{\mathcal{P}}')(\boldsymbol{k}-\boldsymbol{\mathcal{P}})}^{\theta=0} \sigma_{y}^{\null}
    + \tau \cos \bar{\phi}_{(\boldsymbol{k}-\boldsymbol{\mathcal{P}}')(\boldsymbol{k}-\boldsymbol{\mathcal{P}})}^{j} \sigma_{z}^{\null}
    \\
    &= \left[ 1 - 2 \delta_{s_{\boldsymbol{k}-\boldsymbol{\mathcal{P}}'}^{\phi},-s_{\boldsymbol{k}-\boldsymbol{\mathcal{P}}}^{\phi}} \right]
    \left[ \cos \tilde{\phi}_{(\boldsymbol{\mathcal{P}}'-\boldsymbol{k})(\boldsymbol{\mathcal{P}}-\boldsymbol{k})}^{\theta=0} \sigma_{0}^{\null}
    - i \tau \sin \tilde{\phi}_{(\boldsymbol{\mathcal{P}}'-\boldsymbol{k})(\boldsymbol{\mathcal{P}}-\boldsymbol{k})}^{\theta=0} \sigma_{y}^{\null} \right]
    \\
    &\qquad\qquad\qquad- \left[ 1 - 2 \delta_{s_{\boldsymbol{k}-\boldsymbol{\mathcal{P}}'}^{\phi},s_{\boldsymbol{k}-\boldsymbol{\mathcal{P}}}^{\phi}} \right]
    \left[ \sin \bar{\phi}_{(\boldsymbol{\mathcal{P}}'-\boldsymbol{k})(\boldsymbol{\mathcal{P}}-\boldsymbol{k})}^{j} \sigma_{x}^{\null}
    - \tau \cos \bar{\phi}_{(\boldsymbol{\mathcal{P}}'-\boldsymbol{k})(\boldsymbol{\mathcal{P}}-\boldsymbol{k})}^{j} \sigma_{z}^{\null} \right]
    \end{split} \label{T_matrix_PHS_transformation}
\end{equation}
Next, the phase factor may be rewritten as
\begin{equation}
    \bar{\Phi}_{\boldsymbol{k}-\boldsymbol{\mathcal{P}},\boldsymbol{k}-\boldsymbol{\mathcal{P}}'}^{\tau} = \bar{\Phi}_{\boldsymbol{k}-\boldsymbol{\mathcal{P}}',\boldsymbol{k}-\boldsymbol{\mathcal{P}}}^{\tau*} = \left[ 1 - 2 \delta_{s_{\boldsymbol{k}-\boldsymbol{\mathcal{P}}'}^{\phi},-s_{\boldsymbol{k}-\boldsymbol{\mathcal{P}}}^{\phi}} \right] \bar{\Phi}_{\boldsymbol{\mathcal{P}}'-\boldsymbol{k},\boldsymbol{\mathcal{P}}-\boldsymbol{k}}^{\tau*}.
    \label{Gauge_fixing_factor_PHS_transformation}
\end{equation}
By noting that
\begin{equation}
    \left[ 1 - 2 \delta_{s_{\boldsymbol{k}-\boldsymbol{\mathcal{P}}'}^{\phi},-s_{\boldsymbol{k}-\boldsymbol{\mathcal{P}}}^{\phi}} \right]^{2} = 1,
    \qquad
    \left[ 1 - 2 \delta_{s_{\boldsymbol{k}-\boldsymbol{\mathcal{P}}'}^{\phi},-s_{\boldsymbol{k}-\boldsymbol{\mathcal{P}}}^{\phi}} \right]\left[ 1 - 2 \delta_{s_{\boldsymbol{k}-\boldsymbol{\mathcal{P}}'}^{\phi},s_{\boldsymbol{k}-\boldsymbol{\mathcal{P}}}^{\phi}} \right] = -1,
\end{equation}
we may immediately write, using Eqs. (\ref{T_matrix_PHS_transformation}) and (\ref{Gauge_fixing_factor_PHS_transformation}),
\begin{equation}
\begin{split}
    &\bar{\Phi}_{\boldsymbol{k}-\boldsymbol{\mathcal{P}},\boldsymbol{k}-\boldsymbol{\mathcal{P}}'}^{\tau} \sigma_{y}^{\null} \bar{T}_{j,\boldsymbol{k}-\boldsymbol{\mathcal{P}},\boldsymbol{k}-\boldsymbol{\mathcal{P}}'}^{\tau} \sigma_{y}^{\null}
    \\
    &= \bar{\Phi}_{\boldsymbol{\mathcal{P}}'-\boldsymbol{k},\boldsymbol{\mathcal{P}}-\boldsymbol{k}}^{\tau*} \left[ \cos \tilde{\phi}_{(\boldsymbol{\mathcal{P}}'-\boldsymbol{k})(\boldsymbol{\mathcal{P}}-\boldsymbol{k})}^{\theta=0} \sigma_{0}^{\null}
    + \sin \bar{\phi}_{(\boldsymbol{\mathcal{P}}'-\boldsymbol{k})(\boldsymbol{\mathcal{P}}-\boldsymbol{k})}^{j} \sigma_{x}^{\null}
    - i \tau \sin \tilde{\phi}_{(\boldsymbol{\mathcal{P}}'-\boldsymbol{k})(\boldsymbol{\mathcal{P}}-\boldsymbol{k})}^{\theta=0} \sigma_{y}^{\null}
    - \tau \cos \bar{\phi}_{(\boldsymbol{\mathcal{P}}'-\boldsymbol{k})(\boldsymbol{\mathcal{P}}-\boldsymbol{k})}^{j} \sigma_{z}^{\null} \right]
    \\
    &= \bar{\Phi}_{\boldsymbol{\mathcal{P}}'-\boldsymbol{k},\boldsymbol{\mathcal{P}}-\boldsymbol{k}}^{\tau*} \bar{T}_{j,-\boldsymbol{k}+\boldsymbol{\mathcal{P}}',-\boldsymbol{k}+\boldsymbol{\mathcal{P}}}^{\tau \, \dagger},
\end{split}
\end{equation}
and hence
\begin{equation}
\begin{split}
     &\Sigma_{\boldsymbol{\mathcal{P}}\boldsymbol{\mathcal{Q}}}^{-1} \left( \delta_{\boldsymbol{\mathcal{Q}}-\boldsymbol{\mathcal{Q}}',\mathbf{q}_{j}}^{\null} \bar{T}_{j,\boldsymbol{k}+\boldsymbol{\mathcal{Q}},\boldsymbol{k}+\boldsymbol{\mathcal{Q}}'}^{\tau} + \delta_{\boldsymbol{\mathcal{Q}}'-\boldsymbol{\mathcal{Q}},\mathbf{q}_{j}}^{\null} \bar{T}_{j,\boldsymbol{k}+\boldsymbol{\mathcal{Q}}',\boldsymbol{k}+\boldsymbol{\mathcal{Q}}}^{\tau \, \dagger} \right) \Sigma_{\boldsymbol{\mathcal{Q}}'\boldsymbol{\mathcal{P}}'}^{\null}
     \\
     &\qquad\qquad\qquad\qquad\qquad\qquad\qquad\qquad\qquad\qquad= -\left( \delta_{\boldsymbol{\mathcal{P}}-\boldsymbol{\mathcal{P}}',\mathbf{q}_{j}}^{\null} \bar{T}_{j,-\boldsymbol{k}+\boldsymbol{\mathcal{P}},-\boldsymbol{k}+\boldsymbol{\mathcal{P}}'}^{\tau} + \delta_{\boldsymbol{\mathcal{P}}'-\boldsymbol{\mathcal{P}},\mathbf{q}_{j}}^{\null} \bar{T}_{j,-\boldsymbol{k}+\boldsymbol{\mathcal{P}}',-\boldsymbol{k}+\boldsymbol{\mathcal{P}}}^{\tau \, \dagger} \right),
\end{split}
\end{equation}
where we noted that $\xi_{\boldsymbol{\mathcal{P}}}\xi_{\boldsymbol{\mathcal{P}}'} \delta_{\boldsymbol{\mathcal{P}}-\boldsymbol{\mathcal{P}}',\mathbf{q}_{j}} = -\delta_{\boldsymbol{\mathcal{P}}-\boldsymbol{\mathcal{P}}',\mathbf{q}_{j}}$ and $\xi_{\boldsymbol{\mathcal{P}}}\xi_{\boldsymbol{\mathcal{P}}'} \delta_{\boldsymbol{\mathcal{P}}'-\boldsymbol{\mathcal{P}},\mathbf{q}_{j}} = -\delta_{\boldsymbol{\mathcal{P}}'-\boldsymbol{\mathcal{P}},\mathbf{q}_{j}}$, since the Kronecker delta's enforce $\boldsymbol{\mathcal{P}}$ and $\boldsymbol{\mathcal{P}}'$ to connect neighbouring $q$-lattice sites. Thus, we have proven that the approximate TBK Hamiltonian has particle-hole symmetry. Clearly, the reintroduction of $\theta$ into the tunnelling matrices beyond the $q$-lattice will break this particle-hole symmetry. Given that $\theta \ll 1$ for the twist angles we consider here, we therefore arrive at the conclusion that $\mathcal{O}(\theta)$ corrections lift the particle-hole symmetry.

\subsubsection{Antiunitary Particle-Hole Operators}

We now introduce two possible options for an antiunitary particle-hole operator $\mathscr{P}_{i} = \Sigma C_{2i}\mathcal{T}$ ($i \in\{y,z\}$).
\begin{equation}
    \mathscr{P}_{y,\boldsymbol{\mathcal{Q}}\boldsymbol{\mathcal{Q}}'}^{\null} = -i \sigma_{x} \delta_{\mathcal{Q}_{x},-\mathcal{Q}_{x}'} \delta_{\mathcal{Q}_{y},\mathcal{Q}_{y}'} \xi_{\boldsymbol{\mathcal{Q}}} \mathcal{K},
    \qquad
    \mathscr{P}_{z,\boldsymbol{k};\boldsymbol{\mathcal{Q}}\boldsymbol{\mathcal{Q}}'}^{\tau} = \sigma_{y} \delta_{\boldsymbol{\mathcal{Q}},-\boldsymbol{\mathcal{Q}}'} \xi_{\boldsymbol{\mathcal{Q}}'} e^{-i \tau (\phi_{\boldsymbol{k}+\boldsymbol{\mathcal{Q}}'} + \xi_{\boldsymbol{\mathcal{Q}}'} \frac{\theta}{2})} \mathcal{K},
\end{equation}
where we note that the anti-unitary PHS opreator constructed from $C_{2z} \mathcal{T}$ is momentum-dependent. Unlike TBG, the momentum-dependence of $\mathscr{P}_{z,\boldsymbol{k};\boldsymbol{\mathcal{Q}}\boldsymbol{\mathcal{Q}}'}^{\tau}$ prevents it from possessing the simple property of squaring to negative unity, $\mathscr{P}_{z,\boldsymbol{k}}^{\tau \, 2}\neq -1$, responsible for the Euler class topological index appearing in TBG. Instead, we must consider the generalised property $\mathscr{P}_{z,-\mathbf{k}}^{\tau}\mathscr{P}_{i,\mathbf{k}}^{\tau} = -1$ as the signal for topology defined by the Euler class

To understand how we arrive at this relation, let us consider the pair of states, $\ket{\psi_{n,\boldsymbol{k}}^{\text{I}}}$ and $\ket{\psi_{n,\boldsymbol{k}}^{\text{II}}}$, connected by some anti-unitary ``time-reversal'' operator, $\Theta$, as in appendix B of Ref. \cite{Fu2006} and section 3C of Ref. \cite{Song2021}. In these original works, the two states were related according to
\begin{equation}
    \ket{\psi_{n,-\boldsymbol{k}}^{\text{II}}} = \Theta \ket{\psi_{n,\boldsymbol{k}}^{\text{I}}}, \qquad \ket{\psi_{n,-\boldsymbol{k}}^{\text{I}}} = -\Theta \ket{\psi_{n,\boldsymbol{k}}^{\text{II}}}, \qquad \Rightarrow \qquad \Theta^{2} = -1,
\end{equation}
which was shown to generate non-trivial topology defined by the Euler class. We can then apply some momentum-dependent change of basis, $\ket{\psi_{n,\boldsymbol{k}}^{\null}} = M_{\boldsymbol{k}} \ket{\bar{\psi}_{n,\boldsymbol{k}}^{\null}}$, to rewrite these relations as
\begin{equation}
    M_{-\boldsymbol{k}}^{\null} \ket{\bar{\psi}_{n,-\boldsymbol{k}}^{\text{II}}} = \Theta M_{\boldsymbol{k}}^{\null} \ket{\bar{\psi}_{n,\boldsymbol{k}}^{\text{I}}}, \qquad M_{-\boldsymbol{k}}^{\null}\ket{\bar{\psi}_{n,-\boldsymbol{k}}^{\text{I}}} = -\Theta M_{\boldsymbol{k}}^{\null} \ket{\bar{\psi}_{n,\boldsymbol{k}}^{\text{II}}},
\end{equation}
from which, we obtain the momentum-dependent ``time-reversal'' operator $\Theta_{\boldsymbol{k}}^{\null} = M_{-\boldsymbol{k}}^{\dagger} \Theta M_{\boldsymbol{k}}^{\null}$ that connects the two states according to
\begin{equation}
    \ket{\bar{\psi}_{n,-\boldsymbol{k}}^{\text{II}}} = \Theta_{\boldsymbol{k}}^{\null} \ket{\bar{\psi}_{n,\boldsymbol{k}}^{\text{I}}}, \qquad \ket{\bar{\psi}_{n,-\boldsymbol{k}}^{\text{I}}} = -\Theta_{\boldsymbol{k}}^{\null} \ket{\bar{\psi}_{n,\boldsymbol{k}}^{\text{II}}}.
\end{equation}
Clearly, this leads to $\ket{\bar{\psi}_{n,\boldsymbol{k}}^{\text{I}}} = -\Theta_{-\boldsymbol{k}}^{\null} \Theta_{\boldsymbol{k}}^{\null} \ket{\bar{\psi}_{n,\boldsymbol{k}}^{\text{I}}}$, and hence the generalised requirement that $\Theta_{-\boldsymbol{k}}^{\null} \Theta_{\boldsymbol{k}}^{\null} = -1$ for the emergence of topology defined by the Euler class. For the case of TBK, this anti-unitary operation is $\mathscr{P}_{z,\boldsymbol{k}}^{\tau}$, whilst for TBG this operator was momentum-independent.

\subsubsection{Measuring the Quality of Particle-Hole Symmetry}
In the main text, we quoted the error associated to assuming the particle-hole symmetry to be an exact symmetry of the system rather than an approximate one. We define this error as done so in Ref. \cite{Song2021}, which looks at the overlap of the particle-hole transformed state with its approximate particle-hole related partner,
\begin{equation}
    \text{err}(\Sigma C_{2z}^{\null}\mathcal{T}) = \int_{\text{MBZ}} \frac{d^{2}k}{\Omega_{\text{MBZ}}} \left( 1 - |\bra{\Xi_{-1,-\boldsymbol{k}}^{\tau}} \mathscr{P}_{z,\boldsymbol{k}}^{\tau} \ket{\Xi_{1,\boldsymbol{k}}^{\tau}}|^{2} \right),
\end{equation}
where the integral is performed over the area of the MBZ, $\Omega_{\text{MBZ}}$ is the area of the MBZ, and $\ket{\Xi_{n,\boldsymbol{k}}^{\tau}}$ is the eigenstate of the TBK Hamiltonian with band label $n$. When calculating the errors quoted in the main text, we focused on the $K$-valley ($\tau = 1$) and opted to use a TBK Hamiltonian built using the $\Gamma_{\text{M}}$-centric construction method due to its symmetric inclusion of sites around the $q$-lattice. For the first magic angle ($\theta \simeq 2.85^{\circ}$) we retained all points in $\mathcal{Q}_{\pm}^{\tau}$ satisfying $|\boldsymbol{\mathcal{Q}}| \leq 7 |\mathbf{K}_{\text{M}}|$, whilst for the third magic angle ($\theta \simeq 0.946^{\circ}$) we retained all points in $\mathcal{Q}_{\pm}^{\tau}$ satisfying $|\boldsymbol{\mathcal{Q}}| \leq 11 |\mathbf{K}_{\text{M}}|$.

\subsection{The Overlap Integral} \label{Subsec_Overlap_Integral}

The behaviour of the overlap integral are determined by the material details including the intralayer atomic separation, interlayer distance, and active orbitals. These parameters set the scales for the decay rate of the overlap integral with distance and the nature of hopping electrons employ to traverse the lattice. A common means to model electrons tunnelling between orbitals is via Slater-Koster parameterisation \cite{Slater1954}. For example, the overlap integral for bernal bilayer graphene is given by a $p_{z}$-orbital hopping model \cite{Koshino2018},
\begin{equation}
    t(\mathbf{r}) = V_{pp\pi}^{0} e^{-(r-a_{\text{cc}})/r_{0}} \left(1 - \frac{z^{2}}{r^{2}} \right) + V_{pp\sigma}^{0} e^{-(r-d_{0})/r_{0}} \frac{z^{2}}{r^{2}},
    \label{G_SK_model}
\end{equation}
where $a_{\text{cc}} = a_{\text{G}}/\sqrt{3}$ is the carbon-carbon bond length, $a_{\text{G}} = 0.246$nm is the graphene lattice constant, $r_{0} = 0.184 a_{\text{G}}^{\null}$, $d_{0} = 0.335$nm (interlayer), $V_{pp\pi}^{0} = 2.7$eV (parameterises intralayer $\pi$ overlap), and $V_{pp\sigma}^{0} = -0.48$eV (parameterises interlayer $\sigma$ overlap). In the case of the a van der Waals (vdW) heterostructure based upon the kagome lattice, there is room for choosing an appropriate model for the overlap integral since the first kagome-based vdW heterostructures are yet to be realised.

With this in mind, we briefly discuss current kagome materials to determine possible choices for $t(\mathbf{r})$. First, we appreciate that the lattice constant for the kagome lattice will naturally be larger than that of graphene due to possessing three sublattice sites per unit cell as opposed to two, typically in the region of $a \sim 0.5 - 0.7$nm: 0.5338nm in Fe${}_{3}$Sn${}_{2}$ \cite{Ye2018}, 0.665nm in LaTl${}_{3}$ \cite{Vekovshinin2025}, and 0.54949nm in CsV${}_{3-x}$Ta${}_{x}$Sb${}_{5}$ \cite{Luo2023}. We note that true kagome monolayers (i.e. a network of atoms occupying a kagome lattice without an atomic cage) have only been very recently synthesised in rare-earth based systems $X$Pb${}_{3}$ ($X =$ La, Ce, Eu, Gd, Yb) \cite{Mihalyuk2022,Vekovshinin2024,Denisov2025,Denisov2026} and LaTl${}_{3}$ \cite{Vekovshinin2025}, which host perfect kagome monolayers of Pb or Tl, stabalised by a rare-earth atom at the kagome hexagon center and situated on top of a Si(111) substrate. The interlayer separation characterising a vdW kagome bilayer is therefore not yet known with certainty. However, kagome bilayers do appear in materials with more complex atomic registries, such as Fe${}_{3}$Sn${}_{2}$ \cite{Ye2018}, Nb${}_{3}$X${}_{8}$ ($X = $ Cl, Br, or I) \cite{Wang2023,Sante2026}, and $A$V${}_{6}$Sb${}_{6}$ ($A = $ K, Rb, or Cs) \cite{Shi2022}, which display a range of separations between the individual kagome layers. Given the weak nature of van der Waals forces in comparison to standard chemical bonding (covalent, ionic, and metallic), van der Waals bonds are generally longer in contrast to complete chemical bonds. For example, in bernal bilayer graphene, the ratio of the interlayer separation to carbon-carbon bond length is $d_{\perp}/a_{\text{cc}} \simeq 2.36$. Using this ratio as an estimate for the typical variation between chemical bonds and van der Waals bonds, we estimate $d_{\perp} \sim 0.59 - 0.83$ nm as an upper range for the interlayer separation of a vdW kagome bilayer heterostructure. For the main text, we chose to work with $d_{\perp} = 0.6596$ nm as in Ref. \cite{Perkins2025b}, which lies in this range.

In modelling the overlap integral of TBK, Ref. \cite{Lima2019} employed an $s$-orbital model with the form ($\gamma = 20$),
\begin{equation}
    t(\mathbf{r}) = -t_{0} e^{-2\gamma(r-a/2)/a} \delta_{z,0} - 0.3 t_{0} e^{-\gamma(r-d_{\perp})/d_{\perp}} (1-\delta_{z,0}).
    \label{Lima_model}
\end{equation}
Given the 2D in-plane rotational symmetry of the overlap integral, the 2D Fourier transform for interlayer tunnelling can be found quickly using
\begin{equation}
    t_{\perp}(\mathbf{q}) = 2\pi \int_{0}^{\infty} d\rho \, r J_{0}(q\rho) t(\rho,z),
\end{equation}
where $J_{0}(x)$ is the zeroth-order Bessel function of the first-kind. This yields $t_{\perp}(\mathbf{q} \in \mathcal{D}_{2}) \simeq 0.13 t_{\perp}(\mathbf{K}_{\tau})$, where $\mathcal{D}_{2}$ is the second closet set of Dirac point to the origin of the monolayer first BZ, and so we can argue that $t_{\perp}(\mathbf{q})$ decays sufficiently quickly to truncate the sum over reciprocal lattice vectors in Eq. (\ref{TBK_onethird_LE_Hamiltonian}) to account only for $\mathcal{D}_{1}$. We justify this on the basis that next-nearest-neighbour intralayer tunnelling is ignored when it is $10$\% of $t_{0}$. Nonetheless, changing these parameters and the tunnelling model can significantly change the size and sign of $t(\mathbf{q} \in \mathcal{D}_{2})$.

\begin{figure}
    \centering
    \includegraphics[width=\linewidth]{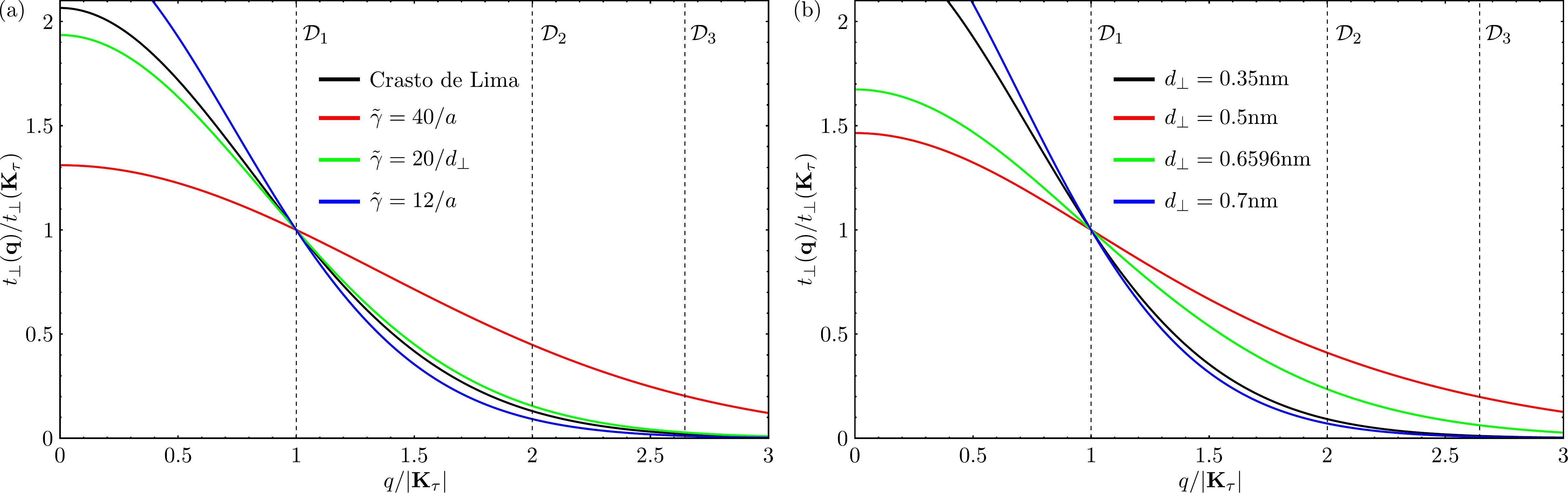}
    \caption{Comparison of the Fourier transformed overlap integral for interlayer tunnelling for various models (a) and interlayer separation (b). In both panels we take $a = 0.5338$nm in-line with Refs. \cite{Ye2018,Lima2019}. (a): Variation of $t_{\perp}(\mathbf{q})$ using the model of Ref. \cite{Lima2019} summarised in Eq. (\ref{Lima_model}) (black) and our $p_{z}$-oribtal model in Eq. (\ref{Kagome_SK_model}). We take $d_{\perp} = 0.6596$nm here. (b): Illustrating how our Eq. (\ref{Kagome_SK_model}) depends upon the choice of $d_{\perp}$ with $\tilde{\gamma} = 12/a$.}
    \label{Tunnelling_models}
\end{figure}

Taking inspiration from the tunnelling model of Ref. \cite{Lima2019}, we can try to recast Eq. (\ref{Lima_model}) in terms of Slater-Koster parameters. To do so, we first need to know what orbitals lie on the KSL sites. To this end, we draw attention towards the previously mentioned rare earth-based systems \cite{Mihalyuk2022,Vekovshinin2024,Denisov2025,Denisov2026,Vekovshinin2025}. The kagome lattice is formed of Pb or Tl atoms in these systems, both of whose outermost electrons are $p$-orbitals, so we shall use a modified version of the overlap integral for graphene as a toy model for tunnelling in TBK,
\begin{equation}
    t(\mathbf{r}) = -t_{0} e^{-\tilde{\gamma}(r-a/2)} \left(1 - \frac{z^{2}}{r^{2}} \right) - 0.3 t_{0} e^{-\tilde{\gamma}(r-d_{\perp})} \frac{z^{2}}{r^{2}}.
    \label{Kagome_SK_model}
\end{equation}
Comparing the 2D Fourier transforms of Eqs. (\ref{Lima_model}) and (\ref{Kagome_SK_model}), we see quite large changes in their behaviour and a sensitivity to the details of the tunnelling model in Fig. \figRef{Tunnelling_models}{a}. In particular, we see that for choosing $\tilde{\gamma} = 40/a$ to match the intralayer decay rate of Ref. \cite{Lima2019}, the profile of $t_{\perp}(\mathbf{q})$ is broadened drastically and $t_{\perp}(\mathbf{q} \in \mathcal{D}_{2}) \simeq 0.448 t_{\perp}(\mathbf{K}_{\tau})$. Taking $\tilde{\gamma} = 20/d_{\perp}$ instead to match the interlayer decay rate of Ref. \cite{Lima2019}, the behaviour of $t_{\perp}(\mathbf{q})$ is only slightly modified compared to Eq. (\ref{Lima_model}) and $t_{\perp}(\mathbf{q} \in \mathcal{D}_{2}) \simeq 0.155 t_{\perp}(\mathbf{K}_{\tau})$. If we instead select take $\tilde{\gamma} = 12/a \simeq 1/r_{0}$ to mimic a decay rate comparable to graphene, which broadens the real-space profile of the overlap integral whilst yielding $t_{0}$ for nearest-neighbour intralayer hopping and $\sim 0.01 t_{0}$ for next-nearest-neighbour intralayer hopping, we obtain a more rapidly decaying Fourier transform with $t_{\perp}(\mathbf{q} \in \mathcal{D}_{2}) \simeq 0.092 t_{\perp}(\mathbf{K}_{\tau})$.

In addition to altering the decay rate set by $\tilde{\gamma}$, we may also consider changes to the interlayer separation. Whilst we originally chose $d_{\perp} = 0.6596$nm, we note that in vdW constructed systems this distance has been observed to vary over the range $0.3\text{nm} \lesssim d_{\perp} \lesssim 0.7$nm in a plethora of bilayer systems \cite{Dai2014,Uchida2014,Li2017,Shi2022,Xia2025}. We demonstrate how changes in $d_{\perp}$ alter $t_{\perp}(\mathbf{q})$ in Fig. \figRef{Tunnelling_models}{b} based upon the model in Eq. (\ref{Kagome_SK_model}), where we see that smaller layer separations broaden $t_{\perp}(\mathbf{q})$ and allow $\mathcal{D}_{n>1}$ to contribute non-negligible corrections to the $\mathcal{D}_{1}$ terms in the tunnelling Hamiltonian.

\subsubsection{Including the Second and Third Sets of Dirac Points}

Let us consider the case where $\mathcal{D}_{2}$ cannot be neglected but the sets $\mathcal{D}_{n>2}$ can be. In this case, we must account for an additional set of terms to those already listed in Eq. (\ref{BM_tunnelling_minimum}) whose reciprocal lattice vectors are $\tau \mathbf{G}_{3} = \tau \mathbf{b}_{2} = -\tau \mathbf{G}_{4}$ and $\tau \mathbf{G}_{5} = -\tau (2\mathbf{b}_{1} + \mathbf{b}_{2})$. These corrections yield
\begin{subequations}
\begin{gather}
    \mathscr{H}_{\text{T},\mathbf{k}\mathbf{p}}' = T_{t,\mathbf{k}}^{\null} \delta_{\mathbf{p} - \mathbf{k}, \tau \mathbf{q}_{t}} + T_{br,\mathbf{k}}^{\null} \delta_{\mathbf{p} - \mathbf{k}, \tau \mathbf{q}_{br}} + T_{bl,\mathbf{k}}^{\null} \delta_{\mathbf{p} - \mathbf{k}, \tau \mathbf{q}_{bl}},
    \\
    T_{t,\mathbf{k}\mathbf{p}}^{\tau} = \omega_{1} \sum_{\chi,\eta} \begin{pmatrix}
            \psi_{-,\mathbf{k}}^{\tau(\chi)*} \psi_{-,\mathbf{p}}^{\tau(\eta)} & \psi_{-,\mathbf{k}}^{\tau(\chi)*} \psi_{+,\mathbf{p}}^{\tau(\eta)} \\
            \psi_{+,\mathbf{k}}^{\tau(\chi)*} \psi_{-,\mathbf{p}}^{\tau(\eta)} & \psi_{+,\mathbf{k}}^{\tau(\chi)*} \psi_{+,\mathbf{p}}^{\tau(\eta)}
    \end{pmatrix}
    e^{-i (2\mathbf{b}_{1}+\mathbf{b}_{2}) \cdot (\boldsymbol{\delta}_{\chi}-\boldsymbol{\delta}_{\eta})}
    \\
    T_{br,\mathbf{k}\mathbf{p}}^{\tau} = \omega_{1} \sum_{\chi,\eta} \begin{pmatrix}
            \psi_{-,\mathbf{k}}^{\tau(\chi)*} \psi_{-,\mathbf{p}}^{\tau(\eta)} & \psi_{-,\mathbf{k}}^{\tau(\chi)*} \psi_{+,\mathbf{p}}^{\tau(\eta)} \\
            \psi_{+,\mathbf{k}}^{\tau(\chi)*} \psi_{-,\mathbf{p}}^{\tau(\eta)} & \psi_{+,\mathbf{k}}^{\tau(\chi)*} \psi_{+,\mathbf{p}}^{\tau(\eta)}
    \end{pmatrix}
    e^{i \tau \mathbf{b}_{2} \cdot (\boldsymbol{\delta}_{\chi}-\boldsymbol{\delta}_{\eta})},
    \\
    T_{bl,\mathbf{k}\mathbf{p}}^{\tau} = \omega_{1} \sum_{\chi,\eta} \begin{pmatrix}
            \psi_{-,\mathbf{k}}^{\tau(\chi)*} \psi_{-,\mathbf{p}}^{\tau(\eta)} & \psi_{-,\mathbf{k}}^{\tau(\chi)*} \psi_{+,\mathbf{p}}^{\tau(\eta)} \\
            \psi_{+,\mathbf{k}}^{\tau(\chi)*} \psi_{-,\mathbf{p}}^{\tau(\eta)} & \psi_{+,\mathbf{k}}^{\tau(\chi)*} \psi_{+,\mathbf{p}}^{\tau(\eta)}
    \end{pmatrix}
    e^{-i \tau \mathbf{b}_{2} \cdot (\boldsymbol{\delta}_{\chi}-\boldsymbol{\delta}_{\eta})},
\end{gather}
\label{BM_tunnelling_2ndSet}
\end{subequations}
where $\omega_{1} = t_{\perp}(\mathbf{K}_{\tau} + \tau \mathbf{G}_{3})$ and 
\begin{subequations}
\begin{gather}
    \mathbf{q}_{t} = (R_{\bar{\vartheta}} - R_{\vartheta}) (\mathbf{K}_{+} - 2 \mathbf{b}_{1} - \mathbf{b}_{2}) = \frac{16\pi}{3a} \sin\left(\frac{\theta}{2}\right)
    \begin{pmatrix}
        0 \\ 1
    \end{pmatrix} = -2\mathbf{q}_{b},
    \\
    \mathbf{q}_{br} = (R_{\bar{\vartheta}} - R_{\vartheta}) (\mathbf{K}_{+} + \mathbf{b}_{2}) = \frac{8\pi}{3a} \sin\left(\frac{\theta}{2}\right)
    \begin{pmatrix}
        \sqrt{3} \\ -1
    \end{pmatrix} = -2\mathbf{q}_{tl},
    \\
    \mathbf{q}_{bl} = (R_{\bar{\vartheta}} - R_{\vartheta}) (\mathbf{K}_{+} - \mathbf{b}_{2}) = \frac{8\pi}{3a} \sin\left(\frac{\theta}{2}\right)
    \begin{pmatrix}
        -\sqrt{3} \\ -1
    \end{pmatrix} = -2\mathbf{q}_{tr}.
\end{gather}
\end{subequations}

Returning to the $q$-lattice depiction of momentum conserving interlayer tunnelling processes, we see that the $\mathcal{D}_{2}$ Dirac points connect third-nearest-neighbours in the $q$-lattice, see Fig. \figRef{q_lattice_illustration}{b}. At the level of the continuum Hamiltonian, the effect of the $\mathcal{D}_{2}$ Dirac points can only be seen when including shells beyond the second in the $q$-lattice, i.e. $\mathcal{S}_{n>2}$. Upon the inclusion of $\mathcal{S}_{3}$, the $\mathcal{D}_{2}$ generate contributions connecting the points in $\mathcal{S}_{3}$ to both $\mathcal{S}_{0}$ and $\mathcal{S}_{2}$.

We note that whilst the expressions provided in Eq. (\ref{BM_tunnelling_2ndSet}) are general for a twisted system governed by two states near the monolayer BZ corners, the kagome lattice proves to be unperturbed by this first set of correction terms. Specifically, for the kagome lattice Dirac states expanded about a given monolayer valley, the matrices yielded by Eq. (\ref{BM_tunnelling_2ndSet}) individually vanish, suggesting that the physics of the kagome lattice is robust against overlap integrals whose Fourier transforms do not decay rapidly in momentum. If the overlap integral Fourier transform decays sufficiently slowly, then corrections due to fifth-nearest-neighbour hopping on the $q$-lattice may become relevant, i.e. the valleys belonging to $\mathcal{D}_{3}$ must be accounted for in this scenario. Consequently, a new set of $q$-vectors will be needed in order to describe this process, of which we find there to be six possible vectors,
\begin{subequations}
\begin{gather}
    \mathbf{q}_{bl}' = (R_{\bar{\vartheta}}^{\null} - R_{\vartheta}^{\null}) (\mathbf{K}_{+}^{\null} + \mathbf{b}_{1}^{\null}) = \frac{4\pi}{3a} \sin\left(\frac{\theta}{2}\right) \begin{pmatrix}
        -\sqrt{3} \\ -5
    \end{pmatrix},
    \\
    \mathbf{q}_{br}' = (R_{\bar{\vartheta}}^{\null} - R_{\vartheta}^{\null}) (\mathbf{K}_{+}^{\null} + \mathbf{b}_{1}^{\null} + \mathbf{b}_{2}^{\null}) = \frac{4\pi}{3a} \sin\left(\frac{\theta}{2}\right) \begin{pmatrix}
        \sqrt{3} \\ -5
    \end{pmatrix},
    \\
    \mathbf{q}_{tr}^{\prime(1)} = (R_{\bar{\vartheta}}^{\null} - R_{\vartheta}^{\null}) (\mathbf{K}_{+}^{\null} + \mathbf{b}_{2}^{\null} - \mathbf{b}_{1}^{\null}) = \frac{4\pi}{3a} \sin\left(\frac{\theta}{2}\right) \begin{pmatrix}
        3\sqrt{3} \\ 1
    \end{pmatrix},
    \\
    \mathbf{q}_{tr}^{\prime(2)} = (R_{\bar{\vartheta}}^{\null} - R_{\vartheta}^{\null}) (\mathbf{K}_{+}^{\null} - 2\mathbf{b}_{1}^{\null}) = \frac{4\pi}{3a} \sin\left(\frac{\theta}{2}\right) \begin{pmatrix}
        2\sqrt{3} \\ 4
    \end{pmatrix},
    \\
    \mathbf{q}_{tl}^{\prime(1)} = (R_{\bar{\vartheta}}^{\null} - R_{\vartheta}^{\null}) (\mathbf{K}_{+}^{\null} - 2 \mathbf{b}_{2}^{\null} - \mathbf{b}_{1}^{\null}) = \frac{4\pi}{3a} \sin\left(\frac{\theta}{2}\right) \begin{pmatrix}
        -3\sqrt{3} \\ 1
    \end{pmatrix},
    \\
    \mathbf{q}_{tl}^{\prime(2)} = (R_{\bar{\vartheta}}^{\null} - R_{\vartheta}^{\null}) (\mathbf{K}_{+}^{\null} - 2(\mathbf{b}_{1}^{\null} + \mathbf{b}_{2}^{\null})) = \frac{4\pi}{3a} \sin\left(\frac{\theta}{2}\right) \begin{pmatrix}
        -2\sqrt{3} \\ 4
    \end{pmatrix}.
\end{gather}
\end{subequations}

\begin{figure}[t]
    \centering
    \includegraphics[width=\linewidth]{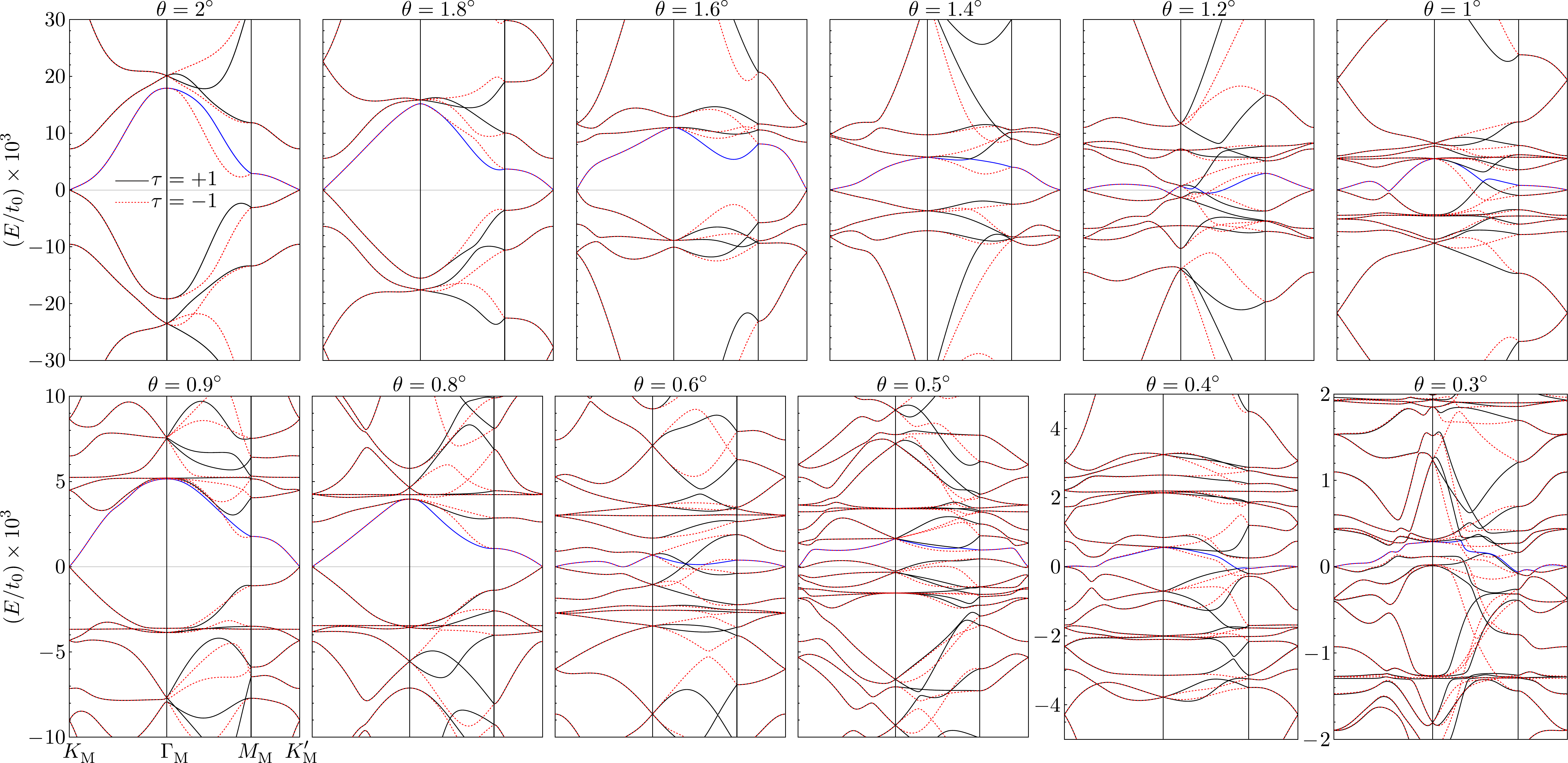}
    \caption{Band structures taken along the high-symmetry path in the MBZ for various twist angles. The tunnelling model applied is the $p_{z}$-orbital model described in Eq. (\ref{Kagome_SK_model}) with $\tilde{\gamma} = 20/d_{\perp}$, $d_{\perp} = 0.6596$nm, and $a = 0.5338$nm yielding $\omega_{0} \simeq -0.0819 t_{0}$. Solid (dashed) lines are bands calculated through expansion around $\mathbf{K}_{+}$ ($\mathbf{K}_{-}$) in the monolayer. The solid blue line indicates the first conduction band for the $\mathbf{K}_{+}$ expansion.}
    \label{Band_structs_twist_var}
\end{figure}

\subsection{Sublattice Projection in Twisted Bilayer Kagome Near $1/3$ Filling}

To acquire the sublattice projection of a band in TBK near $1/3$ filling, we will make use of the monolayer sublattice polarisation operators introduced in Eq. (\ref{KSL_downfold_expansion}). Specifically, given that sublattice projection is a purely intralayer operator, it will be block diagonal in the basis of the TBK Hamiltonian defined in Eq. (\ref{TBK_Hamiltonian}), such that the sublattice projector in a given block will have a momentum matching the monolayer Hamiltonian of that block, i.e. $\widetilde{\mathscr{P}}_{\eta,\mathbf{k}+n_{1}\mathbf{q}_{b}+n_{2}\mathbf{q}_{tr}+n_{3}\mathbf{q}_{tl}}^{\text{D},\tau}$ will appear in the same block as $\mathscr{H}_{\mathbf{k}+n_{1}\mathbf{q}_{b}+n_{2}\mathbf{q}_{tr}+n_{3}\mathbf{q}_{tl}}^{\text{D},\tau}$. Moreover, the sublattice projection of a given layer will only contain non-zero block diagonal components in the blocks matching the layer of interest. To be clear, let us consider the Hamiltonian containing $\mathcal{S}_{1}$ and $\mathcal{S}_{0}$: the sublattice projection operators may be written as follows
\begin{equation}
\begin{gathered}
    \widetilde{\mathscr{P}}_{\eta_{1}^{\null},\mathbf{k}}^{\text{TBK},\tau} = \begin{pmatrix}
        \widetilde{\mathscr{P}}_{\eta_{1}^{\null},\mathbf{k}}^{\text{D},\tau} & 0 & 0 & 0 \\
        0 & 0 & 0 & 0 \\
        0 & 0 & 0 & 0 \\
        0 & 0 & 0 & 0
    \end{pmatrix},
    \qquad
    \widetilde{\mathscr{P}}_{\eta_{2}^{\null},\mathbf{k}}^{\text{TBK},\tau} = \begin{pmatrix}
        0 & 0 & 0 & 0 \\
        0 & \widetilde{\mathscr{P}}_{\eta_{2}^{\null},\mathbf{k}+\mathbf{q}_{b}}^{\text{D},\tau} & 0 & 0 \\
        0 & 0 & \widetilde{\mathscr{P}}_{\eta_{2}^{\null},\mathbf{k}+\mathbf{q}_{tr}}^{\text{D},\tau} & 0 \\
        0 & 0 & 0 & \widetilde{\mathscr{P}}_{\eta_{2}^{\null},\mathbf{k}+\mathbf{q}_{tl}}^{\text{D},\tau}
    \end{pmatrix},
    \\
    \widetilde{\mathscr{P}}_{\eta,\mathbf{k}}^{\text{TBK},\tau} = \widetilde{\mathscr{P}}_{\eta_{1}^{\null},\mathbf{k}}^{\text{TBK},\tau} + \widetilde{\mathscr{P}}_{\eta_{2}^{\null},\mathbf{k}}^{\text{TBK},\tau},
    \qquad
    \widetilde{\mathscr{P}}_{l,\mathbf{k}}^{\text{TBK},\tau} = \sum_{\eta} \widetilde{\mathscr{P}}_{\eta_{l}^{\null},\mathbf{k}}^{\text{TBK},\tau},
\end{gathered}
\end{equation}
with $\widetilde{\mathscr{P}}_{l,\mathbf{k}}^{\text{TBK},\tau}$ being the layer projection operator.

\section{Electronic Structure and Sublattice Projection}

Let us now explore the band structure and role of sublattice interference in TBK near $1/3$ filling. In Fig. \ref{Band_structs_twist_var} we show the band structure for TBK along the high-symmetry path for various twist angles. For the smallest twist angles, $\theta < 1^{\circ}$, we see that whilst the bands become more densely packed, as expected, some bands begin to exhibit extremely flat regions. For all cases shown with $\theta < 1^{\circ}$, such flat regions primarily appear in bands situated away from the first conduction/valence bands. However, for $\theta \simeq 0.95^{\circ}$ (see main text), large areas of flatness develop in the conduction and valence bands, corresponding to the onset of a HOVHS. Though the lack of PHS prevents a mathematically exact monkey saddle manifesting and helps preserve the Dirac cone at the MBZ edges, the effects of a monkey saddle singularity will still be observed in experimental measurements due to the drastic suppression of the renormalised Fermi velocity yielding a Dirac cone of negligible depth relative to the band width. To break these Dirac cones, we would need to break inversion or time-reversal symmetry. Structurally, the breaking of inversion symmetry is the more readily achieved approach since breathing kagome lattices (differing up and down triangle sizes) already exist naturally in nature. Therefore, it is reasonable to infer that monkey saddle singularities may be tunable through simple twisting when the monolayers are breathing kagome lattices.

\subsection{Renormalised Fermi Velocity}

\subsubsection{Calculation}

To establish the existence of magic angles, we calculate the Fermi velocity associated to the Dirac cones appearing at the MBZ edges. We refer to this as the renormalised Fermi velocity, $v_{F}^{*}$, which is generally reduced in comparison to the Fermi velocity of the kagome monolayer Dirac cones. Our calculation of $v_{F}^{*}$ is performed using the velocity operator, $\hat{v}_{i}^{\null} = \partial_{k_{i}}^{\null} H$, which gives the velocity for the electrons of band $n$ at a given momentum, $\mathbf{v}_{n}^{\null}(\mathbf{k}) = \nabla_{\mathbf{k}}^{\null} E_{n}^{\null}(\mathbf{k})$. A Dirac cone exhibits a local $C_{\infty}$ symmetry and so we choose to calculate $v_{F}^{*}$ along the $x$-axis without loss of generality. This calculation is done numerically according to
\begin{equation}
    v_{F}^{*} = \frac{E_{1}(\Delta k_{x}(\theta) \mathbf{e}_{x}) - E_{1}(\mathbf{0})}{\Delta k_{x} (\theta)}, \qquad \Delta k_{x}^{\null}(\theta) = 10^{-7} |\mathbf{q}_{b}^{\null}(\theta)| = \frac{8\pi}{3a} \sin\left(\frac{\theta}{2}\right) \times 10^{-7}.
\end{equation}
We note that as the Dirac cone flattens, $v_{F}^{*} \rightarrow 0$, the $C_{3}^{\null}$ nature of the MBZ corner becomes more apparent, wherein smaller choices of $\Delta k_{x}(\theta)$ will be needed to keep the calculation situated within the region exhibiting local $C_{\infty}^{\null}$.

\begin{table}[t]
    \centering
    \begin{tabular}{c|| c | c | c | c | c | c }
        & \multicolumn{6}{c}{$\omega_{0} = -0.1 t_{0}$} \\
        \hline
        Twist Range & $-$ & $[0.4^{\circ},0.6^{\circ})$ & $[0.6^{\circ},0.8^{\circ})$ & $[0.8^{\circ},1.1^{\circ})$ & $[1.1^{\circ},1.5^{\circ})$ & $[1.5^{\circ},2^{\circ}]$ \\
        Shell number & $-$ & 32 & 24 & 20 & 18 & 15 \\
        \hline\hline
        & \multicolumn{6}{c}{$\omega_{0} = -0.0819 t_{0}$} \\
        \hline
        Twist Range & $-$ & $[0.4^{\circ},0.6^{\circ})$ & $[0.6^{\circ},0.8^{\circ})$ & $[0.8^{\circ},1.1^{\circ})$ & $[1.1^{\circ},1.5^{\circ})$ & $[1.5^{\circ},2^{\circ}]$ \\
        Shell number & $-$ & 22 & 18 & 15 & 12 & 10 \\
        \hline\hline
        & \multicolumn{6}{c}{$\omega_{0} = -0.05 t_{0}$} \\
        \hline
        Twist Range & $[0^{\circ},0.3^{\circ})$ & $[0.3^{\circ},0.4^{\circ})$ & $[0.4^{\circ},0.8^{\circ})$ & $[0.8^{\circ},1.1^{\circ})$ & $[1.1^{\circ},1.5^{\circ})$ & $[1.5^{\circ},2^{\circ}]$ \\
        Shell number & 30 & 24 & 18 & 15 & 12 & 10 \\
        \hline\hline
        & \multicolumn{6}{c}{$\omega_{0} = -0.02 t_{0}$} \\
        \hline
        Twist Range & $-$ & $[0^{\circ},0.2^{\circ})$ & $[0.2^{\circ},0.8^{\circ})$ & $[0.8^{\circ},1.1^{\circ})$ & $[1.1^{\circ},1.5^{\circ})$ & $[1.5^{\circ},2^{\circ}]$ \\
        Shell number & $-$ & 27 & 18 & 15 & 12 & 10 \\
        \hline\hline
        & \multicolumn{6}{c}{$\omega_{0} = -0.01 t_{0}, \, -0.005 t_{0}$} \\
        \hline
        Twist Range & $-$ & $-$ & $[0^{\circ},0.2^{\circ})$ & $[0.2^{\circ},0.8^{\circ})$ & $[0.8^{\circ},1.5^{\circ})$ & $[1.5^{\circ},2^{\circ}]$ \\
        Shell number & $-$ & $-$ & 18 & 15 & 10 & 8
    \end{tabular}
    \caption{Shell numbers used in the construction of the continuum Hamiltonian for the different choices of $\omega_{0}$ and the range of angles for which they were used.}
    \label{Shell_number_table}
\end{table}

\begin{figure}[b]
    \centering
    \includegraphics[width=\linewidth]{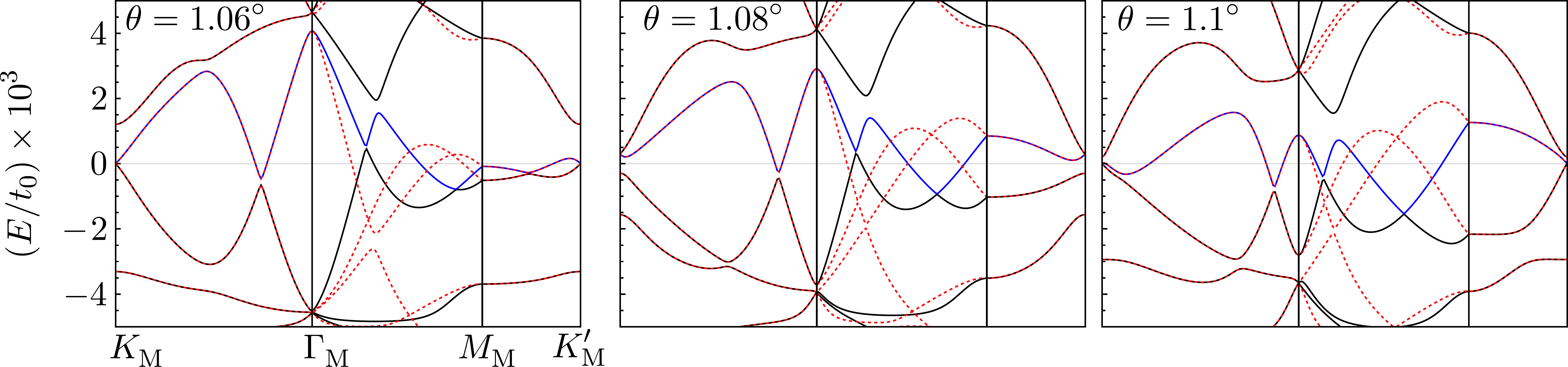}
    \caption{Changing of the bands closest to $1/3$ filling as the twist is varied across a change in degeneracy involving the first conduction band at the MBZ corners.}
    \label{Degeneracy_lifting}
\end{figure}

Naturally, as the twist angle is reduced, the number of shells required in the $q$-lattice increases. The shell numbers we used and the range of angles for which they were applied are listed in Table \ref{Shell_number_table}. These shell numbers were picked so that the resulting band structure exhibited the expected $C_{3z}$, $C_{2x}$, and $\mathcal{T}$ symmetries of the TBK system to reasonable accuracy for the innermost bands (i.e. those with $|n| \sim 1$). We appreciate that the required number of shells is also sensitive to the choice of $\omega_{0}$, hence the additional row in Table \ref{Shell_number_table}.

\subsubsection{Discontinuities and the Lack of Particle-Hole Symmetry}

We saw in the main text that the renormalised Fermi velocity exhibited apparent discontinuities in its twist dependence when particle-hole symmetry was not enforced. This arises due to the degeneracy between the $|n| = 1$ bands which yield the Dirac cone lifting and being replaced by a degeneracy between the $n = 1$ and $n = 2$ bands to create a new Dirac cone. We illustrate this changing of degeneracy in Fig. \ref{Degeneracy_lifting}.

\subsection{Sublattice Projection}

We plot the sublattice projection of the $|n| = 1$ bands for $\theta = 0.7^{\circ}$ and $\theta = 1.4^{\circ}$ in Fig. \figRef{SLP_plots}{a,b} as further examples of how the bands are not as strongly polarized as in large twist commensurate TBK \cite{Perkins2025b}, thus reducing the effect of sublattice interference. We further plot the sublattice projections for near-flat bands (i.e. those with minimal bandwidth) with $|n| > 1$ in Fig. \figRef{SLP_plots}{c--e}, where we see similar sublattice mixing features to the dispersive bands which inhibit sublattice interference. We note that for $\theta = 0.7^{\circ}$ the $n = 3$ flat band is delocalised equally across all three sublattices throughout the majority of the MBZ, only gaining a slight sublattice pereference near the MBZ edges.

\begin{figure}[b]
    \centering
    \includegraphics[width=\linewidth]{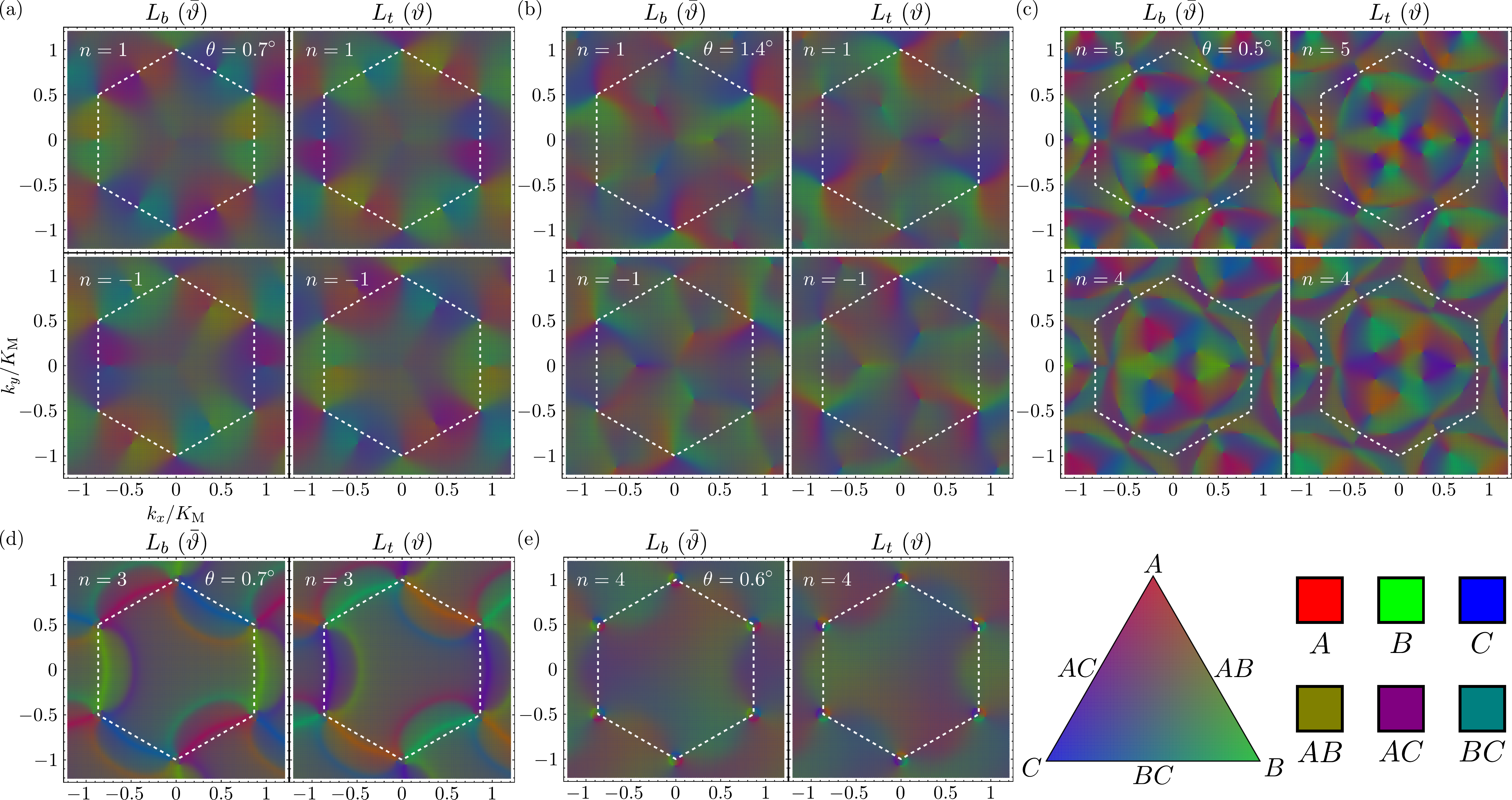}
    \caption{Sublattice projections of various bands for different twist angles. (a),(b): Projections of the first conduction ($n = 1$) and first valence ($n = -1$) bands for $\theta = 0.7^{\circ}$ (a) and $1.4^{\circ}$ (b). (c)--(e): Projections for near-flat bands appearing due to twisting for $\theta = 0.5^{\circ}$ (c), $0.6^{\circ}$ (d), and $0.7^{\circ}$ (e).}
    \label{SLP_plots}
\end{figure}

\begin{figure}[h!]
    \centering
    \includegraphics[width=\linewidth]{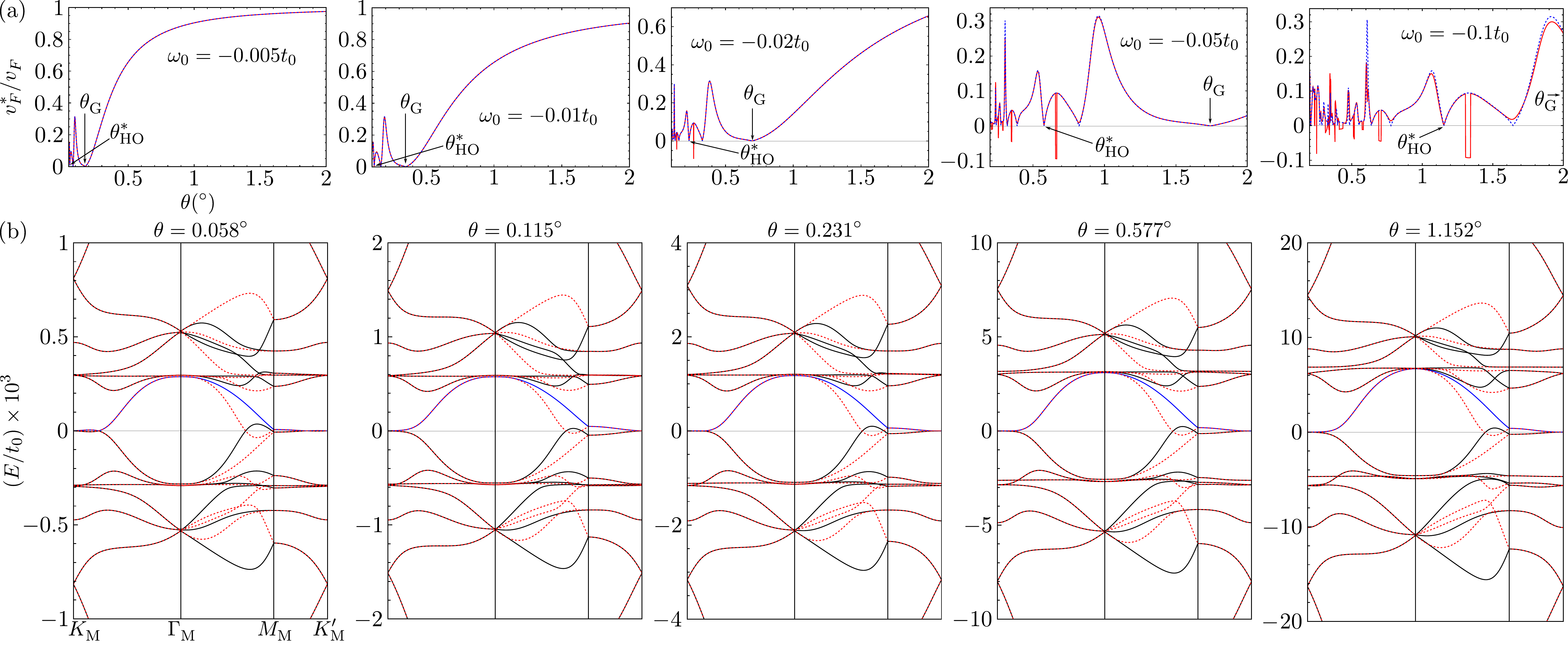}
    \caption{Demonstration of the robustness of magic angles against the characteristic twisted interlayer tunnelling energy. (a): Renormalised Fermi velocity for various choices of $\omega_{0}$, recalling that $\omega_{0} \simeq -0.0819 t_{0}$ was used in the main text. We see that the general features of $v_{F}^{*}$ remain unchanged with variation in $\omega_{0}$ aside from the magic angles reducing to smaller values as $|\omega_{0}|$ is decreased. The first and third magic angles as indicated by $\theta_{1}^{*}$ and $\theta_{3}^{*}$, respectively. (b): Electronic band structure at the third magic angle with $\omega_{0}/t_{0} = -0.005$, $-0.01$, $-0.02$, $-0.05$, $-0.1$ from left to right.}
    \label{Renorm_vF_w0}
\end{figure}

\begin{figure}[h!]
    \centering
    \includegraphics[width=\linewidth]{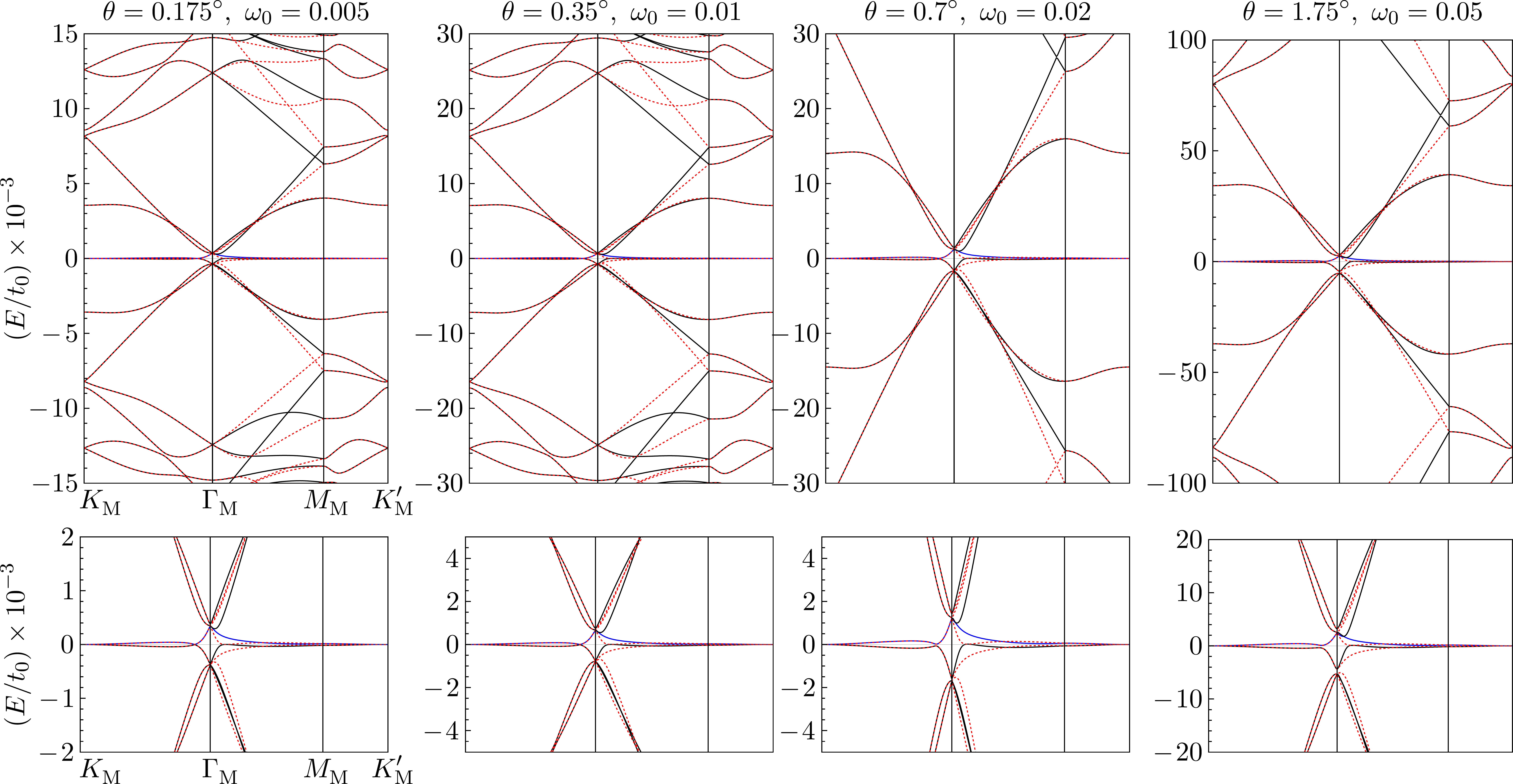}
    \caption{Band structure of TBK for various choices of $\omega_{0}$ and the twist taken close to the first magic angle. The top panels show the band structures of an extended energy range whilst the bottom panels focus on the flat bands reminiscent of magic angle TBG when lattice relaxation is ignored (i.e. when $AA$ tunnelling and $AB$ tunnelling are treated equally).}
    \label{w0_variation_th1}
\end{figure}

\subsection{Robustness of Magic Angles}

In the main text and all prior analysis of the electronic structure and properties of TBK, we assumed $a = 0.5338$ nm, $d_{\perp} = 0.6596$ nm and $\tilde{\gamma} = 20/d_{\perp}$ when employing Eq. (\ref{Kagome_SK_model}) to determine the twisted tunnelling energy scales, $\omega_{0}$ and $\omega_{1}$. This choice of parameters gave $\omega_{0} = -0.819 t_{0}$ and $\omega_{1} = -0.0127 t_{0}$. However, as already shown in Section \ref{Subsec_Overlap_Integral}, these energy scales depend greatly upon the details of the tunnelling model. Nonetheless, we find that the qualitative structure and existence of magic angles is robust against changes in $\omega_{0}$ whilst assuming $t_{\perp}(\mathbf{q})$ decays sufficiently quickly that we may neglect $\omega_{1}$. Specifically, we see in Fig. \figRef{Renorm_vF_w0}{a} that in reducing $|\omega_{0}|$, the magic angles and degeneracy switching angles shift towards smaller twist angles. Moreover, the magic angles appear to yield similar band structures with their shape appearing largely unchanged but the bandwidth being reduced for smaller choices of $\omega_{0}$, see Figs. \figRef{Renorm_vF_w0}{b} and \ref{w0_variation_th1}. This suggests that a system with a larger $\omega_{0}$ may exhibit clearer signatures of HOVHS physics. We also note that the magic angles we highlight hear appear to scale approximately linearly with the interlayer tunnelling energy, $\theta_{n}^{*} \sim \alpha \omega_{0}^{\null}$.

\end{widetext}


\begin{thebibliography}{70}%
\makeatletter
\providecommand \@ifxundefined [1]{%
 \@ifx{#1\undefined}
}%
\providecommand \@ifnum [1]{%
 \ifnum #1\expandafter \@firstoftwo
 \else \expandafter \@secondoftwo
 \fi
}%
\providecommand \@ifx [1]{%
 \ifx #1\expandafter \@firstoftwo
 \else \expandafter \@secondoftwo
 \fi
}%
\providecommand \natexlab [1]{#1}%
\providecommand \enquote  [1]{``#1''}%
\providecommand \bibnamefont  [1]{#1}%
\providecommand \bibfnamefont [1]{#1}%
\providecommand \citenamefont [1]{#1}%
\providecommand \href@noop [0]{\@secondoftwo}%
\providecommand \href [0]{\begingroup \@sanitize@url \@href}%
\providecommand \@href[1]{\@@startlink{#1}\@@href}%
\providecommand \@@href[1]{\endgroup#1\@@endlink}%
\providecommand \@sanitize@url [0]{\catcode `\\12\catcode `\$12\catcode `\&12\catcode `\#12\catcode `\^12\catcode `\_12\catcode `\%12\relax}%
\providecommand \@@startlink[1]{}%
\providecommand \@@endlink[0]{}%
\providecommand \url  [0]{\begingroup\@sanitize@url \@url }%
\providecommand \@url [1]{\endgroup\@href {#1}{\urlprefix }}%
\providecommand \urlprefix  [0]{URL }%
\providecommand \Eprint [0]{\href }%
\providecommand \doibase [0]{https://doi.org/}%
\providecommand \selectlanguage [0]{\@gobble}%
\providecommand \bibinfo  [0]{\@secondoftwo}%
\providecommand \bibfield  [0]{\@secondoftwo}%
\providecommand \translation [1]{[#1]}%
\providecommand \BibitemOpen [0]{}%
\providecommand \bibitemStop [0]{}%
\providecommand \bibitemNoStop [0]{.\EOS\space}%
\providecommand \EOS [0]{\spacefactor3000\relax}%
\providecommand \BibitemShut  [1]{\csname bibitem#1\endcsname}%
\let\auto@bib@innerbib\@empty
\bibitem [{\citenamefont {Bistritzer}\ and\ \citenamefont {MacDonald}(2011)}]{Bistritzer2011}%
  \BibitemOpen
  \bibfield  {author} {\bibinfo {author} {\bibfnamefont {R.}~\bibnamefont {Bistritzer}}\ and\ \bibinfo {author} {\bibfnamefont {A.~H.}\ \bibnamefont {MacDonald}},\ }\bibfield  {title} {\bibinfo {title} {{Moir\'{e} bands in twisted double-layer graphene}},\ }\href {https://doi.org/10.1073/pnas.1108174108} {\bibfield  {journal} {\bibinfo  {journal} {Proc. Natl. Acad. Sci. U.S.A.}\ }\textbf {\bibinfo {volume} {108}},\ \bibinfo {pages} {12233} (\bibinfo {year} {2011})}\BibitemShut {NoStop}%
\bibitem [{\citenamefont {Classen}\ and\ \citenamefont {Betouras}(2025)}]{Classen2025}%
  \BibitemOpen
  \bibfield  {author} {\bibinfo {author} {\bibfnamefont {L.}~\bibnamefont {Classen}}\ and\ \bibinfo {author} {\bibfnamefont {J.~J.}\ \bibnamefont {Betouras}},\ }\bibfield  {title} {\bibinfo {title} {{High-Order Van Hove Singularities and Their Connection to Flat Bands}},\ }\href {https://doi.org/10.1146/annurev-conmatphys-042924-015000} {\bibfield  {journal} {\bibinfo  {journal} {Annu. Rev. Condens. Matter Phys.}\ }\textbf {\bibinfo {volume} {16}},\ \bibinfo {pages} {229} (\bibinfo {year} {2025})}\BibitemShut {NoStop}%
\bibitem [{\citenamefont {Efremov}\ \emph {et~al.}(2019)\citenamefont {Efremov}, \citenamefont {Shtyk}, \citenamefont {Rost}, \citenamefont {Chamon}, \citenamefont {Mackenzie},\ and\ \citenamefont {Betouras}}]{Efremov2019}%
  \BibitemOpen
  \bibfield  {author} {\bibinfo {author} {\bibfnamefont {D.~V.}\ \bibnamefont {Efremov}}, \bibinfo {author} {\bibfnamefont {A.}~\bibnamefont {Shtyk}}, \bibinfo {author} {\bibfnamefont {A.~W.}\ \bibnamefont {Rost}}, \bibinfo {author} {\bibfnamefont {C.}~\bibnamefont {Chamon}}, \bibinfo {author} {\bibfnamefont {A.~P.}\ \bibnamefont {Mackenzie}},\ and\ \bibinfo {author} {\bibfnamefont {J.~J.}\ \bibnamefont {Betouras}},\ }\bibfield  {title} {\bibinfo {title} {{Multicritical Fermi Surface Topological Transitions}},\ }\href {https://doi.org/10.1103/PhysRevLett.123.207202} {\bibfield  {journal} {\bibinfo  {journal} {Phys. Rev. Lett.}\ }\textbf {\bibinfo {volume} {123}},\ \bibinfo {pages} {207202} (\bibinfo {year} {2019})}\BibitemShut {NoStop}%
\bibitem [{\citenamefont {Chandrasekaran}\ \emph {et~al.}(2020)\citenamefont {Chandrasekaran}, \citenamefont {Shtyk}, \citenamefont {Betouras},\ and\ \citenamefont {Chamon}}]{Chandrasekaran_catastrophe_2020}%
  \BibitemOpen
  \bibfield  {author} {\bibinfo {author} {\bibfnamefont {A.}~\bibnamefont {Chandrasekaran}}, \bibinfo {author} {\bibfnamefont {A.}~\bibnamefont {Shtyk}}, \bibinfo {author} {\bibfnamefont {J.~J.}\ \bibnamefont {Betouras}},\ and\ \bibinfo {author} {\bibfnamefont {C.}~\bibnamefont {Chamon}},\ }\bibfield  {title} {\bibinfo {title} {{Catastrophe theory classification of Fermi surface topological transitions in two dimensions}},\ }\href@noop {} {\bibfield  {journal} {\bibinfo  {journal} {Phys. Rev. Res.}\ }\textbf {\bibinfo {volume} {2}},\ \bibinfo {pages} {013355} (\bibinfo {year} {2020})}\BibitemShut {NoStop}%
\bibitem [{\citenamefont {Yuan}\ \emph {et~al.}(2019)\citenamefont {Yuan}, \citenamefont {Isobe},\ and\ \citenamefont {Fu}}]{Yuan2019}%
  \BibitemOpen
  \bibfield  {author} {\bibinfo {author} {\bibfnamefont {N.~F.~Q.}\ \bibnamefont {Yuan}}, \bibinfo {author} {\bibfnamefont {H.}~\bibnamefont {Isobe}},\ and\ \bibinfo {author} {\bibfnamefont {L.}~\bibnamefont {Fu}},\ }\bibfield  {title} {\bibinfo {title} {{Magic of high-order van Hove singularity}},\ }\href {https://doi.org/10.1038/s41467-019-13670-9} {\bibfield  {journal} {\bibinfo  {journal} {Nat. Commun.}\ }\textbf {\bibinfo {volume} {10}},\ \bibinfo {pages} {5769} (\bibinfo {year} {2019})}\BibitemShut {NoStop}%
\bibitem [{\citenamefont {Kang}\ \emph {et~al.}(2022)\citenamefont {Kang}, \citenamefont {Fang}, \citenamefont {Kim}, \citenamefont {Ortiz}, \citenamefont {Ryu}, \citenamefont {Kim}, \citenamefont {Yoo}, \citenamefont {Sangiovanni}, \citenamefont {Di~Sante}, \citenamefont {Park}, \citenamefont {Jozwiak}, \citenamefont {Bostwick}, \citenamefont {Rotenberg}, \citenamefont {Kaxiras}, \citenamefont {Wilson}, \citenamefont {Park},\ and\ \citenamefont {Comin}}]{Kang2022}%
  \BibitemOpen
  \bibfield  {author} {\bibinfo {author} {\bibfnamefont {M.}~\bibnamefont {Kang}}, \bibinfo {author} {\bibfnamefont {S.}~\bibnamefont {Fang}}, \bibinfo {author} {\bibfnamefont {J.-K.}\ \bibnamefont {Kim}}, \bibinfo {author} {\bibfnamefont {B.~R.}\ \bibnamefont {Ortiz}}, \bibinfo {author} {\bibfnamefont {S.~H.}\ \bibnamefont {Ryu}}, \bibinfo {author} {\bibfnamefont {J.}~\bibnamefont {Kim}}, \bibinfo {author} {\bibfnamefont {J.}~\bibnamefont {Yoo}}, \bibinfo {author} {\bibfnamefont {G.}~\bibnamefont {Sangiovanni}}, \bibinfo {author} {\bibfnamefont {D.}~\bibnamefont {Di~Sante}}, \bibinfo {author} {\bibfnamefont {B.-G.}\ \bibnamefont {Park}}, \bibinfo {author} {\bibfnamefont {C.}~\bibnamefont {Jozwiak}}, \bibinfo {author} {\bibfnamefont {A.}~\bibnamefont {Bostwick}}, \bibinfo {author} {\bibfnamefont {E.}~\bibnamefont {Rotenberg}}, \bibinfo {author} {\bibfnamefont {E.}~\bibnamefont {Kaxiras}}, \bibinfo {author} {\bibfnamefont {S.~D.}\ \bibnamefont {Wilson}}, \bibinfo {author} {\bibfnamefont {J.-H.}\ \bibnamefont
  {Park}},\ and\ \bibinfo {author} {\bibfnamefont {R.}~\bibnamefont {Comin}},\ }\bibfield  {title} {\bibinfo {title} {{Twofold van Hove singularity and origin of charge order in topological kagome superconductor {CsV$_3$Sb$_5$}}},\ }\href {https://doi.org/10.1038/s41567-021-01451-5} {\bibfield  {journal} {\bibinfo  {journal} {Nat. Phys.}\ }\textbf {\bibinfo {volume} {18}},\ \bibinfo {pages} {301} (\bibinfo {year} {2022})}\BibitemShut {NoStop}%
\bibitem [{\citenamefont {Hu}\ \emph {et~al.}(2022)\citenamefont {Hu}, \citenamefont {Wu}, \citenamefont {Ortiz}, \citenamefont {Ju}, \citenamefont {Han}, \citenamefont {Ma}, \citenamefont {Plumb}, \citenamefont {Radovic}, \citenamefont {Thomale}, \citenamefont {Wilson}, \citenamefont {Schnyder},\ and\ \citenamefont {Shi}}]{Hu2022}%
  \BibitemOpen
  \bibfield  {author} {\bibinfo {author} {\bibfnamefont {Y.}~\bibnamefont {Hu}}, \bibinfo {author} {\bibfnamefont {X.}~\bibnamefont {Wu}}, \bibinfo {author} {\bibfnamefont {B.~R.}\ \bibnamefont {Ortiz}}, \bibinfo {author} {\bibfnamefont {S.}~\bibnamefont {Ju}}, \bibinfo {author} {\bibfnamefont {X.}~\bibnamefont {Han}}, \bibinfo {author} {\bibfnamefont {J.}~\bibnamefont {Ma}}, \bibinfo {author} {\bibfnamefont {N.~C.}\ \bibnamefont {Plumb}}, \bibinfo {author} {\bibfnamefont {M.}~\bibnamefont {Radovic}}, \bibinfo {author} {\bibfnamefont {R.}~\bibnamefont {Thomale}}, \bibinfo {author} {\bibfnamefont {S.~D.}\ \bibnamefont {Wilson}}, \bibinfo {author} {\bibfnamefont {A.~P.}\ \bibnamefont {Schnyder}},\ and\ \bibinfo {author} {\bibfnamefont {M.}~\bibnamefont {Shi}},\ }\bibfield  {title} {\bibinfo {title} {{Rich nature of Van Hove singularities in Kagome superconductor {CsV$_3$Sb$_5$}}},\ }\href {https://doi.org/10.1038/s41467-022-29828-x} {\bibfield  {journal} {\bibinfo  {journal} {Nat. Commun.}\ }\textbf {\bibinfo
  {volume} {13}},\ \bibinfo {pages} {2220} (\bibinfo {year} {2022})}\BibitemShut {NoStop}%
\bibitem [{\citenamefont {Cho}\ \emph {et~al.}(2021)\citenamefont {Cho}, \citenamefont {Ma}, \citenamefont {Xia}, \citenamefont {Yang}, \citenamefont {Liu}, \citenamefont {Huang}, \citenamefont {Jiang}, \citenamefont {Lu}, \citenamefont {Liu}, \citenamefont {Liu}, \citenamefont {Li}, \citenamefont {Wang}, \citenamefont {Liu}, \citenamefont {Jia}, \citenamefont {Guo}, \citenamefont {Liu},\ and\ \citenamefont {Shen}}]{Cho2021}%
  \BibitemOpen
  \bibfield  {author} {\bibinfo {author} {\bibfnamefont {S.}~\bibnamefont {Cho}}, \bibinfo {author} {\bibfnamefont {H.}~\bibnamefont {Ma}}, \bibinfo {author} {\bibfnamefont {W.}~\bibnamefont {Xia}}, \bibinfo {author} {\bibfnamefont {Y.}~\bibnamefont {Yang}}, \bibinfo {author} {\bibfnamefont {Z.}~\bibnamefont {Liu}}, \bibinfo {author} {\bibfnamefont {Z.}~\bibnamefont {Huang}}, \bibinfo {author} {\bibfnamefont {Z.}~\bibnamefont {Jiang}}, \bibinfo {author} {\bibfnamefont {X.}~\bibnamefont {Lu}}, \bibinfo {author} {\bibfnamefont {J.}~\bibnamefont {Liu}}, \bibinfo {author} {\bibfnamefont {Z.}~\bibnamefont {Liu}}, \bibinfo {author} {\bibfnamefont {J.}~\bibnamefont {Li}}, \bibinfo {author} {\bibfnamefont {J.}~\bibnamefont {Wang}}, \bibinfo {author} {\bibfnamefont {Y.}~\bibnamefont {Liu}}, \bibinfo {author} {\bibfnamefont {J.}~\bibnamefont {Jia}}, \bibinfo {author} {\bibfnamefont {Y.}~\bibnamefont {Guo}}, \bibinfo {author} {\bibfnamefont {J.}~\bibnamefont {Liu}},\ and\ \bibinfo {author} {\bibfnamefont
  {D.}~\bibnamefont {Shen}},\ }\bibfield  {title} {\bibinfo {title} {{Emergence of New van Hove Singularities in the Charge Density Wave State of a Topological Kagome Metal ${\mathrm{RbV}}_{3}{\mathrm{Sb}}_{5}$}},\ }\href {https://doi.org/10.1103/PhysRevLett.127.236401} {\bibfield  {journal} {\bibinfo  {journal} {Phys. Rev. Lett.}\ }\textbf {\bibinfo {volume} {127}},\ \bibinfo {pages} {236401} (\bibinfo {year} {2021})}\BibitemShut {NoStop}%
\bibitem [{\citenamefont {Chandrasekaran}\ \emph {et~al.}(2024)\citenamefont {Chandrasekaran}, \citenamefont {Rhodes}, \citenamefont {Morales}, \citenamefont {Marques}, \citenamefont {King}, \citenamefont {Wahl},\ and\ \citenamefont {Betouras}}]{Chandrasekaran_NatComms_2024}%
  \BibitemOpen
  \bibfield  {author} {\bibinfo {author} {\bibfnamefont {A.}~\bibnamefont {Chandrasekaran}}, \bibinfo {author} {\bibfnamefont {L.~C.}\ \bibnamefont {Rhodes}}, \bibinfo {author} {\bibfnamefont {E.~A.}\ \bibnamefont {Morales}}, \bibinfo {author} {\bibfnamefont {C.~A.}\ \bibnamefont {Marques}}, \bibinfo {author} {\bibfnamefont {P.~D.~C.}\ \bibnamefont {King}}, \bibinfo {author} {\bibfnamefont {P.}~\bibnamefont {Wahl}},\ and\ \bibinfo {author} {\bibfnamefont {J.~J.}\ \bibnamefont {Betouras}},\ }\bibfield  {title} {\bibinfo {title} {{On the engineering of higher-order Van Hove singularities in two dimensions}},\ }\href {https://doi.org/10.1038/s41467-024-53650-2} {\bibfield  {journal} {\bibinfo  {journal} {Nat. Commun.}\ }\textbf {\bibinfo {volume} {15}},\ \bibinfo {pages} {9521} (\bibinfo {year} {2024})}\BibitemShut {NoStop}%
\bibitem [{\citenamefont {Cao}\ \emph {et~al.}(2018)\citenamefont {Cao}, \citenamefont {Fatemi}, \citenamefont {Fang}, \citenamefont {Watanabe}, \citenamefont {Taniguchi}, \citenamefont {Kaxiras},\ and\ \citenamefont {Jarillo-Herrero}}]{Cao2018}%
  \BibitemOpen
  \bibfield  {author} {\bibinfo {author} {\bibfnamefont {Y.}~\bibnamefont {Cao}}, \bibinfo {author} {\bibfnamefont {V.}~\bibnamefont {Fatemi}}, \bibinfo {author} {\bibfnamefont {S.}~\bibnamefont {Fang}}, \bibinfo {author} {\bibfnamefont {K.}~\bibnamefont {Watanabe}}, \bibinfo {author} {\bibfnamefont {T.}~\bibnamefont {Taniguchi}}, \bibinfo {author} {\bibfnamefont {E.}~\bibnamefont {Kaxiras}},\ and\ \bibinfo {author} {\bibfnamefont {P.}~\bibnamefont {Jarillo-Herrero}},\ }\bibfield  {title} {\bibinfo {title} {{Unconventional superconductivity in magic-angle graphene superlattices}},\ }\href {https://doi.org/10.1038/nature26160} {\bibfield  {journal} {\bibinfo  {journal} {Nature}\ }\textbf {\bibinfo {volume} {556}},\ \bibinfo {pages} {43} (\bibinfo {year} {2018})}\BibitemShut {NoStop}%
\bibitem [{\citenamefont {Song}\ \emph {et~al.}(2019)\citenamefont {Song}, \citenamefont {Wang}, \citenamefont {Shi}, \citenamefont {Li}, \citenamefont {Fang},\ and\ \citenamefont {Bernevig}}]{Song2019}%
  \BibitemOpen
  \bibfield  {author} {\bibinfo {author} {\bibfnamefont {Z.}~\bibnamefont {Song}}, \bibinfo {author} {\bibfnamefont {Z.}~\bibnamefont {Wang}}, \bibinfo {author} {\bibfnamefont {W.}~\bibnamefont {Shi}}, \bibinfo {author} {\bibfnamefont {G.}~\bibnamefont {Li}}, \bibinfo {author} {\bibfnamefont {C.}~\bibnamefont {Fang}},\ and\ \bibinfo {author} {\bibfnamefont {B.~A.}\ \bibnamefont {Bernevig}},\ }\bibfield  {title} {\bibinfo {title} {{All Magic Angles in Twisted Bilayer Graphene are Topological}},\ }\href {https://doi.org/10.1103/PhysRevLett.123.036401} {\bibfield  {journal} {\bibinfo  {journal} {Phys. Rev. Lett.}\ }\textbf {\bibinfo {volume} {123}},\ \bibinfo {pages} {036401} (\bibinfo {year} {2019})}\BibitemShut {NoStop}%
\bibitem [{\citenamefont {Cao}\ \emph {et~al.}(2020)\citenamefont {Cao}, \citenamefont {Chowdhury}, \citenamefont {Rodan-Legrain}, \citenamefont {Rubies-Bigorda}, \citenamefont {Watanabe}, \citenamefont {Taniguchi}, \citenamefont {Senthil},\ and\ \citenamefont {Jarillo-Herrero}}]{Cao2021}%
  \BibitemOpen
  \bibfield  {author} {\bibinfo {author} {\bibfnamefont {Y.}~\bibnamefont {Cao}}, \bibinfo {author} {\bibfnamefont {D.}~\bibnamefont {Chowdhury}}, \bibinfo {author} {\bibfnamefont {D.}~\bibnamefont {Rodan-Legrain}}, \bibinfo {author} {\bibfnamefont {O.}~\bibnamefont {Rubies-Bigorda}}, \bibinfo {author} {\bibfnamefont {K.}~\bibnamefont {Watanabe}}, \bibinfo {author} {\bibfnamefont {T.}~\bibnamefont {Taniguchi}}, \bibinfo {author} {\bibfnamefont {T.}~\bibnamefont {Senthil}},\ and\ \bibinfo {author} {\bibfnamefont {P.}~\bibnamefont {Jarillo-Herrero}},\ }\bibfield  {title} {\bibinfo {title} {{Strange Metal in Magic-Angle Graphene with near Planckian Dissipation}},\ }\href {https://doi.org/10.1103/PhysRevLett.124.076801} {\bibfield  {journal} {\bibinfo  {journal} {Phys. Rev. Lett.}\ }\textbf {\bibinfo {volume} {124}},\ \bibinfo {pages} {076801} (\bibinfo {year} {2020})}\BibitemShut {NoStop}%
\bibitem [{\citenamefont {Park}\ \emph {et~al.}(2022)\citenamefont {Park}, \citenamefont {Cao}, \citenamefont {Xia}, \citenamefont {Sun}, \citenamefont {Watanabe}, \citenamefont {T.},\ and\ \citenamefont {Jarillo-Herrero}}]{Park2022}%
  \BibitemOpen
  \bibfield  {author} {\bibinfo {author} {\bibfnamefont {J.}~\bibnamefont {Park}}, \bibinfo {author} {\bibfnamefont {Y.}~\bibnamefont {Cao}}, \bibinfo {author} {\bibfnamefont {L.-Q.}\ \bibnamefont {Xia}}, \bibinfo {author} {\bibfnamefont {S.}~\bibnamefont {Sun}}, \bibinfo {author} {\bibfnamefont {K.}~\bibnamefont {Watanabe}}, \bibinfo {author} {\bibfnamefont {T.}~\bibnamefont {T.}},\ and\ \bibinfo {author} {\bibfnamefont {P.}~\bibnamefont {Jarillo-Herrero}},\ }\bibfield  {title} {\bibinfo {title} {{Robust superconductivity in magic-angle multilayer graphene family}},\ }\href {https://doi.org/10.1038/s41563-022-01287-1} {\bibfield  {journal} {\bibinfo  {journal} {Nat. Mater.}\ }\textbf {\bibinfo {volume} {21}},\ \bibinfo {pages} {877} (\bibinfo {year} {2022})}\BibitemShut {NoStop}%
\bibitem [{\citenamefont {Ahn}\ \emph {et~al.}(2018)\citenamefont {Ahn}, \citenamefont {Moon}, \citenamefont {Kim}, \citenamefont {Kim}, \citenamefont {Shin}, \citenamefont {Kim}, \citenamefont {Cha}, \citenamefont {Kahng}, \citenamefont {Kim}, \citenamefont {Koshino}, \citenamefont {Son}, \citenamefont {Yang},\ and\ \citenamefont {Ahn}}]{Ahn2018}%
  \BibitemOpen
  \bibfield  {author} {\bibinfo {author} {\bibfnamefont {S.~J.}\ \bibnamefont {Ahn}}, \bibinfo {author} {\bibfnamefont {P.}~\bibnamefont {Moon}}, \bibinfo {author} {\bibfnamefont {T.-H.}\ \bibnamefont {Kim}}, \bibinfo {author} {\bibfnamefont {H.-W.}\ \bibnamefont {Kim}}, \bibinfo {author} {\bibfnamefont {H.-C.}\ \bibnamefont {Shin}}, \bibinfo {author} {\bibfnamefont {E.~H.}\ \bibnamefont {Kim}}, \bibinfo {author} {\bibfnamefont {H.~W.}\ \bibnamefont {Cha}}, \bibinfo {author} {\bibfnamefont {S.-J.}\ \bibnamefont {Kahng}}, \bibinfo {author} {\bibfnamefont {P.}~\bibnamefont {Kim}}, \bibinfo {author} {\bibfnamefont {M.}~\bibnamefont {Koshino}}, \bibinfo {author} {\bibfnamefont {Y.-W.}\ \bibnamefont {Son}}, \bibinfo {author} {\bibfnamefont {C.-W.}\ \bibnamefont {Yang}},\ and\ \bibinfo {author} {\bibfnamefont {J.~R.}\ \bibnamefont {Ahn}},\ }\bibfield  {title} {\bibinfo {title} {{Dirac electrons in a dodecagonal graphene quasicrystal}},\ }\href {https://doi.org/10.1126/science.aar8412} {\bibfield  {journal}
  {\bibinfo  {journal} {Science}\ }\textbf {\bibinfo {volume} {361}},\ \bibinfo {pages} {782} (\bibinfo {year} {2018})}\BibitemShut {NoStop}%
\bibitem [{\citenamefont {Moon}\ \emph {et~al.}(2019)\citenamefont {Moon}, \citenamefont {Koshino},\ and\ \citenamefont {Son}}]{Moon2019}%
  \BibitemOpen
  \bibfield  {author} {\bibinfo {author} {\bibfnamefont {P.}~\bibnamefont {Moon}}, \bibinfo {author} {\bibfnamefont {M.}~\bibnamefont {Koshino}},\ and\ \bibinfo {author} {\bibfnamefont {Y.-W.}\ \bibnamefont {Son}},\ }\bibfield  {title} {\bibinfo {title} {{Quasicrystalline electronic states in ${30}^{\ensuremath{\circ}}$ rotated twisted bilayer graphene}},\ }\href {https://doi.org/10.1103/PhysRevB.99.165430} {\bibfield  {journal} {\bibinfo  {journal} {Phys. Rev. B}\ }\textbf {\bibinfo {volume} {99}},\ \bibinfo {pages} {165430} (\bibinfo {year} {2019})}\BibitemShut {NoStop}%
\bibitem [{\citenamefont {Yu}\ \emph {et~al.}(2019)\citenamefont {Yu}, \citenamefont {Wu}, \citenamefont {Zhan}, \citenamefont {Katsnelson},\ and\ \citenamefont {Yuan}}]{Yu2019}%
  \BibitemOpen
  \bibfield  {author} {\bibinfo {author} {\bibfnamefont {G.}~\bibnamefont {Yu}}, \bibinfo {author} {\bibfnamefont {Z.}~\bibnamefont {Wu}}, \bibinfo {author} {\bibfnamefont {Z.}~\bibnamefont {Zhan}}, \bibinfo {author} {\bibfnamefont {M.~I.}\ \bibnamefont {Katsnelson}},\ and\ \bibinfo {author} {\bibfnamefont {S.}~\bibnamefont {Yuan}},\ }\bibfield  {title} {\bibinfo {title} {{Dodecagonal bilayer graphene quasicrystal and its approximants}},\ }\href {https://doi.org/10.1038/s41524-019-0258-0} {\bibfield  {journal} {\bibinfo  {journal} {npj Comput. Mater.}\ }\textbf {\bibinfo {volume} {5}},\ \bibinfo {pages} {122} (\bibinfo {year} {2019})}\BibitemShut {NoStop}%
\bibitem [{\citenamefont {Pezzini}\ \emph {et~al.}(2020)\citenamefont {Pezzini}, \citenamefont {Mi{\v s}eikis}, \citenamefont {Piccinini}, \citenamefont {Forti}, \citenamefont {Pace}, \citenamefont {Engelke}, \citenamefont {Rossella}, \citenamefont {Watanabe}, \citenamefont {Taniguchi}, \citenamefont {Kim},\ and\ \citenamefont {Coletti}}]{Pezzini2020}%
  \BibitemOpen
  \bibfield  {author} {\bibinfo {author} {\bibfnamefont {S.}~\bibnamefont {Pezzini}}, \bibinfo {author} {\bibfnamefont {V.}~\bibnamefont {Mi{\v s}eikis}}, \bibinfo {author} {\bibfnamefont {G.}~\bibnamefont {Piccinini}}, \bibinfo {author} {\bibfnamefont {S.}~\bibnamefont {Forti}}, \bibinfo {author} {\bibfnamefont {S.}~\bibnamefont {Pace}}, \bibinfo {author} {\bibfnamefont {R.}~\bibnamefont {Engelke}}, \bibinfo {author} {\bibfnamefont {F.}~\bibnamefont {Rossella}}, \bibinfo {author} {\bibfnamefont {K.}~\bibnamefont {Watanabe}}, \bibinfo {author} {\bibfnamefont {T.}~\bibnamefont {Taniguchi}}, \bibinfo {author} {\bibfnamefont {P.}~\bibnamefont {Kim}},\ and\ \bibinfo {author} {\bibfnamefont {C.}~\bibnamefont {Coletti}},\ }\bibfield  {title} {\bibinfo {title} {{30$\,^{\circ}$-Twisted Bilayer Graphene Quasicrystals from Chemical Vapor Deposition}},\ }\href {https://doi.org/10.1021/acs.nanolett.0c00172} {\bibfield  {journal} {\bibinfo  {journal} {Nano Lett.}\ }\textbf {\bibinfo {volume} {20}},\ \bibinfo {pages}
  {3313} (\bibinfo {year} {2020})}\BibitemShut {NoStop}%
\bibitem [{\citenamefont {Li}\ and\ \citenamefont {Koshino}(2019)}]{Li2019}%
  \BibitemOpen
  \bibfield  {author} {\bibinfo {author} {\bibfnamefont {Y.}~\bibnamefont {Li}}\ and\ \bibinfo {author} {\bibfnamefont {M.}~\bibnamefont {Koshino}},\ }\bibfield  {title} {\bibinfo {title} {{Twist-angle dependence of the proximity spin-orbit coupling in graphene on transition-metal dichalcogenides}},\ }\href {https://doi.org/10.1103/PhysRevB.99.075438} {\bibfield  {journal} {\bibinfo  {journal} {Phys. Rev. B}\ }\textbf {\bibinfo {volume} {99}},\ \bibinfo {pages} {075438} (\bibinfo {year} {2019})}\BibitemShut {NoStop}%
\bibitem [{\citenamefont {Sousa}\ \emph {et~al.}(2022)\citenamefont {Sousa}, \citenamefont {Perkins},\ and\ \citenamefont {Ferreira}}]{Sousa2022}%
  \BibitemOpen
  \bibfield  {author} {\bibinfo {author} {\bibfnamefont {F.}~\bibnamefont {Sousa}}, \bibinfo {author} {\bibfnamefont {D.~T.~S.}\ \bibnamefont {Perkins}},\ and\ \bibinfo {author} {\bibfnamefont {A.}~\bibnamefont {Ferreira}},\ }\bibfield  {title} {\bibinfo {title} {{Weak localisation driven by pseudospin-spin entanglement}},\ }\href {https://doi.org/10.1038/s42005-022-01066-z} {\bibfield  {journal} {\bibinfo  {journal} {Commun. Phys.}\ }\textbf {\bibinfo {volume} {5}},\ \bibinfo {pages} {291} (\bibinfo {year} {2022})}\BibitemShut {NoStop}%
\bibitem [{\citenamefont {P\'eterfalvi}\ \emph {et~al.}(2022)\citenamefont {P\'eterfalvi}, \citenamefont {David}, \citenamefont {Rakyta}, \citenamefont {Burkard},\ and\ \citenamefont {Korm\'anyos}}]{Peterfalvi2022}%
  \BibitemOpen
  \bibfield  {author} {\bibinfo {author} {\bibfnamefont {C.}~\bibnamefont {P\'eterfalvi}}, \bibinfo {author} {\bibfnamefont {A.}~\bibnamefont {David}}, \bibinfo {author} {\bibfnamefont {P.}~\bibnamefont {Rakyta}}, \bibinfo {author} {\bibfnamefont {G.}~\bibnamefont {Burkard}},\ and\ \bibinfo {author} {\bibfnamefont {A.}~\bibnamefont {Korm\'anyos}},\ }\bibfield  {title} {\bibinfo {title} {{Quantum interference tuning of spin-orbit coupling in twisted van der Waals trilayers}},\ }\href {https://doi.org/10.1103/PhysRevResearch.4.L022049} {\bibfield  {journal} {\bibinfo  {journal} {Phys. Rev. Res.}\ }\textbf {\bibinfo {volume} {4}},\ \bibinfo {pages} {L022049} (\bibinfo {year} {2022})}\BibitemShut {NoStop}%
\bibitem [{\citenamefont {Veneri}\ \emph {et~al.}(2022)\citenamefont {Veneri}, \citenamefont {Perkins}, \citenamefont {P\'eterfalvi},\ and\ \citenamefont {Ferreira}}]{Veneri2022}%
  \BibitemOpen
  \bibfield  {author} {\bibinfo {author} {\bibfnamefont {A.}~\bibnamefont {Veneri}}, \bibinfo {author} {\bibfnamefont {D.~T.~S.}\ \bibnamefont {Perkins}}, \bibinfo {author} {\bibfnamefont {C.~G.}\ \bibnamefont {P\'eterfalvi}},\ and\ \bibinfo {author} {\bibfnamefont {A.}~\bibnamefont {Ferreira}},\ }\bibfield  {title} {\bibinfo {title} {{Twist angle controlled collinear Edelstein effect in van der Waals heterostructures}},\ }\href {https://doi.org/10.1103/PhysRevB.106.L081406} {\bibfield  {journal} {\bibinfo  {journal} {Phys. Rev. B}\ }\textbf {\bibinfo {volume} {106}},\ \bibinfo {pages} {L081406} (\bibinfo {year} {2022})}\BibitemShut {NoStop}%
\bibitem [{\citenamefont {Perkins}\ \emph {et~al.}(2024)\citenamefont {Perkins}, \citenamefont {Veneri},\ and\ \citenamefont {Ferreira}}]{Perkins2024}%
  \BibitemOpen
  \bibfield  {author} {\bibinfo {author} {\bibfnamefont {D.}~\bibnamefont {Perkins}}, \bibinfo {author} {\bibfnamefont {A.}~\bibnamefont {Veneri}},\ and\ \bibinfo {author} {\bibfnamefont {A.}~\bibnamefont {Ferreira}},\ }\bibfield  {title} {\bibinfo {title} {{Spin Hall effect: Symmetry breaking, twisting, and giant disorder renormalization}},\ }\href {https://doi.org/10.1103/PhysRevB.109.L241404} {\bibfield  {journal} {\bibinfo  {journal} {Phys. Rev. B}\ }\textbf {\bibinfo {volume} {109}},\ \bibinfo {pages} {L241404} (\bibinfo {year} {2024})}\BibitemShut {NoStop}%
\bibitem [{\citenamefont {Wojciechowska}\ and\ \citenamefont {Dyrda{\l}}(2025)}]{Wojciechowska2025}%
  \BibitemOpen
  \bibfield  {author} {\bibinfo {author} {\bibfnamefont {I.}~\bibnamefont {Wojciechowska}}\ and\ \bibinfo {author} {\bibfnamefont {A.}~\bibnamefont {Dyrda{\l}}},\ }\bibfield  {title} {\bibinfo {title} {{Twist tunable spin to charge conversion and valley contrasting effects in graphene on 2D transition metal dichalcogenides}},\ }\href {https://doi.org/10.1038/s41598-025-23786-2} {\bibfield  {journal} {\bibinfo  {journal} {Scientific Reports}\ }\textbf {\bibinfo {volume} {15}},\ \bibinfo {pages} {39156} (\bibinfo {year} {2025})}\BibitemShut {NoStop}%
\bibitem [{\citenamefont {Moon}\ and\ \citenamefont {Koshino}(2014)}]{Moon2014}%
  \BibitemOpen
  \bibfield  {author} {\bibinfo {author} {\bibfnamefont {P.}~\bibnamefont {Moon}}\ and\ \bibinfo {author} {\bibfnamefont {M.}~\bibnamefont {Koshino}},\ }\bibfield  {title} {\bibinfo {title} {{Electronic properties of graphene/hexagonal-boron-nitride moir\'e superlattice}},\ }\href {https://doi.org/10.1103/PhysRevB.90.155406} {\bibfield  {journal} {\bibinfo  {journal} {Phys. Rev. B}\ }\textbf {\bibinfo {volume} {90}},\ \bibinfo {pages} {155406} (\bibinfo {year} {2014})}\BibitemShut {NoStop}%
\bibitem [{\citenamefont {Lieb}(1989)}]{Lieb1989}%
  \BibitemOpen
  \bibfield  {author} {\bibinfo {author} {\bibfnamefont {E.~H.}\ \bibnamefont {Lieb}},\ }\bibfield  {title} {\bibinfo {title} {{Two theorems on the Hubbard model}},\ }\href {https://doi.org/10.1103/PhysRevLett.62.1201} {\bibfield  {journal} {\bibinfo  {journal} {Phys. Rev. Lett.}\ }\textbf {\bibinfo {volume} {62}},\ \bibinfo {pages} {1201} (\bibinfo {year} {1989})}\BibitemShut {NoStop}%
\bibitem [{\citenamefont {Koll{\'a}r}\ \emph {et~al.}(2020)\citenamefont {Koll{\'a}r}, \citenamefont {Fitzpatrick}, \citenamefont {Sarnak},\ and\ \citenamefont {Houck}}]{Kollar2020}%
  \BibitemOpen
  \bibfield  {author} {\bibinfo {author} {\bibfnamefont {A.~J.}\ \bibnamefont {Koll{\'a}r}}, \bibinfo {author} {\bibfnamefont {M.}~\bibnamefont {Fitzpatrick}}, \bibinfo {author} {\bibfnamefont {P.}~\bibnamefont {Sarnak}},\ and\ \bibinfo {author} {\bibfnamefont {A.~A.}\ \bibnamefont {Houck}},\ }\bibfield  {title} {\bibinfo {title} {{Line-Graph Lattices: Euclidean and Non-Euclidean Flat Bands, and Implementations in Circuit Quantum Electrodynamics}},\ }\href {https://doi.org/10.1007/s00220-019-03645-8} {\bibfield  {journal} {\bibinfo  {journal} {Commun. Math. Phys.}\ }\textbf {\bibinfo {volume} {376}},\ \bibinfo {pages} {1909} (\bibinfo {year} {2020})}\BibitemShut {NoStop}%
\bibitem [{\citenamefont {Ma}\ \emph {et~al.}(2020)\citenamefont {Ma}, \citenamefont {Xu}, \citenamefont {Chiu}, \citenamefont {Regnault}, \citenamefont {Houck}, \citenamefont {Song},\ and\ \citenamefont {Bernevig}}]{Ma2020}%
  \BibitemOpen
  \bibfield  {author} {\bibinfo {author} {\bibfnamefont {D.-S.}\ \bibnamefont {Ma}}, \bibinfo {author} {\bibfnamefont {Y.}~\bibnamefont {Xu}}, \bibinfo {author} {\bibfnamefont {C.~S.}\ \bibnamefont {Chiu}}, \bibinfo {author} {\bibfnamefont {N.}~\bibnamefont {Regnault}}, \bibinfo {author} {\bibfnamefont {A.~A.}\ \bibnamefont {Houck}}, \bibinfo {author} {\bibfnamefont {Z.}~\bibnamefont {Song}},\ and\ \bibinfo {author} {\bibfnamefont {B.~A.}\ \bibnamefont {Bernevig}},\ }\bibfield  {title} {\bibinfo {title} {{Spin-Orbit-Induced Topological Flat Bands in Line and Split Graphs of Bipartite Lattices}},\ }\href {https://doi.org/10.1103/PhysRevLett.125.266403} {\bibfield  {journal} {\bibinfo  {journal} {Phys. Rev. Lett.}\ }\textbf {\bibinfo {volume} {125}},\ \bibinfo {pages} {266403} (\bibinfo {year} {2020})}\BibitemShut {NoStop}%
\bibitem [{\citenamefont {Di~Sante}\ \emph {et~al.}(2026)\citenamefont {Di~Sante}, \citenamefont {Neupert}, \citenamefont {Sangiovanni}, \citenamefont {Thomale}, \citenamefont {Comin}, \citenamefont {Checkelsky}, \citenamefont {Zeljkovic},\ and\ \citenamefont {Wilson}}]{Sante2026}%
  \BibitemOpen
  \bibfield  {author} {\bibinfo {author} {\bibfnamefont {D.}~\bibnamefont {Di~Sante}}, \bibinfo {author} {\bibfnamefont {T.}~\bibnamefont {Neupert}}, \bibinfo {author} {\bibfnamefont {G.}~\bibnamefont {Sangiovanni}}, \bibinfo {author} {\bibfnamefont {R.}~\bibnamefont {Thomale}}, \bibinfo {author} {\bibfnamefont {R.}~\bibnamefont {Comin}}, \bibinfo {author} {\bibfnamefont {J.~G.}\ \bibnamefont {Checkelsky}}, \bibinfo {author} {\bibfnamefont {I.}~\bibnamefont {Zeljkovic}},\ and\ \bibinfo {author} {\bibfnamefont {S.~D.}\ \bibnamefont {Wilson}},\ }\bibfield  {title} {\bibinfo {title} {{Kagome metals}},\ }\href {https://doi.org/10.1103/1g9n-wm38} {\bibfield  {journal} {\bibinfo  {journal} {Rev. Mod. Phys.}\ }\textbf {\bibinfo {volume} {98}},\ \bibinfo {pages} {015002} (\bibinfo {year} {2026})}\BibitemShut {NoStop}%
\bibitem [{\citenamefont {Regnault}\ \emph {et~al.}(2022)\citenamefont {Regnault}, \citenamefont {Xu}, \citenamefont {Li}, \citenamefont {Ma}, \citenamefont {Jovanovic}, \citenamefont {Yazdani}, \citenamefont {Parkin}, \citenamefont {Felser}, \citenamefont {Schoop}, \citenamefont {Ong}, \citenamefont {Cava}, \citenamefont {Elcoro}, \citenamefont {Song},\ and\ \citenamefont {Bernevig}}]{Regnault2022}%
  \BibitemOpen
  \bibfield  {author} {\bibinfo {author} {\bibfnamefont {N.}~\bibnamefont {Regnault}}, \bibinfo {author} {\bibfnamefont {Y.}~\bibnamefont {Xu}}, \bibinfo {author} {\bibfnamefont {M.-R.}\ \bibnamefont {Li}}, \bibinfo {author} {\bibfnamefont {D.-S.}\ \bibnamefont {Ma}}, \bibinfo {author} {\bibfnamefont {M.}~\bibnamefont {Jovanovic}}, \bibinfo {author} {\bibfnamefont {A.}~\bibnamefont {Yazdani}}, \bibinfo {author} {\bibfnamefont {S.~S.~P.}\ \bibnamefont {Parkin}}, \bibinfo {author} {\bibfnamefont {C.}~\bibnamefont {Felser}}, \bibinfo {author} {\bibfnamefont {L.~M.}\ \bibnamefont {Schoop}}, \bibinfo {author} {\bibfnamefont {N.~P.}\ \bibnamefont {Ong}}, \bibinfo {author} {\bibfnamefont {R.~J.}\ \bibnamefont {Cava}}, \bibinfo {author} {\bibfnamefont {L.}~\bibnamefont {Elcoro}}, \bibinfo {author} {\bibfnamefont {Z.-D.}\ \bibnamefont {Song}},\ and\ \bibinfo {author} {\bibfnamefont {B.~A.}\ \bibnamefont {Bernevig}},\ }\bibfield  {title} {\bibinfo {title} {{Catalogue of flat-band stoichiometric materials}},\ }\href
  {https://doi.org/10.1038/s41586-022-04519-1} {\bibfield  {journal} {\bibinfo  {journal} {Nature}\ }\textbf {\bibinfo {volume} {603}},\ \bibinfo {pages} {824} (\bibinfo {year} {2022})}\BibitemShut {NoStop}%
\bibitem [{\citenamefont {Perkins}\ \emph {et~al.}(2025)\citenamefont {Perkins}, \citenamefont {Chandrasekaran},\ and\ \citenamefont {Betouras}}]{Perkins2025b}%
  \BibitemOpen
  \bibfield  {author} {\bibinfo {author} {\bibfnamefont {D.~T.~S.}\ \bibnamefont {Perkins}}, \bibinfo {author} {\bibfnamefont {A.}~\bibnamefont {Chandrasekaran}},\ and\ \bibinfo {author} {\bibfnamefont {J.~J.}\ \bibnamefont {Betouras}},\ }\bibfield  {title} {\bibinfo {title} {{Designing topological high-order Van Hove singularities: Twisted bilayer kagome}},\ }\href {https://doi.org/10.1103/8y2v-kx2w} {\bibfield  {journal} {\bibinfo  {journal} {Phys. Rev. B}\ }\textbf {\bibinfo {volume} {112}},\ \bibinfo {pages} {235134} (\bibinfo {year} {2025})}\BibitemShut {NoStop}%
\bibitem [{\citenamefont {Wang}\ \emph {et~al.}(2025)\citenamefont {Wang}, \citenamefont {Pullasseri},\ and\ \citenamefont {Santos}}]{Wang2025}%
  \BibitemOpen
  \bibfield  {author} {\bibinfo {author} {\bibfnamefont {E.}~\bibnamefont {Wang}}, \bibinfo {author} {\bibfnamefont {L.}~\bibnamefont {Pullasseri}},\ and\ \bibinfo {author} {\bibfnamefont {L.~H.}\ \bibnamefont {Santos}},\ }\bibfield  {title} {\bibinfo {title} {{Higher-order Van Hove singularities in kagome topological bands}},\ }\href {https://doi.org/10.1103/PhysRevB.111.075114} {\bibfield  {journal} {\bibinfo  {journal} {Phys. Rev. B}\ }\textbf {\bibinfo {volume} {111}},\ \bibinfo {pages} {075114} (\bibinfo {year} {2025})}\BibitemShut {NoStop}%
\bibitem [{\citenamefont {Hung}\ \emph {et~al.}(2026)\citenamefont {Hung}, \citenamefont {Zhou},\ and\ \citenamefont {Bansil}}]{Hung2026arxiv}%
  \BibitemOpen
  \bibfield  {author} {\bibinfo {author} {\bibfnamefont {Y.-C.}\ \bibnamefont {Hung}}, \bibinfo {author} {\bibfnamefont {X.}~\bibnamefont {Zhou}},\ and\ \bibinfo {author} {\bibfnamefont {A.}~\bibnamefont {Bansil}},\ }\href@noop {} {\bibinfo {title} {{Twist-Induced Quantum Geometry Reconfiguration in Moir\'{e} Flat Bands}}} (\bibinfo {year} {2026}),\ \bibinfo {note} {\hyperlink{https://doi.org/10.48550/arXiv.2603.20849}{arXiv:2603.20849 [cond-mat.mes-hall]}}\BibitemShut {NoStop}%
\bibitem [{\citenamefont {Lopes~dos Santos}\ \emph {et~al.}(2007)\citenamefont {Lopes~dos Santos}, \citenamefont {Peres},\ and\ \citenamefont {Castro~Neto}}]{Lopes_dos_Santos2007}%
  \BibitemOpen
  \bibfield  {author} {\bibinfo {author} {\bibfnamefont {J.~M.~B.}\ \bibnamefont {Lopes~dos Santos}}, \bibinfo {author} {\bibfnamefont {N.~M.~R.}\ \bibnamefont {Peres}},\ and\ \bibinfo {author} {\bibfnamefont {A.~H.}\ \bibnamefont {Castro~Neto}},\ }\bibfield  {title} {\bibinfo {title} {{Graphene Bilayer with a Twist: Electronic Structure}},\ }\href {https://doi.org/10.1103/PhysRevLett.99.256802} {\bibfield  {journal} {\bibinfo  {journal} {Phys. Rev. Lett.}\ }\textbf {\bibinfo {volume} {99}},\ \bibinfo {pages} {256802} (\bibinfo {year} {2007})}\BibitemShut {NoStop}%
\bibitem [{\citenamefont {Lopes~dos Santos}\ \emph {et~al.}(2012)\citenamefont {Lopes~dos Santos}, \citenamefont {Peres},\ and\ \citenamefont {Castro~Neto}}]{Lopes_dos_Santos2012}%
  \BibitemOpen
  \bibfield  {author} {\bibinfo {author} {\bibfnamefont {J.~M.~B.}\ \bibnamefont {Lopes~dos Santos}}, \bibinfo {author} {\bibfnamefont {N.~M.~R.}\ \bibnamefont {Peres}},\ and\ \bibinfo {author} {\bibfnamefont {A.~H.}\ \bibnamefont {Castro~Neto}},\ }\bibfield  {title} {\bibinfo {title} {Continuum model of the twisted graphene bilayer},\ }\href {https://doi.org/10.1103/PhysRevB.86.155449} {\bibfield  {journal} {\bibinfo  {journal} {Phys. Rev. B}\ }\textbf {\bibinfo {volume} {86}},\ \bibinfo {pages} {155449} (\bibinfo {year} {2012})}\BibitemShut {NoStop}%
\bibitem [{\citenamefont {Koshino}\ \emph {et~al.}(2018)\citenamefont {Koshino}, \citenamefont {Yuan}, \citenamefont {Koretsune}, \citenamefont {Ochi}, \citenamefont {Kuroki},\ and\ \citenamefont {Fu}}]{Koshino2018}%
  \BibitemOpen
  \bibfield  {author} {\bibinfo {author} {\bibfnamefont {M.}~\bibnamefont {Koshino}}, \bibinfo {author} {\bibfnamefont {N.~F.~Q.}\ \bibnamefont {Yuan}}, \bibinfo {author} {\bibfnamefont {T.}~\bibnamefont {Koretsune}}, \bibinfo {author} {\bibfnamefont {M.}~\bibnamefont {Ochi}}, \bibinfo {author} {\bibfnamefont {K.}~\bibnamefont {Kuroki}},\ and\ \bibinfo {author} {\bibfnamefont {L.}~\bibnamefont {Fu}},\ }\bibfield  {title} {\bibinfo {title} {{Maximally Localized Wannier Orbitals and the Extended Hubbard Model for Twisted Bilayer Graphene}},\ }\href {https://doi.org/10.1103/PhysRevX.8.031087} {\bibfield  {journal} {\bibinfo  {journal} {Phys. Rev. X}\ }\textbf {\bibinfo {volume} {8}},\ \bibinfo {pages} {031087} (\bibinfo {year} {2018})}\BibitemShut {NoStop}%
\bibitem [{\citenamefont {Catarina}\ \emph {et~al.}(2019)\citenamefont {Catarina}, \citenamefont {Amorim}, \citenamefont {Castro}, \citenamefont {Lopes}, \citenamefont {Lopes},\ and\ \citenamefont {Peres}}]{Catarina2019}%
  \BibitemOpen
  \bibfield  {author} {\bibinfo {author} {\bibfnamefont {G.}~\bibnamefont {Catarina}}, \bibinfo {author} {\bibfnamefont {B.}~\bibnamefont {Amorim}}, \bibinfo {author} {\bibfnamefont {E.~V.}\ \bibnamefont {Castro}}, \bibinfo {author} {\bibfnamefont {J.~M. V.~P.}\ \bibnamefont {Lopes}}, \bibinfo {author} {\bibfnamefont {J.~M. V.~P.}\ \bibnamefont {Lopes}},\ and\ \bibinfo {author} {\bibfnamefont {N.}~\bibnamefont {Peres}},\ }\bibinfo {title} {{Twisted Bilayer Graphene: Low-Energy Physics, Electronic and Optical Properties}},\ in\ \href {https://doi.org/https://doi.org/10.1002/9781119468455.ch44} {\emph {\bibinfo {booktitle} {Handbook of Graphene Set}}}\ (\bibinfo  {publisher} {John Wiley \& Sons, Ltd},\ \bibinfo {year} {2019})\ Chap.~\bibinfo {chapter} {6}, pp.\ \bibinfo {pages} {177--231}\BibitemShut {NoStop}%
\bibitem [{\citenamefont {Koshino}(2019)}]{Koshino2019}%
  \BibitemOpen
  \bibfield  {author} {\bibinfo {author} {\bibfnamefont {M.}~\bibnamefont {Koshino}},\ }\bibfield  {title} {\bibinfo {title} {{Band structure and topological properties of twisted double bilayer graphene}},\ }\href {https://doi.org/10.1103/PhysRevB.99.235406} {\bibfield  {journal} {\bibinfo  {journal} {Phys. Rev. B}\ }\textbf {\bibinfo {volume} {99}},\ \bibinfo {pages} {235406} (\bibinfo {year} {2019})}\BibitemShut {NoStop}%
\bibitem [{\citenamefont {Wu}\ \emph {et~al.}(2019)\citenamefont {Wu}, \citenamefont {Lovorn}, \citenamefont {Tutuc}, \citenamefont {Martin},\ and\ \citenamefont {MacDonald}}]{Wu2019}%
  \BibitemOpen
  \bibfield  {author} {\bibinfo {author} {\bibfnamefont {F.}~\bibnamefont {Wu}}, \bibinfo {author} {\bibfnamefont {T.}~\bibnamefont {Lovorn}}, \bibinfo {author} {\bibfnamefont {E.}~\bibnamefont {Tutuc}}, \bibinfo {author} {\bibfnamefont {I.}~\bibnamefont {Martin}},\ and\ \bibinfo {author} {\bibfnamefont {A.~H.}\ \bibnamefont {MacDonald}},\ }\bibfield  {title} {\bibinfo {title} {{Topological Insulators in Twisted Transition Metal Dichalcogenide Homobilayers}},\ }\href {https://doi.org/10.1103/PhysRevLett.122.086402} {\bibfield  {journal} {\bibinfo  {journal} {Phys. Rev. Lett.}\ }\textbf {\bibinfo {volume} {122}},\ \bibinfo {pages} {086402} (\bibinfo {year} {2019})}\BibitemShut {NoStop}%
\bibitem [{\citenamefont {Bernevig}\ \emph {et~al.}(2021)\citenamefont {Bernevig}, \citenamefont {Song}, \citenamefont {Regnault},\ and\ \citenamefont {Lian}}]{Bernevig2021}%
  \BibitemOpen
  \bibfield  {author} {\bibinfo {author} {\bibfnamefont {B.~A.}\ \bibnamefont {Bernevig}}, \bibinfo {author} {\bibfnamefont {Z.-D.}\ \bibnamefont {Song}}, \bibinfo {author} {\bibfnamefont {N.}~\bibnamefont {Regnault}},\ and\ \bibinfo {author} {\bibfnamefont {B.}~\bibnamefont {Lian}},\ }\bibfield  {title} {\bibinfo {title} {{Twisted bilayer graphene. I. Matrix elements, approximations, perturbation theory, and a $k\ifmmode\cdot\else\textperiodcentered\fi{}p$ two-band model}},\ }\href {https://doi.org/10.1103/PhysRevB.103.205411} {\bibfield  {journal} {\bibinfo  {journal} {Phys. Rev. B}\ }\textbf {\bibinfo {volume} {103}},\ \bibinfo {pages} {205411} (\bibinfo {year} {2021})}\BibitemShut {NoStop}%
\bibitem [{\citenamefont {Scheer}\ \emph {et~al.}(2022)\citenamefont {Scheer}, \citenamefont {Gu},\ and\ \citenamefont {Lian}}]{Scheer2022}%
  \BibitemOpen
  \bibfield  {author} {\bibinfo {author} {\bibfnamefont {M.~G.}\ \bibnamefont {Scheer}}, \bibinfo {author} {\bibfnamefont {K.}~\bibnamefont {Gu}},\ and\ \bibinfo {author} {\bibfnamefont {B.}~\bibnamefont {Lian}},\ }\bibfield  {title} {\bibinfo {title} {{Magic angles in twisted bilayer graphene near commensuration: Towards a hypermagic regime}},\ }\href {https://doi.org/10.1103/PhysRevB.106.115418} {\bibfield  {journal} {\bibinfo  {journal} {Phys. Rev. B}\ }\textbf {\bibinfo {volume} {106}},\ \bibinfo {pages} {115418} (\bibinfo {year} {2022})}\BibitemShut {NoStop}%
\bibitem [{\citenamefont {Ma}\ \emph {et~al.}(2024)\citenamefont {Ma}, \citenamefont {Chen}, \citenamefont {Yu},\ and\ \citenamefont {Luo}}]{Ma2024}%
  \BibitemOpen
  \bibfield  {author} {\bibinfo {author} {\bibfnamefont {D.}~\bibnamefont {Ma}}, \bibinfo {author} {\bibfnamefont {Y.-G.}\ \bibnamefont {Chen}}, \bibinfo {author} {\bibfnamefont {Y.}~\bibnamefont {Yu}},\ and\ \bibinfo {author} {\bibfnamefont {X.}~\bibnamefont {Luo}},\ }\bibfield  {title} {\bibinfo {title} {{Moir\'e semiconductors on the twisted bilayer dice lattice}},\ }\href {https://doi.org/10.1103/PhysRevB.109.155159} {\bibfield  {journal} {\bibinfo  {journal} {Phys. Rev. B}\ }\textbf {\bibinfo {volume} {109}},\ \bibinfo {pages} {155159} (\bibinfo {year} {2024})}\BibitemShut {NoStop}%
\bibitem [{\citenamefont {Zhou}\ \emph {et~al.}(2024)\citenamefont {Zhou}, \citenamefont {Hung}, \citenamefont {Wang},\ and\ \citenamefont {Bansil}}]{Zhou2024}%
  \BibitemOpen
  \bibfield  {author} {\bibinfo {author} {\bibfnamefont {X.}~\bibnamefont {Zhou}}, \bibinfo {author} {\bibfnamefont {Y.-C.}\ \bibnamefont {Hung}}, \bibinfo {author} {\bibfnamefont {B.}~\bibnamefont {Wang}},\ and\ \bibinfo {author} {\bibfnamefont {A.}~\bibnamefont {Bansil}},\ }\bibfield  {title} {\bibinfo {title} {{Generation of Isolated Flat Bands with Tunable Numbers through Moir\'e Engineering}},\ }\href {https://doi.org/10.1103/PhysRevLett.133.236401} {\bibfield  {journal} {\bibinfo  {journal} {Phys. Rev. Lett.}\ }\textbf {\bibinfo {volume} {133}},\ \bibinfo {pages} {236401} (\bibinfo {year} {2024})}\BibitemShut {NoStop}%
\bibitem [{\citenamefont {C{\u a}lug{\u a}ru}\ \emph {et~al.}(2025)\citenamefont {C{\u a}lug{\u a}ru}, \citenamefont {Jiang}, \citenamefont {Hu}, \citenamefont {Pi}, \citenamefont {Yu}, \citenamefont {Vergniory}, \citenamefont {Shan}, \citenamefont {Felser}, \citenamefont {Schoop}, \citenamefont {Efetov}, \citenamefont {Mak},\ and\ \citenamefont {Bernevig}}]{Calugaru2025}%
  \BibitemOpen
  \bibfield  {author} {\bibinfo {author} {\bibfnamefont {D.}~\bibnamefont {C{\u a}lug{\u a}ru}}, \bibinfo {author} {\bibfnamefont {Y.}~\bibnamefont {Jiang}}, \bibinfo {author} {\bibfnamefont {H.}~\bibnamefont {Hu}}, \bibinfo {author} {\bibfnamefont {H.}~\bibnamefont {Pi}}, \bibinfo {author} {\bibfnamefont {J.}~\bibnamefont {Yu}}, \bibinfo {author} {\bibfnamefont {M.~G.}\ \bibnamefont {Vergniory}}, \bibinfo {author} {\bibfnamefont {J.}~\bibnamefont {Shan}}, \bibinfo {author} {\bibfnamefont {C.}~\bibnamefont {Felser}}, \bibinfo {author} {\bibfnamefont {L.~M.}\ \bibnamefont {Schoop}}, \bibinfo {author} {\bibfnamefont {D.~K.}\ \bibnamefont {Efetov}}, \bibinfo {author} {\bibfnamefont {K.~F.}\ \bibnamefont {Mak}},\ and\ \bibinfo {author} {\bibfnamefont {B.~A.}\ \bibnamefont {Bernevig}},\ }\bibfield  {title} {\bibinfo {title} {{Moir{\'e} materials based on M-point twisting}},\ }\href {https://doi.org/10.1038/s41586-025-09187-5} {\bibfield  {journal} {\bibinfo  {journal} {Nature}\ }\textbf {\bibinfo {volume} {643}},\
  \bibinfo {pages} {376} (\bibinfo {year} {2025})}\BibitemShut {NoStop}%
\bibitem [{SMr()}]{SMref}%
  \BibitemOpen
  \href@noop {} {}\bibinfo {note} {See the Supplemental Material where we provide a summary of the Hamiltonian and electronic structure for the monolayer kagome system, show how the BM model can be generalized to TBK, demonstrate approximate particle-hole symmetry, provide additional band structures and sublattice projection plots, discuss the rapid changes in $v_{F}^{*}$ that only appear when particle-hole symmetry is broken, show the robustness of magic angles to the choice of $\omega_{0}$, and provide details about the number of shells included in the $q$-lattice as the twist angle is varied. We include Refs. \cite{Slater1954,Dai2014,Uchida2014,Li2017,Ye2018,Lima2019,Shi2022,Luo2023,Wang2023,Xia2025,Zhou2025} as additional references here.}\BibitemShut {Stop}%
\bibitem [{\citenamefont {Mihalyuk}\ \emph {et~al.}(2022)\citenamefont {Mihalyuk}, \citenamefont {Gruznev}, \citenamefont {Bondarenko}, \citenamefont {Tupchuya}, \citenamefont {Vekovshinin}, \citenamefont {Eremeev}, \citenamefont {Zotov},\ and\ \citenamefont {Saranin}}]{Mihalyuk2022}%
  \BibitemOpen
  \bibfield  {author} {\bibinfo {author} {\bibfnamefont {A.~N.}\ \bibnamefont {Mihalyuk}}, \bibinfo {author} {\bibfnamefont {D.~V.}\ \bibnamefont {Gruznev}}, \bibinfo {author} {\bibfnamefont {L.~V.}\ \bibnamefont {Bondarenko}}, \bibinfo {author} {\bibfnamefont {A.~Y.}\ \bibnamefont {Tupchuya}}, \bibinfo {author} {\bibfnamefont {Y.~E.}\ \bibnamefont {Vekovshinin}}, \bibinfo {author} {\bibfnamefont {S.~V.}\ \bibnamefont {Eremeev}}, \bibinfo {author} {\bibfnamefont {A.~V.}\ \bibnamefont {Zotov}},\ and\ \bibinfo {author} {\bibfnamefont {A.~A.}\ \bibnamefont {Saranin}},\ }\bibfield  {title} {\bibinfo {title} {{A 2D heavy fermion CePb${}_{3}$ kagome material on silicon: emergence of unique spin polarized states for spintronics}},\ }\href {https://doi.org/10.1039/D2NR04280K} {\bibfield  {journal} {\bibinfo  {journal} {Nanoscale}\ }\textbf {\bibinfo {volume} {14}},\ \bibinfo {pages} {14732} (\bibinfo {year} {2022})}\BibitemShut {NoStop}%
\bibitem [{\citenamefont {Vekovshinin}\ \emph {et~al.}(2024)\citenamefont {Vekovshinin}, \citenamefont {Bondarenko}, \citenamefont {Tupchaya}, \citenamefont {Mihalyuk}, \citenamefont {Denisov}, \citenamefont {Gruznev}, \citenamefont {Zotov},\ and\ \citenamefont {Saranin}}]{Vekovshinin2024}%
  \BibitemOpen
  \bibfield  {author} {\bibinfo {author} {\bibfnamefont {Y.~E.}\ \bibnamefont {Vekovshinin}}, \bibinfo {author} {\bibfnamefont {L.~V.}\ \bibnamefont {Bondarenko}}, \bibinfo {author} {\bibfnamefont {A.~Y.}\ \bibnamefont {Tupchaya}}, \bibinfo {author} {\bibfnamefont {A.~N.}\ \bibnamefont {Mihalyuk}}, \bibinfo {author} {\bibfnamefont {N.~V.}\ \bibnamefont {Denisov}}, \bibinfo {author} {\bibfnamefont {D.~V.}\ \bibnamefont {Gruznev}}, \bibinfo {author} {\bibfnamefont {A.~V.}\ \bibnamefont {Zotov}},\ and\ \bibinfo {author} {\bibfnamefont {A.~A.}\ \bibnamefont {Saranin}},\ }\bibfield  {title} {\bibinfo {title} {{Lifshitz Transition in a Single-Atom-Thick Gd${}_{x}$Yb${}_{1–x}$Pb${}_{3}$ Kagome Compound on Si(111)}},\ }\href {https://doi.org/10.1021/acs.nanolett.4c02420} {\bibfield  {journal} {\bibinfo  {journal} {Nano Lett.}\ }\textbf {\bibinfo {volume} {24}},\ \bibinfo {pages} {9931} (\bibinfo {year} {2024})}\BibitemShut {NoStop}%
\bibitem [{\citenamefont {Denisov}\ \emph {et~al.}(2025)\citenamefont {Denisov}, \citenamefont {Bondarenko}, \citenamefont {Vekovshinin}, \citenamefont {Mihalyuk}, \citenamefont {Eremeev}, \citenamefont {Gruznev}, \citenamefont {Zotov},\ and\ \citenamefont {Saranin}}]{Denisov2025}%
  \BibitemOpen
  \bibfield  {author} {\bibinfo {author} {\bibfnamefont {N.~V.}\ \bibnamefont {Denisov}}, \bibinfo {author} {\bibfnamefont {L.~V.}\ \bibnamefont {Bondarenko}}, \bibinfo {author} {\bibfnamefont {Y.~E.}\ \bibnamefont {Vekovshinin}}, \bibinfo {author} {\bibfnamefont {A.~N.}\ \bibnamefont {Mihalyuk}}, \bibinfo {author} {\bibfnamefont {S.~V.}\ \bibnamefont {Eremeev}}, \bibinfo {author} {\bibfnamefont {D.~V.}\ \bibnamefont {Gruznev}}, \bibinfo {author} {\bibfnamefont {A.~V.}\ \bibnamefont {Zotov}},\ and\ \bibinfo {author} {\bibfnamefont {A.~A.}\ \bibnamefont {Saranin}},\ }\bibfield  {title} {\bibinfo {title} {{Magnetotransport properties of a single-atom-thick GdPb${}_{3}$ kagome compound on Si(111)}},\ }\href {https://doi.org/10.1039/D4TC04410J} {\bibfield  {journal} {\bibinfo  {journal} {J. Mater. Chem. C}\ }\textbf {\bibinfo {volume} {13}},\ \bibinfo {pages} {7219} (\bibinfo {year} {2025})}\BibitemShut {NoStop}%
\bibitem [{\citenamefont {Vekovshinin}\ \emph {et~al.}(2025)\citenamefont {Vekovshinin}, \citenamefont {Bondarenko}, \citenamefont {Tupchaya}, \citenamefont {Utas}, \citenamefont {Wang}, \citenamefont {Mihalyuk}, \citenamefont {Gruznev}, \citenamefont {Zotov},\ and\ \citenamefont {Saranin}}]{Vekovshinin2025}%
  \BibitemOpen
  \bibfield  {author} {\bibinfo {author} {\bibfnamefont {Y.~E.}\ \bibnamefont {Vekovshinin}}, \bibinfo {author} {\bibfnamefont {L.~V.}\ \bibnamefont {Bondarenko}}, \bibinfo {author} {\bibfnamefont {A.~Y.}\ \bibnamefont {Tupchaya}}, \bibinfo {author} {\bibfnamefont {T.~V.}\ \bibnamefont {Utas}}, \bibinfo {author} {\bibfnamefont {E.}~\bibnamefont {Wang}}, \bibinfo {author} {\bibfnamefont {A.~N.}\ \bibnamefont {Mihalyuk}}, \bibinfo {author} {\bibfnamefont {D.~V.}\ \bibnamefont {Gruznev}}, \bibinfo {author} {\bibfnamefont {A.~V.}\ \bibnamefont {Zotov}},\ and\ \bibinfo {author} {\bibfnamefont {A.~A.}\ \bibnamefont {Saranin}},\ }\bibfield  {title} {\bibinfo {title} {{High-Order Van Hove Singularities in Atomically Thin Kagome Metal LaTl${}_{3}$}},\ }\href {https://doi.org/10.1021/acsnano.5c11205} {\bibfield  {journal} {\bibinfo  {journal} {ACS Nano}\ }\textbf {\bibinfo {volume} {19}},\ \bibinfo {pages} {36510} (\bibinfo {year} {2025})}\BibitemShut {NoStop}%
\bibitem [{\citenamefont {Denisov}\ \emph {et~al.}(2026)\citenamefont {Denisov}, \citenamefont {Bondarenko}, \citenamefont {Tupchaya}, \citenamefont {Vekovshinin}, \citenamefont {Kotlyar}, \citenamefont {Utas}, \citenamefont {Burkovskaya}, \citenamefont {Brykin}, \citenamefont {Mihalyuk}, \citenamefont {Eremeev}, \citenamefont {Gruznev}, \citenamefont {Zotov},\ and\ \citenamefont {Saranin}}]{Denisov2026}%
  \BibitemOpen
  \bibfield  {author} {\bibinfo {author} {\bibfnamefont {N.~V.}\ \bibnamefont {Denisov}}, \bibinfo {author} {\bibfnamefont {L.~V.}\ \bibnamefont {Bondarenko}}, \bibinfo {author} {\bibfnamefont {A.~Y.}\ \bibnamefont {Tupchaya}}, \bibinfo {author} {\bibfnamefont {Y.~E.}\ \bibnamefont {Vekovshinin}}, \bibinfo {author} {\bibfnamefont {V.~G.}\ \bibnamefont {Kotlyar}}, \bibinfo {author} {\bibfnamefont {T.~V.}\ \bibnamefont {Utas}}, \bibinfo {author} {\bibfnamefont {P.~V.}\ \bibnamefont {Burkovskaya}}, \bibinfo {author} {\bibfnamefont {L.~O.}\ \bibnamefont {Brykin}}, \bibinfo {author} {\bibfnamefont {A.~N.}\ \bibnamefont {Mihalyuk}}, \bibinfo {author} {\bibfnamefont {S.~V.}\ \bibnamefont {Eremeev}}, \bibinfo {author} {\bibfnamefont {D.~V.}\ \bibnamefont {Gruznev}}, \bibinfo {author} {\bibfnamefont {A.~V.}\ \bibnamefont {Zotov}},\ and\ \bibinfo {author} {\bibfnamefont {A.~A.}\ \bibnamefont {Saranin}},\ }\bibfield  {title} {\bibinfo {title} {{Magnetism in the single-atom-thick ${\mathrm{EuPb}}_{3}$ kagome compound on
  Si(111) studied using in situ transport and magnetotransport measurements}},\ }\href {https://doi.org/10.1103/hyjd-f1rw} {\bibfield  {journal} {\bibinfo  {journal} {Phys. Rev. B}\ }\textbf {\bibinfo {volume} {113}},\ \bibinfo {pages} {035415} (\bibinfo {year} {2026})}\BibitemShut {NoStop}%
\bibitem [{\citenamefont {Perkins}(2025)}]{Perkins2025a}%
  \BibitemOpen
  \bibfield  {author} {\bibinfo {author} {\bibfnamefont {D.~T.~S.}\ \bibnamefont {Perkins}},\ }\bibfield  {title} {\bibinfo {title} {{Symmetry preservation in commensurate twisted bilayers}},\ }\href {https://doi.org/10.1103/r61z-jhrn} {\bibfield  {journal} {\bibinfo  {journal} {Phys. Rev. B}\ }\textbf {\bibinfo {volume} {112}},\ \bibinfo {pages} {035410} (\bibinfo {year} {2025})}\BibitemShut {NoStop}%
\bibitem [{\citenamefont {Sheffer}\ \emph {et~al.}(2023)\citenamefont {Sheffer}, \citenamefont {Queiroz},\ and\ \citenamefont {Stern}}]{Sheffer2023}%
  \BibitemOpen
  \bibfield  {author} {\bibinfo {author} {\bibfnamefont {Y.}~\bibnamefont {Sheffer}}, \bibinfo {author} {\bibfnamefont {R.}~\bibnamefont {Queiroz}},\ and\ \bibinfo {author} {\bibfnamefont {A.}~\bibnamefont {Stern}},\ }\bibfield  {title} {\bibinfo {title} {{Symmetries as the Guiding Principle for Flattening Bands of Dirac Fermions}},\ }\href {https://doi.org/10.1103/PhysRevX.13.021012} {\bibfield  {journal} {\bibinfo  {journal} {Phys. Rev. X}\ }\textbf {\bibinfo {volume} {13}},\ \bibinfo {pages} {021012} (\bibinfo {year} {2023})}\BibitemShut {NoStop}%
\bibitem [{\citenamefont {Fu}\ and\ \citenamefont {Kane}(2006)}]{Fu2006}%
  \BibitemOpen
  \bibfield  {author} {\bibinfo {author} {\bibfnamefont {L.}~\bibnamefont {Fu}}\ and\ \bibinfo {author} {\bibfnamefont {C.~L.}\ \bibnamefont {Kane}},\ }\bibfield  {title} {\bibinfo {title} {Time reversal polarization and a ${Z}_{2}$ adiabatic spin pump},\ }\href {https://doi.org/10.1103/PhysRevB.74.195312} {\bibfield  {journal} {\bibinfo  {journal} {Phys. Rev. B}\ }\textbf {\bibinfo {volume} {74}},\ \bibinfo {pages} {195312} (\bibinfo {year} {2006})}\BibitemShut {NoStop}%
\bibitem [{\citenamefont {Song}\ \emph {et~al.}(2021)\citenamefont {Song}, \citenamefont {Lian}, \citenamefont {Regnault},\ and\ \citenamefont {Bernevig}}]{Song2021}%
  \BibitemOpen
  \bibfield  {author} {\bibinfo {author} {\bibfnamefont {Z.-D.}\ \bibnamefont {Song}}, \bibinfo {author} {\bibfnamefont {B.}~\bibnamefont {Lian}}, \bibinfo {author} {\bibfnamefont {N.}~\bibnamefont {Regnault}},\ and\ \bibinfo {author} {\bibfnamefont {B.~A.}\ \bibnamefont {Bernevig}},\ }\bibfield  {title} {\bibinfo {title} {Twisted bilayer graphene. ii. stable symmetry anomaly},\ }\href {https://doi.org/10.1103/PhysRevB.103.205412} {\bibfield  {journal} {\bibinfo  {journal} {Phys. Rev. B}\ }\textbf {\bibinfo {volume} {103}},\ \bibinfo {pages} {205412} (\bibinfo {year} {2021})}\BibitemShut {NoStop}%
\bibitem [{\citenamefont {Kiesel}\ and\ \citenamefont {Thomale}(2012)}]{Kiesel2012}%
  \BibitemOpen
  \bibfield  {author} {\bibinfo {author} {\bibfnamefont {M.~L.}\ \bibnamefont {Kiesel}}\ and\ \bibinfo {author} {\bibfnamefont {R.}~\bibnamefont {Thomale}},\ }\bibfield  {title} {\bibinfo {title} {{Sublattice interference in the kagome Hubbard model}},\ }\href {https://doi.org/10.1103/PhysRevB.86.121105} {\bibfield  {journal} {\bibinfo  {journal} {Phys. Rev. B}\ }\textbf {\bibinfo {volume} {86}},\ \bibinfo {pages} {121105} (\bibinfo {year} {2012})}\BibitemShut {NoStop}%
\bibitem [{\citenamefont {Nag}\ \emph {et~al.}(2024)\citenamefont {Nag}, \citenamefont {Batabyal}, \citenamefont {Ingham}, \citenamefont {Morali}, \citenamefont {Tan}, \citenamefont {Koo}, \citenamefont {Consiglio}, \citenamefont {Liu}, \citenamefont {Avraham}, \citenamefont {Queiroz}, \citenamefont {Thomale}, \citenamefont {Yan}, \citenamefont {Felser},\ and\ \citenamefont {Beidenkopf}}]{Nag2024arxiv}%
  \BibitemOpen
  \bibfield  {author} {\bibinfo {author} {\bibfnamefont {P.~K.}\ \bibnamefont {Nag}}, \bibinfo {author} {\bibfnamefont {R.}~\bibnamefont {Batabyal}}, \bibinfo {author} {\bibfnamefont {J.}~\bibnamefont {Ingham}}, \bibinfo {author} {\bibfnamefont {N.}~\bibnamefont {Morali}}, \bibinfo {author} {\bibfnamefont {H.}~\bibnamefont {Tan}}, \bibinfo {author} {\bibfnamefont {J.}~\bibnamefont {Koo}}, \bibinfo {author} {\bibfnamefont {A.}~\bibnamefont {Consiglio}}, \bibinfo {author} {\bibfnamefont {E.}~\bibnamefont {Liu}}, \bibinfo {author} {\bibfnamefont {N.}~\bibnamefont {Avraham}}, \bibinfo {author} {\bibfnamefont {R.}~\bibnamefont {Queiroz}}, \bibinfo {author} {\bibfnamefont {R.}~\bibnamefont {Thomale}}, \bibinfo {author} {\bibfnamefont {B.}~\bibnamefont {Yan}}, \bibinfo {author} {\bibfnamefont {C.}~\bibnamefont {Felser}},\ and\ \bibinfo {author} {\bibfnamefont {H.}~\bibnamefont {Beidenkopf}},\ }\href@noop {} {\bibinfo {title} {{Pomeranchuk Instability Induced by an Emergent Higher-Order van {Hove} Singularity on the
  Distorted Kagome Surface of {Co${}_{3}$Sn${}_{2}$S${}_{2}$}}}} (\bibinfo {year} {2024}),\ \bibinfo {note} {\hyperlink{https://doi.org/10.48550/arXiv.2410.01994}{arXiv:2410.01994 [cond-mat.str-el]}}\BibitemShut {NoStop}%
\bibitem [{\citenamefont {Beck}\ \emph {et~al.}(2026)\citenamefont {Beck}, \citenamefont {Bodky}, \citenamefont {D\"urrnagel}, \citenamefont {Thomale}, \citenamefont {Ingham}, \citenamefont {Klebl},\ and\ \citenamefont {Hohmann}}]{Beck2026}%
  \BibitemOpen
  \bibfield  {author} {\bibinfo {author} {\bibfnamefont {J.}~\bibnamefont {Beck}}, \bibinfo {author} {\bibfnamefont {J.}~\bibnamefont {Bodky}}, \bibinfo {author} {\bibfnamefont {M.}~\bibnamefont {D\"urrnagel}}, \bibinfo {author} {\bibfnamefont {R.}~\bibnamefont {Thomale}}, \bibinfo {author} {\bibfnamefont {J.}~\bibnamefont {Ingham}}, \bibinfo {author} {\bibfnamefont {L.}~\bibnamefont {Klebl}},\ and\ \bibinfo {author} {\bibfnamefont {H.}~\bibnamefont {Hohmann}},\ }\bibfield  {title} {\bibinfo {title} {{Kekul\'e Order from Diffuse Nesting near Higher-Order Van Hove Points}},\ }\href {https://doi.org/10.1103/gwc6-1vv1} {\bibfield  {journal} {\bibinfo  {journal} {Phys. Rev. Lett.}\ }\textbf {\bibinfo {volume} {136}},\ \bibinfo {pages} {106503} (\bibinfo {year} {2026})}\BibitemShut {NoStop}%
\bibitem [{\citenamefont {Rao}\ \emph {et~al.}(2023)\citenamefont {Rao}, \citenamefont {Kang}, \citenamefont {Xue}, \citenamefont {Ye}, \citenamefont {Feng}, \citenamefont {Watanabe}, \citenamefont {Taniguchi}, \citenamefont {Wang}, \citenamefont {Liu},\ and\ \citenamefont {Ki}}]{Rao2023}%
  \BibitemOpen
  \bibfield  {author} {\bibinfo {author} {\bibfnamefont {Q.}~\bibnamefont {Rao}}, \bibinfo {author} {\bibfnamefont {W.-H.}\ \bibnamefont {Kang}}, \bibinfo {author} {\bibfnamefont {H.}~\bibnamefont {Xue}}, \bibinfo {author} {\bibfnamefont {Z.}~\bibnamefont {Ye}}, \bibinfo {author} {\bibfnamefont {X.}~\bibnamefont {Feng}}, \bibinfo {author} {\bibfnamefont {K.}~\bibnamefont {Watanabe}}, \bibinfo {author} {\bibfnamefont {T.}~\bibnamefont {Taniguchi}}, \bibinfo {author} {\bibfnamefont {N.}~\bibnamefont {Wang}}, \bibinfo {author} {\bibfnamefont {M.-H.}\ \bibnamefont {Liu}},\ and\ \bibinfo {author} {\bibfnamefont {D.-K.}\ \bibnamefont {Ki}},\ }\bibfield  {title} {\bibinfo {title} {{Ballistic transport spectroscopy of spin-orbit-coupled bands in monolayer graphene on WSe${}_{2}$}},\ }\href {https://doi.org/10.1038/s41467-023-41826-1} {\bibfield  {journal} {\bibinfo  {journal} {Nat. Commun.}\ }\textbf {\bibinfo {volume} {14}},\ \bibinfo {pages} {6124} (\bibinfo {year} {2023})}\BibitemShut {NoStop}%
\bibitem [{\citenamefont {Sun}\ \emph {et~al.}(2023)\citenamefont {Sun}, \citenamefont {Rademaker}, \citenamefont {Mauro}, \citenamefont {Scarfato}, \citenamefont {P{\'a}sztor}, \citenamefont {Guti{\'e}rrez-Lezama}, \citenamefont {Wang}, \citenamefont {Martinez-Castro}, \citenamefont {Morpurgo},\ and\ \citenamefont {Renner}}]{Sun2023}%
  \BibitemOpen
  \bibfield  {author} {\bibinfo {author} {\bibfnamefont {L.}~\bibnamefont {Sun}}, \bibinfo {author} {\bibfnamefont {L.}~\bibnamefont {Rademaker}}, \bibinfo {author} {\bibfnamefont {D.}~\bibnamefont {Mauro}}, \bibinfo {author} {\bibfnamefont {A.}~\bibnamefont {Scarfato}}, \bibinfo {author} {\bibfnamefont {{\'A}.}~\bibnamefont {P{\'a}sztor}}, \bibinfo {author} {\bibfnamefont {I.}~\bibnamefont {Guti{\'e}rrez-Lezama}}, \bibinfo {author} {\bibfnamefont {Z.}~\bibnamefont {Wang}}, \bibinfo {author} {\bibfnamefont {J.}~\bibnamefont {Martinez-Castro}}, \bibinfo {author} {\bibfnamefont {A.~F.}\ \bibnamefont {Morpurgo}},\ and\ \bibinfo {author} {\bibfnamefont {C.}~\bibnamefont {Renner}},\ }\bibfield  {title} {\bibinfo {title} {Determining spin-orbit coupling in graphene by quasiparticle interference imaging},\ }\href {https://doi.org/10.1038/s41467-023-39453-x} {\bibfield  {journal} {\bibinfo  {journal} {Nat. Commun.}\ }\textbf {\bibinfo {volume} {14}},\ \bibinfo {pages} {3771} (\bibinfo {year} {2023})}\BibitemShut
  {NoStop}%
\bibitem [{\citenamefont {Offidani}\ and\ \citenamefont {Ferreira}(2018)}]{Offidani2018}%
  \BibitemOpen
  \bibfield  {author} {\bibinfo {author} {\bibfnamefont {M.}~\bibnamefont {Offidani}}\ and\ \bibinfo {author} {\bibfnamefont {A.}~\bibnamefont {Ferreira}},\ }\bibfield  {title} {\bibinfo {title} {Anomalous hall effect in 2d dirac materials},\ }\href {https://doi.org/10.1103/PhysRevLett.121.126802} {\bibfield  {journal} {\bibinfo  {journal} {Phys. Rev. Lett.}\ }\textbf {\bibinfo {volume} {121}},\ \bibinfo {pages} {126802} (\bibinfo {year} {2018})}\BibitemShut {NoStop}%
\bibitem [{\citenamefont {Slater}\ and\ \citenamefont {Koster}(1954)}]{Slater1954}%
  \BibitemOpen
  \bibfield  {author} {\bibinfo {author} {\bibfnamefont {J.~C.}\ \bibnamefont {Slater}}\ and\ \bibinfo {author} {\bibfnamefont {G.~F.}\ \bibnamefont {Koster}},\ }\bibfield  {title} {\bibinfo {title} {{Simplified LCAO Method for the Periodic Potential Problem}},\ }\href {https://doi.org/10.1103/PhysRev.94.1498} {\bibfield  {journal} {\bibinfo  {journal} {Phys. Rev.}\ }\textbf {\bibinfo {volume} {94}},\ \bibinfo {pages} {1498} (\bibinfo {year} {1954})}\BibitemShut {NoStop}%
\bibitem [{\citenamefont {Dai}\ and\ \citenamefont {Zeng}(2014)}]{Dai2014}%
  \BibitemOpen
  \bibfield  {author} {\bibinfo {author} {\bibfnamefont {J.}~\bibnamefont {Dai}}\ and\ \bibinfo {author} {\bibfnamefont {X.~C.}\ \bibnamefont {Zeng}},\ }\bibfield  {title} {\bibinfo {title} {Bilayer phosphorene: Effect of stacking order on bandgap and its potential applications in thin-film solar cells},\ }\href {https://doi.org/10.1021/jz500409m} {\bibfield  {journal} {\bibinfo  {journal} {J. Phys. Chem. Lett.}\ }\textbf {\bibinfo {volume} {5}},\ \bibinfo {pages} {1289} (\bibinfo {year} {2014})}\BibitemShut {NoStop}%
\bibitem [{\citenamefont {Uchida}\ \emph {et~al.}(2014)\citenamefont {Uchida}, \citenamefont {Furuya}, \citenamefont {Iwata},\ and\ \citenamefont {Oshiyama}}]{Uchida2014}%
  \BibitemOpen
  \bibfield  {author} {\bibinfo {author} {\bibfnamefont {K.}~\bibnamefont {Uchida}}, \bibinfo {author} {\bibfnamefont {S.}~\bibnamefont {Furuya}}, \bibinfo {author} {\bibfnamefont {J.-I.}\ \bibnamefont {Iwata}},\ and\ \bibinfo {author} {\bibfnamefont {A.}~\bibnamefont {Oshiyama}},\ }\bibfield  {title} {\bibinfo {title} {Atomic corrugation and electron localization due to moir\'e patterns in twisted bilayer graphenes},\ }\href {https://doi.org/10.1103/PhysRevB.90.155451} {\bibfield  {journal} {\bibinfo  {journal} {Phys. Rev. B}\ }\textbf {\bibinfo {volume} {90}},\ \bibinfo {pages} {155451} (\bibinfo {year} {2014})}\BibitemShut {NoStop}%
\bibitem [{\citenamefont {Li}\ \emph {et~al.}(2017)\citenamefont {Li}, \citenamefont {Moldovan}, \citenamefont {Xu},\ and\ \citenamefont {Peeters}}]{Li2017}%
  \BibitemOpen
  \bibfield  {author} {\bibinfo {author} {\bibfnamefont {L.~L.}\ \bibnamefont {Li}}, \bibinfo {author} {\bibfnamefont {D.}~\bibnamefont {Moldovan}}, \bibinfo {author} {\bibfnamefont {W.}~\bibnamefont {Xu}},\ and\ \bibinfo {author} {\bibfnamefont {F.~M.}\ \bibnamefont {Peeters}},\ }\bibfield  {title} {\bibinfo {title} {Electronic properties of bilayer phosphorene quantum dots in the presence of perpendicular electric and magnetic fields},\ }\href {https://doi.org/10.1103/PhysRevB.96.155425} {\bibfield  {journal} {\bibinfo  {journal} {Phys. Rev. B}\ }\textbf {\bibinfo {volume} {96}},\ \bibinfo {pages} {155425} (\bibinfo {year} {2017})}\BibitemShut {NoStop}%
\bibitem [{\citenamefont {Ye}\ \emph {et~al.}(2018)\citenamefont {Ye}, \citenamefont {Kang}, \citenamefont {Liu}, \citenamefont {von Cube}, \citenamefont {Wicker}, \citenamefont {Suzuki}, \citenamefont {Jozwiak}, \citenamefont {Bostwick}, \citenamefont {Rotenberg}, \citenamefont {Bell}, \citenamefont {Fu}, \citenamefont {Comin},\ and\ \citenamefont {Checkelsky}}]{Ye2018}%
  \BibitemOpen
  \bibfield  {author} {\bibinfo {author} {\bibfnamefont {L.}~\bibnamefont {Ye}}, \bibinfo {author} {\bibfnamefont {M.}~\bibnamefont {Kang}}, \bibinfo {author} {\bibfnamefont {J.}~\bibnamefont {Liu}}, \bibinfo {author} {\bibfnamefont {F.}~\bibnamefont {von Cube}}, \bibinfo {author} {\bibfnamefont {C.~R.}\ \bibnamefont {Wicker}}, \bibinfo {author} {\bibfnamefont {T.}~\bibnamefont {Suzuki}}, \bibinfo {author} {\bibfnamefont {C.}~\bibnamefont {Jozwiak}}, \bibinfo {author} {\bibfnamefont {A.}~\bibnamefont {Bostwick}}, \bibinfo {author} {\bibfnamefont {E.}~\bibnamefont {Rotenberg}}, \bibinfo {author} {\bibfnamefont {D.~C.}\ \bibnamefont {Bell}}, \bibinfo {author} {\bibfnamefont {L.}~\bibnamefont {Fu}}, \bibinfo {author} {\bibfnamefont {R.}~\bibnamefont {Comin}},\ and\ \bibinfo {author} {\bibfnamefont {J.~G.}\ \bibnamefont {Checkelsky}},\ }\bibfield  {title} {\bibinfo {title} {{Massive Dirac fermions in a ferromagnetic kagome metal}},\ }\href {https://doi.org/10.1038/nature25987} {\bibfield  {journal} {\bibinfo
  {journal} {Nature}\ }\textbf {\bibinfo {volume} {555}},\ \bibinfo {pages} {638} (\bibinfo {year} {2018})}\BibitemShut {NoStop}%
\bibitem [{\citenamefont {Crasto~de Lima}\ \emph {et~al.}(2019)\citenamefont {Crasto~de Lima}, \citenamefont {Miwa},\ and\ \citenamefont {Su\'arez~Morell}}]{Lima2019}%
  \BibitemOpen
  \bibfield  {author} {\bibinfo {author} {\bibfnamefont {F.}~\bibnamefont {Crasto~de Lima}}, \bibinfo {author} {\bibfnamefont {R.~H.}\ \bibnamefont {Miwa}},\ and\ \bibinfo {author} {\bibfnamefont {E.}~\bibnamefont {Su\'arez~Morell}},\ }\bibfield  {title} {\bibinfo {title} {{Double flat bands in kagome twisted bilayers}},\ }\href {https://doi.org/10.1103/PhysRevB.100.155421} {\bibfield  {journal} {\bibinfo  {journal} {Phys. Rev. B}\ }\textbf {\bibinfo {volume} {100}},\ \bibinfo {pages} {155421} (\bibinfo {year} {2019})}\BibitemShut {NoStop}%
\bibitem [{\citenamefont {Shi}\ \emph {et~al.}(2022)\citenamefont {Shi}, \citenamefont {Shih}, \citenamefont {Rhodes}, \citenamefont {Kim}, \citenamefont {Barmak}, \citenamefont {Watanabe}, \citenamefont {Taniguchi}, \citenamefont {Papi{\'c}}, \citenamefont {Abanin}, \citenamefont {Hone},\ and\ \citenamefont {Dean}}]{Shi2022}%
  \BibitemOpen
  \bibfield  {author} {\bibinfo {author} {\bibfnamefont {Q.}~\bibnamefont {Shi}}, \bibinfo {author} {\bibfnamefont {E.-M.}\ \bibnamefont {Shih}}, \bibinfo {author} {\bibfnamefont {D.}~\bibnamefont {Rhodes}}, \bibinfo {author} {\bibfnamefont {B.}~\bibnamefont {Kim}}, \bibinfo {author} {\bibfnamefont {K.}~\bibnamefont {Barmak}}, \bibinfo {author} {\bibfnamefont {K.}~\bibnamefont {Watanabe}}, \bibinfo {author} {\bibfnamefont {T.}~\bibnamefont {Taniguchi}}, \bibinfo {author} {\bibfnamefont {Z.}~\bibnamefont {Papi{\'c}}}, \bibinfo {author} {\bibfnamefont {D.~A.}\ \bibnamefont {Abanin}}, \bibinfo {author} {\bibfnamefont {J.}~\bibnamefont {Hone}},\ and\ \bibinfo {author} {\bibfnamefont {C.~R.}\ \bibnamefont {Dean}},\ }\bibfield  {title} {\bibinfo {title} {{Bilayer WSe${}_{2}$ as a natural platform for interlayer exciton condensates in the strong coupling limit}},\ }\href {https://doi.org/10.1038/s41565-022-01104-5} {\bibfield  {journal} {\bibinfo  {journal} {Nat. Nanotechnol.}\ }\textbf {\bibinfo {volume} {17}},\
  \bibinfo {pages} {577} (\bibinfo {year} {2022})}\BibitemShut {NoStop}%
\bibitem [{\citenamefont {Luo}\ \emph {et~al.}(2023)\citenamefont {Luo}, \citenamefont {Han}, \citenamefont {Liu}, \citenamefont {Chen}, \citenamefont {Huang}, \citenamefont {Huai}, \citenamefont {Li}, \citenamefont {Wang}, \citenamefont {Shen}, \citenamefont {Ding}, \citenamefont {Li}, \citenamefont {Peng}, \citenamefont {Wei}, \citenamefont {Miao}, \citenamefont {Sun}, \citenamefont {Ou}, \citenamefont {Xiang}, \citenamefont {Hashimoto}, \citenamefont {Lu}, \citenamefont {Yao}, \citenamefont {Yang}, \citenamefont {Chen}, \citenamefont {Gao}, \citenamefont {Qiao}, \citenamefont {Wang},\ and\ \citenamefont {He}}]{Luo2023}%
  \BibitemOpen
  \bibfield  {author} {\bibinfo {author} {\bibfnamefont {Y.}~\bibnamefont {Luo}}, \bibinfo {author} {\bibfnamefont {Y.}~\bibnamefont {Han}}, \bibinfo {author} {\bibfnamefont {J.}~\bibnamefont {Liu}}, \bibinfo {author} {\bibfnamefont {H.}~\bibnamefont {Chen}}, \bibinfo {author} {\bibfnamefont {Z.}~\bibnamefont {Huang}}, \bibinfo {author} {\bibfnamefont {L.}~\bibnamefont {Huai}}, \bibinfo {author} {\bibfnamefont {H.}~\bibnamefont {Li}}, \bibinfo {author} {\bibfnamefont {B.}~\bibnamefont {Wang}}, \bibinfo {author} {\bibfnamefont {J.}~\bibnamefont {Shen}}, \bibinfo {author} {\bibfnamefont {S.}~\bibnamefont {Ding}}, \bibinfo {author} {\bibfnamefont {Z.}~\bibnamefont {Li}}, \bibinfo {author} {\bibfnamefont {S.}~\bibnamefont {Peng}}, \bibinfo {author} {\bibfnamefont {Z.}~\bibnamefont {Wei}}, \bibinfo {author} {\bibfnamefont {Y.}~\bibnamefont {Miao}}, \bibinfo {author} {\bibfnamefont {X.}~\bibnamefont {Sun}}, \bibinfo {author} {\bibfnamefont {Z.}~\bibnamefont {Ou}}, \bibinfo {author} {\bibfnamefont {Z.}~\bibnamefont
  {Xiang}}, \bibinfo {author} {\bibfnamefont {M.}~\bibnamefont {Hashimoto}}, \bibinfo {author} {\bibfnamefont {D.}~\bibnamefont {Lu}}, \bibinfo {author} {\bibfnamefont {Y.}~\bibnamefont {Yao}}, \bibinfo {author} {\bibfnamefont {H.}~\bibnamefont {Yang}}, \bibinfo {author} {\bibfnamefont {X.}~\bibnamefont {Chen}}, \bibinfo {author} {\bibfnamefont {H.-J.}\ \bibnamefont {Gao}}, \bibinfo {author} {\bibfnamefont {Z.}~\bibnamefont {Qiao}}, \bibinfo {author} {\bibfnamefont {Z.}~\bibnamefont {Wang}},\ and\ \bibinfo {author} {\bibfnamefont {J.}~\bibnamefont {He}},\ }\bibfield  {title} {\bibinfo {title} {{A unique van Hove singularity in kagome superconductor CsV${}_{3-x}$Ta${}_{x}$Sb${}_{5}$ with enhanced superconductivity}},\ }\href {https://doi.org/10.1038/s41467-023-39500-7} {\bibfield  {journal} {\bibinfo  {journal} {Nat. Commun.}\ }\textbf {\bibinfo {volume} {14}},\ \bibinfo {pages} {3819} (\bibinfo {year} {2023})}\BibitemShut {NoStop}%
\bibitem [{\citenamefont {Wang}\ \emph {et~al.}(2023)\citenamefont {Wang}, \citenamefont {Wu}, \citenamefont {McCandless}, \citenamefont {Chan},\ and\ \citenamefont {Ali}}]{Wang2023}%
  \BibitemOpen
  \bibfield  {author} {\bibinfo {author} {\bibfnamefont {Y.}~\bibnamefont {Wang}}, \bibinfo {author} {\bibfnamefont {H.}~\bibnamefont {Wu}}, \bibinfo {author} {\bibfnamefont {G.~T.}\ \bibnamefont {McCandless}}, \bibinfo {author} {\bibfnamefont {J.~Y.}\ \bibnamefont {Chan}},\ and\ \bibinfo {author} {\bibfnamefont {M.~N.}\ \bibnamefont {Ali}},\ }\bibfield  {title} {\bibinfo {title} {Quantum states and intertwining phases in kagome materials},\ }\href {https://doi.org/10.1038/s42254-023-00635-7} {\bibfield  {journal} {\bibinfo  {journal} {Nat. Rev. Phys.}\ }\textbf {\bibinfo {volume} {5}},\ \bibinfo {pages} {635} (\bibinfo {year} {2023})}\BibitemShut {NoStop}%
\bibitem [{\citenamefont {Xia}\ \emph {et~al.}(2025)\citenamefont {Xia}, \citenamefont {Han}, \citenamefont {Watanabe}, \citenamefont {Taniguchi}, \citenamefont {Shan},\ and\ \citenamefont {Mak}}]{Xia2025}%
  \BibitemOpen
  \bibfield  {author} {\bibinfo {author} {\bibfnamefont {Y.}~\bibnamefont {Xia}}, \bibinfo {author} {\bibfnamefont {Z.}~\bibnamefont {Han}}, \bibinfo {author} {\bibfnamefont {K.}~\bibnamefont {Watanabe}}, \bibinfo {author} {\bibfnamefont {T.}~\bibnamefont {Taniguchi}}, \bibinfo {author} {\bibfnamefont {J.}~\bibnamefont {Shan}},\ and\ \bibinfo {author} {\bibfnamefont {K.~F.}\ \bibnamefont {Mak}},\ }\bibfield  {title} {\bibinfo {title} {{Superconductivity in twisted bilayer WSe${}_{2}$}},\ }\href {https://doi.org/10.1038/s41586-024-08116-2} {\bibfield  {journal} {\bibinfo  {journal} {Nature}\ }\textbf {\bibinfo {volume} {637}},\ \bibinfo {pages} {833} (\bibinfo {year} {2025})}\BibitemShut {NoStop}%
\bibitem [{\citenamefont {Zhou}\ \emph {et~al.}(2025)\citenamefont {Zhou}, \citenamefont {Wang},\ and\ \citenamefont {Chen}}]{Zhou2025}%
  \BibitemOpen
  \bibfield  {author} {\bibinfo {author} {\bibfnamefont {X.}~\bibnamefont {Zhou}}, \bibinfo {author} {\bibfnamefont {Z.}~\bibnamefont {Wang}},\ and\ \bibinfo {author} {\bibfnamefont {H.}~\bibnamefont {Chen}},\ }\bibfield  {title} {\bibinfo {title} {Topological aspects of dirac fermions in a kagom\'e lattice},\ }\href {https://doi.org/10.1103/5qs7-bdsh} {\bibfield  {journal} {\bibinfo  {journal} {Phys. Rev. B}\ }\textbf {\bibinfo {volume} {112}},\ \bibinfo {pages} {075117} (\bibinfo {year} {2025})}\BibitemShut {NoStop}%
\end{thebibliography}
\end{document}